\documentclass[a4paper,11pt]{article}
\pdfoutput=1 

\usepackage{jcappub} 
                     
\usepackage[T1]{fontenc} 
\usepackage{graphicx}
\usepackage{amsmath}
\usepackage{amssymb}
\usepackage{amsfonts}
\usepackage{bm}
\usepackage{resizegather}
\usepackage{caption}
\usepackage{bm,color}
\definecolor{green}{rgb}{0.3,0.7,0.3}
\definecolor{purple}{rgb}{0.58,0.0,0.83}
\definecolor{dt}{rgb}{0.65, 0.4, 0}
\usepackage{ulem}
\usepackage{cancel}
\usepackage[export]{adjustbox}

\title{Dynamical Systems Analysis in Post-Friedmann Parametrizations of Modified Theories of Gravity}

\author[a]{Abril Su\'arez,}
\author[b]{Ana Avilez,}
\author[a]{David Tamayo,}
\author[c]{Tula Bernal,}
\author[d]{and Jorge L. Cervantes-Cota,}

\affiliation[a]{Depto. de F\'isica, Centro de Investigaci\'on y de Estudios Avanzados del IPN, A.P. 14-740, 07000, Ciudad de M\'exico, M\'exico}
\affiliation[b]{Facultad de Ciencias F\'isico-Matem\'aticas, Ciudad Universitaria, Benem\'erita Universidad Aut\'onoma de Puebla, Av. San Claudio SN, Col. San Manuel, Puebla, M\'exico.}
\affiliation[c]{Universidad Aut\'onoma Chapingo, km.~38.5
Carretera M\'exico-Texcoco, 56230, Texcoco, Estado de M\'exico,	M\'exico}
\affiliation[d]{Depto.~de F\'isica, Instituto Nacional de Investigaciones N\'ucleares, Col.~Escand\'on, Apdo. Postal 18-1027, 11801, Ciudad de M\'exico, M\'exico}

\emailAdd{asuarez@fis.cinvestav.mx}
\emailAdd{aavilez@fcfm.buap.mx}
\emailAdd{dtamayo@fis.cinvestav.mx}
\emailAdd{ac13341@chapingo.mx}
\emailAdd{jorge.cervantes@inin.gob.mx}

\abstract{We carry out a dynamical analysis of first order perturbations for Cold Dark Matter, $\Lambda$ Cold Dark Matter, and a couple of Modified Gravity models using the Parametrized Post-Friedmann formalism. We use normalized variables to set the proper dynamical system of equations through which we make the analysis in order to shed some light on the dynamics of such perturbations inside these models. For Modified Gravity models, we use the scale-independent and -dependent parametrizations, in particular, two $f(R)$ and two Chameleon-like models are considered within the quasi-static approximation. Given the employed formalism, we found that the critical points and stability features of the dynamical systems for Modified Gravity models are the same as those found in the standard $\Lambda$ Cold Dark Matter model. However, the behavior around the critical points suffers important modifications in some specific cases. We explicitly find that signatures of these Modified Gravity models mainly arise on the velocity perturbations, while the density contrast and the curvature potentials turn out to be less sensitive to the parametrization taken into consideration. We also provide a percentage estimation of the extent of modification in the perturbations in the Modified Gravity models considered in comparison to the standard $\Lambda$ Cold Dark Matter model along the expansion history and for a couple of wavenumbers.}

\begin{document}
\maketitle
\flushbottom
\section{Introduction}
Different measurements suggest that our Universe is currently experiencing a phase of accelerated expansion \cite{riess2,perlmutter2,betoule}. For more than a decade, cosmological data of the Cosmic Microwave Background (CMB) anisotropies \citep{komatsu} and supernovae (SN) Ia \citep{sullivan}, plus the mapping of galaxies \cite{reid}, have established that the Universe is prevailed by a new form of energy, called dark energy (DE), that is responsible of such expansion \citep{sahni, peebles,padmanabhan}, making at present around 68\% of the total matter-energy density \citep{planck}. Together with the other ingredients (mainly, dark matter and baryons), DE can fit fairly well the current observational data. The presence of a cosmological constant $\Lambda$ in the Einstein field equations, which is based on the standard model of cosmology $\Lambda$ Cold Dark Matter ($\Lambda$CDM), is the simplest way to explain such accelerated expansion.
However, it still faces severe objections, such as the coincidence and cosmological constant problems \citep{peebles}, and even though it looks as the most straightforward explanation, the presence of such a small, but different from zero constant, is one of the most exciting questions in today's physics.

Such a challenging subject shows an imbalance in the Friedmann equations, and the physics community has addressed such problem either by proposing new sources of matter or by modifying (or extending) Einstein's general relativistic description of gravity. In the framework of standard cosmology, the first one is addressed through DE, while the second one suggests the introduction of some modifications in the gravity part of the equations, known as Modified Gravity (MG) theories \cite{copeland1,amendola2}.

Einstein's General Relativity (GR) is a well founded theory and though there have been different theoretical reasons to expand it, the case is that GR has passed all tests of experimental gravitation, from very small to Solar System scales \citep[see for example][]{Will,Berti}. However, cosmological tests of GR are still an open challenge. Currently, the dynamics of the cosmological background and the properties of the perturbed Universe demand to be tested in consistency to the measurements of the Hubble parameter \citep{moresco}, Baryon Acoustic Oscillations (BAO) distances \citep{gil}, the Redshift Space Distortions expected from the gravitational attraction of cosmological structures \cite{percival} and the CMB \citep{planck}, among others. 

The assumption of DE comes once the Einstein's field equations are established on the Friedmann-Lema\^itre-Robertson-Walker (FLRW) metric, which gives rise to the Friedmann equations. These equations imply that a stress-energy-momentum component with negative pressure is required to describe the cosmic acceleration. This substance is usually interpreted as the vacuum energy of the Universe (i.e., $\Lambda$ in the $\Lambda$CDM model) or a scalar field \citep{copeland1}.

From the theoretical point of view, dynamical DE models that come from a minimally coupled, canonical scalar field give an equation of state $w(z)$ of the form: $P_\text{DE}=w(z)\rho_\text{DE}$, for $P_\text{DE}$ and $\rho_\text{DE}$ the pressure and density of DE, respectively, to achieve the accelerated expansion for redshifts $z\leq 1$ and, in addition, assumptions are made for its perturbed behavior (i.e. sound velocity). Alternatively, in MG models a large-scale modification to GR is achieved by introducing new degrees of freedom to the geometric sector and therefore the cosmic expansion is due to the dynamics of these new modes. Diverse models can be found, such as scalar-tensor theories \cite{brans,amendola1,tsujikawa}, $f(R)$ metric theories \cite{nojiri,nojiri2,defelice}, Horndeski models \cite{Horndeski:1974wa,Deffayet:2009mn,Charmousis:2011bf}, among others, which on one hand pass local gravitation experiments in accordance with GR by means of screening mechanisms \cite{Koyama:2015vza}, and on the other, they alter the laws of gravity at large scales.  

It is well known that the expansion history is degenerated between models and a given MG parametrization can always be mapped to the $\Lambda$CDM background \citep[see e.g.][]{capozziello,nojiri1,song}. Forthcoming and ongoing redshift and weak lensing surveys, such as the Dark Energy Survey (DES) \cite{des}, Dark Energy Spectroscopic Instrument (DESI) \cite{desi}, Euclid \cite{euclid} and the Large Synoptic Survey Telescope (LSST) \cite{lsst}, together with \textit{Planck} measurements \cite{planck} and other cosmological studies, will trace the growth of cosmic structures precisely through different epochs. They will give the opportunity to analyze GR by checking the relationship between the matter distribution, the gravitational potential and the lensing potential on various scales. Such experiments may give clues to the physics generating cosmic acceleration or increase the range of scales over which Einstein's gravity has been proven up to now, and may also give an opportunity to discriminate between DE and MG models.

In this direction, the Parametrized Post-Friedmann (PPF) formalism \cite{hu,Baker:2012zs} is based on formulating a parametrized system that can be used to investigate cosmological linear perturbations in a very general, model-independent way for a broad range of MG models, assuming that their background evolution is equivalent to that of the $\Lambda$CDM model. Unlike the Parametrized Post-Newtonian approach, the PPF framework holds for arbitrary background metrics such as the FLRW metric, given that perturbations to the curvature scalar remain small. Being this approach linear, it is applicable to big length-scales on which matter perturbations have not yet passed the non-linear cutoff ($\delta_{nl}(k)\sim1$) and which lay inside the horizon. The PPF formalism is thus useful to test gravity on cosmological scales and also to discriminate MG from DE. This type of parametric approach is not new; there has been a big amount of effort to understand the physics along these lines \citep{hu,bertschinger1,linder1}.

Departures from GR could be crucial, not only in the linear regime of cosmological perturbations, but also in the nonlinear regime. Nonlinear effects may admit MG to be distinguished from exotic DE assuming that DE fluctuations are small. Although the linear growth of structure can provide means to an analysis of GR versus MG, it must be connected to other investigations. It is foreseen that a mixture of galaxy clustering, peculiar velocities and weak lensing observations will be necessary to get robust constraints on scale-dependent MG theories.

In the present work, we pursuit to study MG models using the numerical phase space technique known as {\itshape dynamical systems} (DS); this method provides a nice and powerful way of analyzing the physical nature of perturbations in such theories \cite{carloni,carloni1} as it gives a rather straightforward way for achieving a qualitative picture of the dynamics of these models. In addition, in the context of MG, it facilitates to qualitatively estimate the sensitivity of the dynamics of perturbations to modifications to the geometric sector. These DS approaches have already been  proved to a large range of space-times (FLRW, Bianchi models, etc.) \cite{bahamonde}, for which the evolution equations can be reduced to a system of autonomous ordinary differential equations (ODE) that describe a self-consistent phase space \citep{wainwright,coley,wainwright1}. Therefore, such tool allows for a preliminary analysis of these theories, suggesting what kind of models deserve further attention or which could be used to identify possible sources of observational signatures. Such techniques have been used to perform analysis in some DE/MG models in various recent investigations (see \cite{bahamonde} for examples and references therein), where in all of them only the dynamics of the background is considered. In the present work we extend the formalism to first order perturbation theory. 

This article is dedicated to numerically contrast the previously mentioned parametrized MG models to the standard cosmological $\Lambda$CDM model through the DS tool. Prosperous MG models are known to satisfy the background dynamics demanded by observations and that turn out to be similar to those in the $\Lambda$CDM model, but whose perturbations are more challenging to treat. In this sense, we give the critical points and eigenvalues for the different models which will describe many interesting results from the phase space analysis of the linear perturbations. Although the method is used for some specific cases in order to study the physics behind the models, the exposed tool is general and can be used to implement dynamical analysis of diverse models of DE/MG (and even DM).

Our main findings are the following: i) The critical points of CDM, $\Lambda$CDM and the MG parametrizations considered in this work are essentially the same and they have the same stability features. This is not surprising since these parametrizations are constructed under the assumption that $\Lambda$CDM is a fiducial model from which modifications arise in specific scale regimes (spatial and temporal), depending on the type of the underlying theory, ii) The dynamical stream-field in the  phase space of scale-independent PPF parametrizations is identical to that of the $\Lambda$CDM model, iii) By comparing the phase-stream-field nearby the critical points of scale-dependent PPF parametrizations against the $\Lambda$CDM model, we analyze the dynamics of the scale dependence of perturbations introduced by different MG parametrizations. We estimate the percentage of modification of each scale-dependent perturbed variable in comparison to the standard $\Lambda$CDM model, numerically showing how the amplitude of such deviations depend mainly on the scale of the perturbation as well as on the MG model considered.

This work is organized as follows: In Section~\ref{TB}, the basic equations of the present paper are given. The cosmological perturbations within the PPF formalism are studied, focusing primarily on the terms and equations that are of significance for this analysis. In Section~\ref{DS}, we briefly revise the DS theory. In Section~\ref{PPF-ds}, the evolution equations of the perturbed PPF system are transformed into an autonomous system by a convenient definition of the dynamical variables. In Section \ref{results2}, we present the dynamical analysis of the standard CDM, $\Lambda$CDM and MG cosmologies, investigating the different trajectories that show up (vector field diagrams).

The critical points are listed and the phase space analysis is given. We also show the numerical solutions and stability of the system in the vicinity of the critical points, and show the deviation in the amplitude of the perturbations of the MG model compared to those in the $\Lambda$CDM model. Finally, Section~\ref{C} is dedicated to discussion and conclusions.
\section{Theoretical Background}
\label{TB} 
Despite that the Universe may look practically homogeneous at scales over $150-300\ \text{Mpc}$, it is definitely inhomogeneous at smaller scales in which structure formation has taken place and hence the use of perturbation theory is needed. Given that MG theories can end up with the same predictions as the $\Lambda$CDM model at the level of the cosmological background, data from large scale structure (LSS) might be able to break such degeneracy between models and new signatures are to be identified by means of perturbations upon a FLRW Universe. The dynamics of subhorizon perturbations is appropriate to describe the large-scale regime of
LSS formation and therefore the PPF formalism provides a convenient framework to encompass a large set of theories. 
In this work we shall use the PPF formalism, which includes parameters 
that are constrained through observational data \citep{clifton}. The starting idea behind the PPF approach is to parametrize the right-hand-side of the $\Lambda$CDM equations for subhorizon perturbations of the metric by using free functions of the wavenumber and scale factor. Any modification of gravity at the level of the background is ignored and only modifications at the level of the perturbations are introduced by means of the PPF free functions defining the parametrization. The formalism is then implemented for a fixed $\Lambda$CDM background history and, from it, we focus on the dynamics of the perturbations. This approach is supported by the fact that the standard cosmological model is in agreement with present constraints of the expansion history, thus, modifications to GR or exotic fluids/DE must be about the same at the unperturbed level. Therefore, models are to be discriminated at the level of perturbations \citep{hu,bertschinger1,Baker:2012zs}. 
\subsection{General Parametrized Post-Friedmann Formalism}
In this work we assume an approximately homogeneous and isotropic Universe which undergoes an accelerated expansion. At the background level, we assume a content of 
the Universe given by standard non-relativistic matter (baryons+DM, as they both behave the same way during the era of structure formation), and an effective cosmological constant that is supposed to be provided by the MG schemes considered, since there is well known degeneracy between possible models of expansion. Structure formation at very large scales in such a Universe can be modeled by the linear density and velocity perturbations of matter together with the scalar perturbations of the metric which can be described by (in the Newtonian gauge):

\begin{equation}
	\text{d}s^2 = a^2(\eta) \left[-(1+2\psi) \text{d}\eta^2 + (1+2\phi) \delta_{ij} \text{d}x^i \text{d}x^j \right],
\end{equation}

where $a$ is the scale factor and $\eta$ is the conformal time (related to the cosmological time $t$ by $\text{d}/\text{d}\eta=a\ \text{d}/\text{d}t$). Perturbations that rule the growth of structure are described by $\phi$ (the Newtonian potential) and $\psi$ (the spatial curvature potential); $\phi$ and $\psi$ alter both the energy of photons (Integrated Sachs-Wolfe [ISW] effect) and their direction of propagation (gravitational lensing).

The observables that characterize the LSS are determined using cosmological perturbation theory in Fourier space. As mentioned before, the relevant variables are the two scalar metric potentials, $\phi$ and $\psi$, along with the matter density contrast $\delta$ and the matter velocity perturbation $v$. Thus, one needs four equations to describe the evolution of these four variables. Two equations are given by the covariant conservation of the matter energy-momentum tensor, the other two equations are assumed by a theory of gravity that defines how the metric responds to the matter stress-energy tensor. The former two are dynamical and the latter correspond to constraint equations.

In the case of the quasi-static (QS) PPF parametrization, modifications to this system of equations due to extra degrees of freedom modifying either the matter or the gravitational sectors  are introduced only by means of the free functions $\mu(a,k)$ and $\gamma(a,k)$ (with wavenumber $k$ dependence) in the Poisson and shear equations. A way to pose this in the Newtonian metric, in an epoch where radiation can be neglected, is by writing the adiabatic fluid equations in the following way, where the modifications are present in Eqs.~\eqref{eq-p1} and~\eqref{eq-p2}\cite{Baker:2012zs}:
\begin{eqnarray}
	\dot\delta_m&=&-\theta_m,\label{delta-dot}\\
	\dot\theta_m &=&-2H\theta_m+\frac{k^2} {a^2}\psi,\label{theta-dot}\\
	k^2\phi&=&4\pi Ga^2\rho_\text{t}\left[\delta_\text{eff}+3a \frac{H}{k^2} (1+ w_\text{eff})\theta_\text{eff}\right]\mu , \label{eq-p1}\\
	\psi&=&-\frac{\phi}{\gamma}. \label{eq-p2}
\end{eqnarray}
The fluid perturbed variables labeled with $m$ refer to non-relativistic matter, the subindex `eff' refers to effective quantities if there is more than one component, dots $\dot{\{\}}$ mean derivative with respect to the cosmological time $t$, $\delta$ is the density contrast, $\theta$ the divergence of the velocities, $H$ the Hubble parameter, $\rho_\text{t}$ the total density and $G$ refers to Newton's gravitational constant.
The units of the variables involved are
\begin{eqnarray}
&[a]=[\delta]=[\phi]=[\psi]=[\mu]=[\gamma]=[\mbox{dimensionless}] , \nonumber \\
&[H]=[\theta]=[k]=\left[\frac{1}{t}\right]=[\text{eV}] , \nonumber
\end{eqnarray}
where we are assuming $c=\hbar=1$.

In Eqs.~\eqref{eq-p1}-\eqref{eq-p2} we used the PPF approach in the QS limit for the gravitational sector only; thus, all effects due to degrees of freedom of MG are encoded by $\mu$ and $\gamma$.  
In Eq.~\eqref{eq-p1}, we have
\begin{eqnarray}
\rho_\text{t}&=&\rho_m+\rho_\text{DE},\\
\delta_\text{eff}&=&\Omega_m\delta_m+\Omega_\text{DE}\delta_\text{DE}=\Omega_m\delta_m,\\
w_\text{eff}&=&\Omega_m w_m+\Omega_\text{DE}w_\text{DE}=-\Omega_\text{DE}=\Omega_m-1,\nonumber\\\\
\theta_\text{eff}&=&\frac{(1+w_m)\Omega_m\theta_m+(1+w_\text{DE})\Omega_\text{DE}\theta_\text{DE}}{1+w_\text{eff}}=\frac{\Omega_m\theta_m}{1+\Omega_\text{DE}w_\text{DE}}=\theta_m,
\end{eqnarray}
where $\rho_m$ and $\rho_\text{DE}$ denote the non-relativistic matter and DE density components, respectively, and $\Omega_m$ and $\Omega_\text{DE}$ are the corresponding density parameters. When $\gamma(a,k)=\mu(a,k)=1$, the case of GR is recovered.
We define $G_\text{eff}\equiv\mu(a,k)G$, which in Eqs.~\eqref{eq-p1} and~\eqref{eq-p2} introduces a gravitational modification at cosmological scales. For simplicity, we will only focus on flat cosmologies ($\kappa=0$) 
and omit entropy perturbations. These definitions infer that anisotropic stress of matter can be ignored at the epoch of interest, although it can be included, if needed, as it was done in \cite{bean,hojjati}.
\subsection{Scale-independent Parametrizations}
\label{sip}
During the radiation-dominated era, the Jeans length of the matter-radiation system is comparable to the Hubble length of the Universe. In this case, GR and MG operate similarly. However, at late times, the Jeans length has fallen to a few Mpc or less, and modifications of gravity come into play for most of the sub-horizon modes of the perturbations. Thus, the time and space dependence of perturbations must factorize for wavelengths longer than the Jeans length \cite{bertschinger1}:
\begin{eqnarray}\label{relpotentials}
	\phi(k,a) = D(a)\zeta(k) ,
\end{eqnarray}
where $D(a)$ is a function of the scale factor only and $\zeta(k)$ is the curvature perturbation, which at the same time corresponds to the transfer function of the potentials. This factorization implies that the factor of both potentials depends only on the scale factor:
\begin{eqnarray}
	\phi(k,a) = \gamma(a)\psi(k,a) + \mathcal{O}(k^2\zeta) .
\end{eqnarray}

In MG theories, scalar perturbations with very long wavelengths are mainly affected by $\gamma$, and $\mu$ is identified to the analysis of the physical gravitational constant and with experiments of the weak equivalence principle. By assuming that the $\mathcal{O}(k^2\zeta)$ terms are negligible on sub-horizon scales larger than the Jeans length, a class of theories is defined: the \textit{scale-independent} MG models. Under the previous hypothesis, modifications of gravity are entirely described by the dynamics of the perturbations at large scales, and $\gamma$ and $\mu$ are given by \citep{bertschinger1}
\begin{equation}
	\gamma(a)=\mu(a)=1+\beta a^s,
\label{brans}
\end{equation}
where $\beta$ and $s$ are constants that can be obtained from observational constraints.
\subsection{Scale-dependent Parametrizations}\label{scdp}
In scale-dependent MG theories, the parameters are functions not just of the scale factor $a$, but also of the wavenumber $k$, and there is no straightforward relationship between them. Thus, more parameters are required to describe such theories \cite{amin}. In the QS approximation, the Poisson and trace shear equations can again be parametrized in terms of both the time- and scale-dependent functions, $\mu(a,k)$ and $\gamma(a,k)$ as shown in \cite{defelice}.  Under quite general conditions, and in the QS approximation ($k/aH\gg1$), $\mu$ and $\gamma$ should always acquire a form of ratio of polynomials in $k$. The coefficients inside the polynomials are functions of the background quantities and can be expected to be gradually changing functions. Technically, the number of these time-dependent coefficients is infinite if one allows for an arbitrary modification of GR. Quantitative differences in the predictions for these coefficients are important to discriminate between MG theories. This form of parametrization has been considered in a large number of works  \citep[e.g.][]{hu,bertschinger1,linder1,Baker:2012zs}, using various functional forms of the scale-dependence of $(\mu,\gamma)$, corresponding to different theories.

GR and MG theories have in common that the curvature perturbation is a conserved quantity at some range of scales. Therefore, the scale-dependence of the transfer functions only appear below the Jeans length or at scales of strong-field sector, for MG theories where the factorization of cosmological perturbations no longer holds. Theories of this kind are called {\itshape scale-dependent} MG models. Examples of this class of models are the $f(R)$ metric theories, where the wavelength of the scalar degree of freedom of the modification characterizes the theory. Eventually, measuring this scale-dependence (at several redshifts) can constrain scale-dependent MG theories.

A suitable scheme to encompass all theories having a modification of gravity with second order equations of motion are included in the Horndeski class \cite{Horndeski:1974wa}, for which the ratio of even polynomials is as follows \citep{silvestri}:
\begin{eqnarray}
	\mu(a,k)&=&\frac{1+p_3(a)k^2}{p_4(a)+p_5(a)k^2} ,\label{mu-pol} \\
    \gamma(a,k)&=&\frac{p_1(a)+p_2(a)k^2}{1+p_3(a)k^2},\label{gamma-pol}
\end{eqnarray}
where the $p_i(a)$'s are, in general, free functions of the scale factor.

Even though Eqs.~\eqref{mu-pol}-\eqref{gamma-pol} come from general arguments, the subspace of possible models to which we are restraining corresponds to the set of models contained in the Horndeski class, which includes many of the possible theories of DE and MG \cite{defelice1,Horndeski:1974wa}. It is important to note that although this ansatz was obtained using the QS approximation, it also allows for near and super-horizon modifications of gravity: $\gamma(a,k\rightarrow0)=p_1(a)\neq1$.

Specific forms of the $p_i$ functions can be obtained from particular MG theories. Having five free functions gives an unnecessary level of indetermination in this analysis. Then, it is good to demand physical arguments in order to assemble an improved parametrization of this type of theories. It is common to accommodate the transition scale, which separates the different regimes in which gravity acts in different ways. This transition can be studied by the following parametrization \cite{zhao}:
\begin{eqnarray}
\mu(a,k)&=&\frac{1+\beta_1\lambda_1^2k^2a^s}{1+\lambda_1^2k^2a^s},\label{mu}\\
\gamma(a,k)&=&\frac{1+\beta_2\lambda_2^2k^2a^s}{1+\lambda_2^2k^2a^s},\label{gamma}
\end{eqnarray}
where the parameters $\lambda_i$'s have dimensions of length, while the $\beta_i$'s serve as dimensionless couplings. According to \citep{zhao}, the above parametrization is suitable to scalar-tensor theories which include a host of cosmologically relevant models. 
Table \ref{tab:ST-models} shows well-known examples found in the literature \citep[see e.g.][]{zhao}.
\begin{table}[h]
\begin{center}
\begin{tabular}{|c|c|c|c|c|c|}
\hline
Model & $\beta_1$ & $\beta_2$ & $\lambda_1^2$ & $\lambda_2^2$&$s$\\
\hline 
$f(R)$ I & $4/3$ & $1/2$ & $\beta_1^{-1}10^3 \ \text{Mpc}^2$ & $10^3\ \text{Mpc}^2$ & $4$\\
\hline
$f(R)$ II & $4/3$ & $1/2$ & $\beta_1^{-1}10^4 \ \text{Mpc}^2$ & $10^4\text{Mpc}^2$ & $4$\\
\hline
Chameleon-like I & $9/8$ & $7/9$ & $\beta_1^{-1}10^3\ \text{Mpc}^2$ & $10^3\text{Mpc}^2$ & $2$\\
\hline
Chameleon-like II & $9/8$ & $7/9$ & $\beta_1^{-1}10^4\ \text{Mpc}^2$ & $10^4\ \text{Mpc}^2$ & $2$\\
\hline
\end{tabular}
\captionof{table}{MG scale-dependent model parameters.}
\label{tab:ST-models}
\end{center}
\end{table}
\section{A brief revision of Dynamical Systems}
\label{DS} 
The qualitative analysis of different cosmological models whose evolution is ruled by a finite-dimensional ODE autonomous system has become a nice tool for the study of the different variables (parameters) contained in such models \cite{percival}. The system is said to be autonomous if for $\dot x_i=f_i(x_1,...,x_n,t)$, the functions $f_i$ do not include explicit time-dependent terms. The DS approach provides the opportunity to study the stability of the solutions of the system in a simple way by investigating the behavior of the system around its critical points. For example, equilibrium points of the reduced system can correspond to dynamically evolving cosmological models \cite{szydlowski}.

Since our intention is to provide a qualitative picture of the perturbations within the PPF formalism, a DS approach is undertaken. Commonly, a dimensionless (logarithmic) time variable, $N$ (e-folding), is introduced so that the evolution is valid for all times (i.e., $N$ accepts all real values), and a normalized set of variables is build for a number of reasons: i) This usually leads to a bounded system, ii) the variables are well behaved and often have a straightforward clear physical explanation, and iii) due to a symmetry in the equations, one of the equations decouple from the others (in GR the expansion rate is used to normalize the variables), and the arising simplified reduced system is studied. The goal of our qualitative analysis is to understand the nature of typical solutions of this kind of systems.

In this direction, the dynamics of the system~\eqref{delta-dot}-\eqref{eq-p2} can be treated by inspecting its evolution around fixed/critical points, i.e.~points $P_i$ fulfilling the stability condition $\text{d}P_i/\text{d}\ln a=0$. After obtaining the fixed points, one proceeds to get the eigenvalues $\lambda_i$ of the Jacobian matrix of the system, in order to linearize it around each critical point. This determines the stability nature of a particular point, in other words, it controls how the system behaves when approaching to such critical point. 

Now, in more than three dimensions it turns out difficult to label all possible critical points based on their eigenvalues. Nevertheless, for an $n$-dimensional system, if one has $n$ eigenvalues for each point, the stability depending on the nature of these eigenvalues can be roughly interpreted given the following classification:
\begin{itemize}
\item All the eigenvalues $\lambda_i$ are real and have the same sign:\\
- Negative eigenvalues: Stable node/Attractor.\\
- Positive eigenvalues: Unstable node.
\item All the eigenvalues $\lambda_i$ are real and at least one is positive and one negative: Saddle points.
\item At least one eigenvalue is real and there are pairs of complex eigenvalues:\\
- All eigenvalues have negative real parts: Stable Focus-Node.\\
- All eigenvalues have positive real parts: Unstable Focus-Node.\\
- At least one positive real part and one negative: Saddle Focus.
\end{itemize} 
For a thorough description of the technique, and for some applications to other cosmological models, please refer to \citep{wainwright,strogatz,copeland}.  
\section{PPF Dynamical System}
\label{PPF-ds}
In this section we give the ODE that are to be solved in the present problem, and the convenient normalization variables are introduced in order to obtain the corresponding DS that will be analyzed in the following sections. Linear perturbation theory is assumed to be valid throughout the paper.
\subsection{The autonomous system}
The system of Eqs.~\eqref{delta-dot}-\eqref{eq-p2} will be written as an autonomous system. The first step in the implementation is the definition of the variables. We introduce the general dimensionless variables:
\begin{eqnarray}
x_1=\delta_m,
\qquad
x_2=\frac{\theta_m}{H} ,
\qquad
x_3=\frac{H^2}{k^2},
\qquad
x_4=\Omega_m ,
\qquad
x_5=a^2,
\end{eqnarray}
where $x_1$ is the density contrast, $x_2$ is a re-definition of the velocity divergence over the Hubble parameter, $x_3$ is the squared Hubble rate in units of the perturbation mode $k$, $x_4$ is the non-relativistic matter density parameter and $x_5$ is the squared scale factor.

For $x_2$ one has that the velocity divergence decreases at early times when $H$ is large, but it  increases with time; $x_3$ conveniently provides a good comparison parameter of the size of a mode at a given time in the expansion history.

The physical bounds of the involved variables are: $-1 \ll x_1 \ll 1$ (linear approximation), $-1 \ll x_2 \ll 1$ (linear approximation), $0 \leq x_3 < \infty$, $0 \leq x_4 \leq 1$ and $0 \leq x_5 \leq 1$.

Physically, all variables $x_i$ are bounded with the exception of $x_3$, which can tend towards infinity since $H$ has no upper bound. It is important then to build a bounded variable for $x_3$ to carry out a reliable search of the critical points in our system. To that aim the following change of variable is made:
\begin{equation}
y_3 \equiv \frac{1}{1+x_3}.
\label{eq-y5}
\end{equation}
Once relation ~\eqref{eq-y5} is introduced, the new variable $y_3$ will be bounded between $[0,1]$, which corresponds to $x_3 \rightarrow \infty$ and $x_3=0$, respectively. 
From a phenomenological point of view, in this work, apart from $x_3$ we shall only determine the dynamics in the neighborhood of finite fixed points which are already physically bounded.

Using Eqs.~\eqref{delta-dot}-\eqref{eq-p2} and  \eqref{eq-y5}, our dimensionless DS reads as follows: 
\begin{eqnarray}
x_1'&=&-\frac{x_2}{x_5^{1/2}},\label{eq-x1}\\
x_2'&=&-x_2\left(1-\frac{3}{2}x_4\right)-\frac{3}{2}x_4x_5^{1/2}(x_1+3x_2x_3x_5^{1/2})\frac{\mu}{\gamma},\nonumber\\\label{eq-x2}\\
y_3'&=&3y_3^2x_3x_4\label{eq-x3},\\
x_4'&=&-3x_4(1-x_4)\label{eq-x4},\\
x_5'&=&2x_5,\label{eq-x5}
\end{eqnarray}
where the unbounded variable $x_3$ has been rewritten in terms of the bounded variable $y_3$ following the relation in Eq.~\eqref{eq-y5} (i.e. $x_3 = 1/y_3-1$), and $'$ means derivative with respect to the e-folding: $\text{d}/(H\text{d}t)=\text{d}/\text{d}N$, with $N=\ln a$.

The generalization to the multi-fluid case is trivial: one has just to add a new variable $\Omega_i$ for each new type of fluid. As a result, the number of dynamical equations increases and then also the dimension of the phase space. 

Let us make an important point: the application of the ``$\mu-\gamma$'' parametrization to the equations of motion in more general cases does not always assume the QS approximation. In the $\Lambda$CDM limit, when $\mu=\gamma=1$, the {\itshape exact} equations of GR are recovered, while the parametrization admits for departures from $\mu=\gamma=1$ at all scales. Relativistic effects may be important for some MG models \cite{yoo,yoo1,challinor}, but our present approach does not consider them.
\section{Results}
\label{results2}
In what follows, the critical points and eigenvalues of system~\eqref{eq-x1}-\eqref{eq-x5} are obtained for different models and their cosmological viability is examined. Since we are interested in perturbations at late times, radiation has decoupled and can be safely ignored. The overall features of the dynamics will be given in the form of vector phase portraits, which reflect the phase space solutions of the problem. Also, when studying MG models we shall integrate out the system with initial conditions close to the critical points in order to analyze their behavior and compare them to the $\Lambda$CDM model. For this purpose, we carry out the corresponding MG-DS analysis of the models considered ($f(R)$ and Chameleon-like of table \ref{tab:ST-models}), finally arriving to the conclusion that their critical points are exactly the same to those found in $\Lambda$CDM (see Eqs.~\eqref{P1-P2} and \eqref{P3-P4} in the following sections).  
The reason for this is that the MG-PPF parametrization introduces multiplicative factors to Eq.(\ref{eq-x2}), so that the $\Lambda$CDM critical points are also a solution of the MG system. Given this though, the stream flows of the phase space in both models are quite different for some of the variables involved, mainly for the peculiar velocities. Also, we point out that the DS approach used here turns to be a powerful tool when it comes to quantify the extent of modification with respect to $\Lambda$CDM. In this direction, we extract useful information about how much $\Lambda$CDM differs from MG solutions, either with the scale $k$ and/or the initial conditions around the critical points. This information can be useful when studying the phenomenology of these models, as we will see below, and it provides valuable criteria to identify which data results convenient to use in order to constrain MG models.
\subsection{CDM system}\label{cdm}
The CDM model consists of a Universe only filled with non-relativistic matter. For the variables of our DS, Eqs.~\eqref{eq-x1}-\eqref{eq-x5}, this implies $x_4=\Omega_m=1$ ($\Omega_\text{DE}=0$) and $\mu=\gamma=1$. 
The corresponding DS is then given by
\begin{eqnarray}
x_1'&=&-\frac{x_2}{x_5^{1/2}},\\
x_2'&=&-x_2\left(1-\frac{3}{2}x_4\right)-\frac{3}{2}x_4x_5^{1/2}\left(x_1+3x_2x_3x_5^{1/2}\right),\nonumber\\
y_3'&=&3y_3^2x_3x_4,\\
x_4'&=&0,\\
x_5'&=&2x_5.
\end{eqnarray}
This system has the following critical points:
\begin{equation}
\begin{split}
P_1=(x_1=0,x_2=0,y_3=0,x_4=1,x_5=0),\\
P_2=(x_1=0,x_2=0,y_3=1,x_4=1,x_5=0).
\label{P1-P2}
\end{split}
\end{equation}

In this case, we are using the fact that in a Universe without anisotropic stress, the Einstein field equations yield $\psi=-\phi$ (the stress-energy tensor is invariant under spatial rotations, or the three principal pressures are identical) and the scalar potentials result the same during the epoch of structure formation. However, this will no longer be true in MG theories, and the potentials need not to be the same; this case will be treated in the next sections. In general, the solution for the matter density satisfies $x_4=$ constant (free parameter), but as we are working on a flat FLRW background Universe with only non-relativistic matter, this implies $x_4=1$.

Note how the value corresponding to the density perturbations is $x_1=\delta=0$ at the critical points. 
Also, the divergence of the velocity perturbation divided by the Hubble parameter $x_2=\theta/H=0$ at the critical points. Given that $\theta=\vec\nabla\cdot\vec v$, in Fourier space this condition results equivalent to $\hat k\cdot\vec v=0$, that is, the velocity projected onto the direction of the wavevector vanishes independently of the magnitude of $k$. Therefore the condition $\vec k\cdot\vec v\ll 1$ has two possibilities: i) the velocity perturbation is always transverse to the wavevector, or ii) it vanishes. The first possibility may not necessarily be true, hence we adopt the second possibility and then the Universe has no peculiar velocities at these critical points.

The value corresponding to the squared ratio of the wavenumber to the Hubble parameter, $x_3=H^2/k^2$, can take two values at the critical points; $P_1$ corresponds to $x_3\rightarrow\infty$ which either means $H$ very large (this case corresponds to the Big Bang singularity) or mathematically $k\rightarrow0$ corresponding to large scales whose Jeans length lays inside the cosmological horizon. $P_2$ corresponding to $x_3=0$ implies $H=0$ (static Universe) or $k \gg 1$ (corresponding to small structures), which cannot be since we are working inside the linear regime. Finally, $a$ turns out to be very small in both cases. 

Furthermore, the system seems to diverge for the critical points in $x_5$; this happens because in the operations involved in the derivation of the autonomous system, terms like $1/x_5$ come out in the denominators. Nevertheless, this is not a real problem for the method but a result of the fact that for these values of the parameter the cosmological equations take an special form.

The eigenvalues $\lambda_i$ corresponding to the Jacobian matrix of CDM evaluated at the critical points are given in Table~\ref{tab-cdm}.
\begin{table}[h]
\begin{center}
\begin{tabular}{|c|c|}
   \hline
   Critical point & Eigenvalues $\lambda$ \\ 
   \hline
   $P_1$  & $\{-4.345,\,3.0,\,3.0,\,2.0,\,0.345\}$ \\
   \hline
   $P_2$  & $\{-3.0,\,3.0,\,2.0,\,1.5,-1.0\}$ \\
   \hline
\end{tabular}
\captionof{table}{Eigenvalues for the two critical points in the CDM case.}
\label{tab-cdm}
\end{center}
\end{table}

For both, $P_1$ and $P_2$, all eigenvalues are real, with at least one of them being negative. 
 From the Hartman-Grobman theorem \citep{coley}, it is known that if {\itshape all} eigenvalues of the Jacobian matrix satisfy $\mathcal{R}e(\lambda)\neq0$, then the point is {\itshape hyperbolic}. In this case, $P_1$ and $P_2$ have non vanishing eigenvalues, so they are hyperbolic in nature, both of them being saddle points. 
As it will be seen in the next subsections, the phase portraits and numerical solutions near $P_1$ and $P_2$ can give us a good understanding of the nature of these points. 
\subsection{$\Lambda$CDM system}
A $\Lambda$CDM expansion history is defined by $1-\Omega_\text{DE}=\Omega_m$. Also, $\dot H=0$ ($H'=0$) corresponds to either a static or a de Sitter solution ($w_\text{eff}=-1$). In addition, we have the energy constraint condition
\begin{equation}
\Omega_m+\Omega_\text{DE} = 1\longrightarrow \Omega_\text{DE}
= 1-\Omega_m=1-x_4,
\end{equation}
which can be used to shorten the dimensions of the system. For the system of equations that describe the $\Lambda$CDM case we have: $w_\text{eff}=-\Omega_\text{DE}=\Omega_m-1$. 
The DS is then given by
\begin{eqnarray}
x_1'&=&-\frac{x_2}{x_5^{1/2}},\label{lcdme1}\\
x_2'&=&-x_2\left(1-\frac{3}{2}x_4\right)-\frac{3}{2}x_4x_5^{1/2}\left(x_1+3x_2x_3x_5^{1/2}\right),\nonumber\\\\
y_3'&=&3y_3^2x_3x_4,\\
x_4'&=&-3x_4(1-x_4),\\
x_5'&=&2x_5.\label{lcdme5}
\end{eqnarray}

The $\Lambda$CDM system has four critical points: two of them corresponding to $P_1$ and $P_2$ from the CDM model (Table~\ref{tab-cdm}); i. e., they correspond to a submanifold of the $\Lambda$CDM case and their physical interpretation is the same as that given in the previous case. There are also two new critical points, $P_3$ and $P_4$, given by
\begin{equation}
\begin{split}
P_3 = (x_1=0,\,x_2=0,\,y_3=0,\,x_4=0,\,x_5=0),\\
P_4 = (x_1=0,\,x_2=0,\,y_3=1,\,x_4=0,\,x_5=0),
\end{split}
\label{P3-P4}
\end{equation}
$P_3$ and $P_4$ are the new critical points which correspond to a DE-only dominated Universe, an empty ($\rho_m=0$) de Sitter Universe where a cosmological constant pervades. 

Now, we obtain the eigenvalues of the autonomous system~\eqref{lcdme1}-\eqref{lcdme5} in order to determine their stability.
\begin{table}[h]
\begin{center}
\begin{tabular}{|c|c|}
\hline
Critical point & Eigenvalues $\lambda$\\ 
\hline
$P_1$&$\{-4.345,\,3.0,\,3.0,\,2.0,\,0.345\}$\\
\hline
$P_2$&$\{-3.0,\,3.0,\,2.0,\,1.5,\,-1.0\}$\\
\hline
$P_3=P_4$&$\{-3.0,\,2.0,\,-1.0,\,0.0,\,0.0\}$\\
\hline
\end{tabular}
\captionof{table}{Eigenvalues for the four critical points in the $\Lambda$CDM case.}
\label{tablcdm}
\end{center}
\end{table}

For the $\Lambda$CDM case, from table \ref{tablcdm} it can be seen how for the new critical points $P_3$ and $P_4$ which have the same eigenvalues, have at least one of them with a value equal to zero, then the point is {\itshape non-hyperbolic}, and no further can be said apart from that. There are several important facts to consider from the critical points and their eigenvalues. 
As in the CDM case, $P_1$ corresponds to the point of {\itshape Big Bang singularity}. In the case of $P_2$, we have again $y_3=1$ which means $x_3=H^2/k^2\rightarrow0$ ($H=0$ is a stationary solution), and both points correspond to the previous results obtained in Sec.\ref{cdm}. $P_3$ and $P_4$ correspond to an empty de Sitter Universe filled up with only DE, infinitely expanding for the case $y_3=0$ ($x_3\rightarrow\infty$) and stationary in the case $y_3=1$ ($x_3=0$). In all the critical points we have $x_2=\theta/H=0$. 

\begin{figure*}[!]
\includegraphics[width=0.45\textwidth]{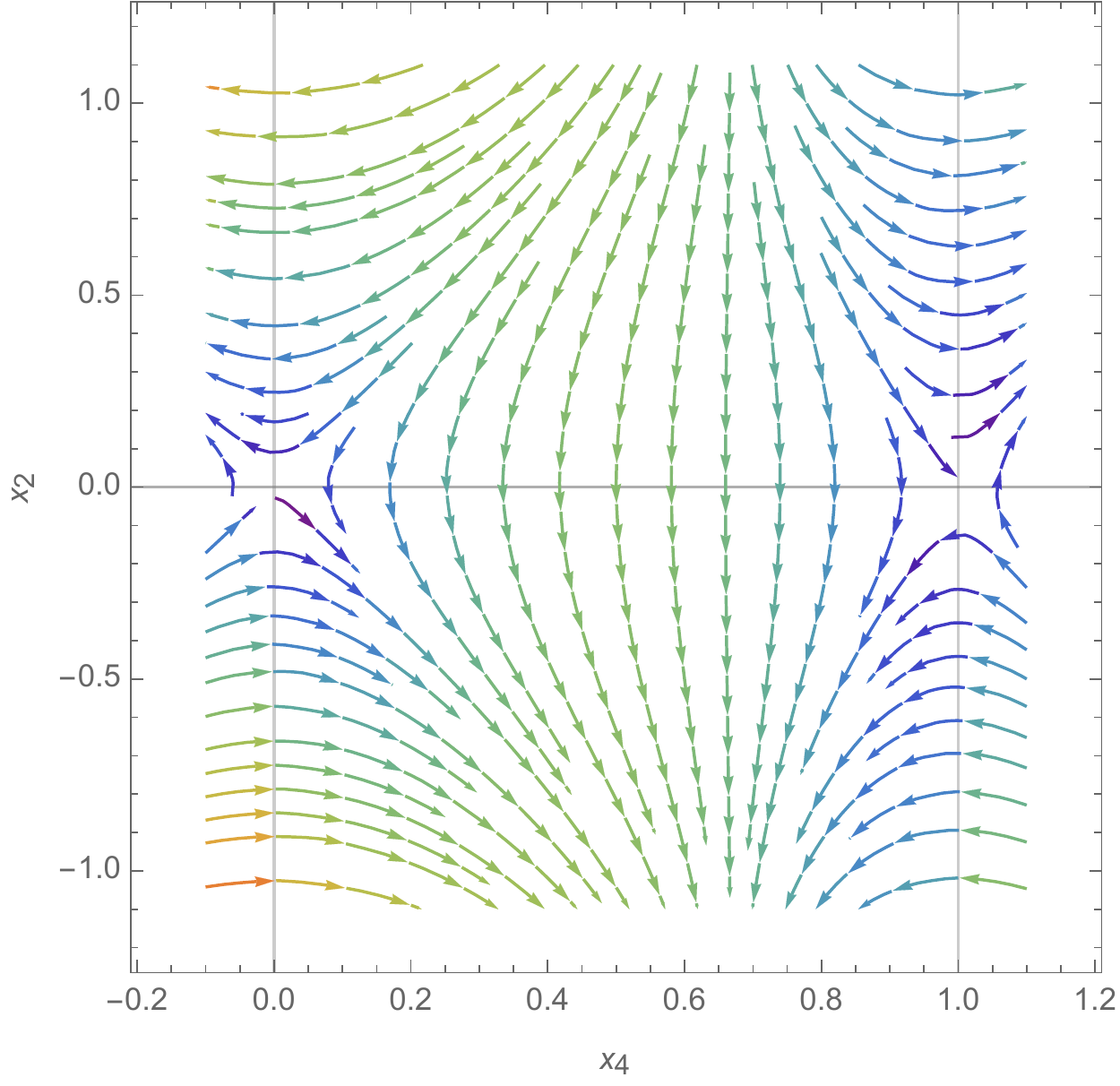}
\includegraphics[width=0.45\textwidth]{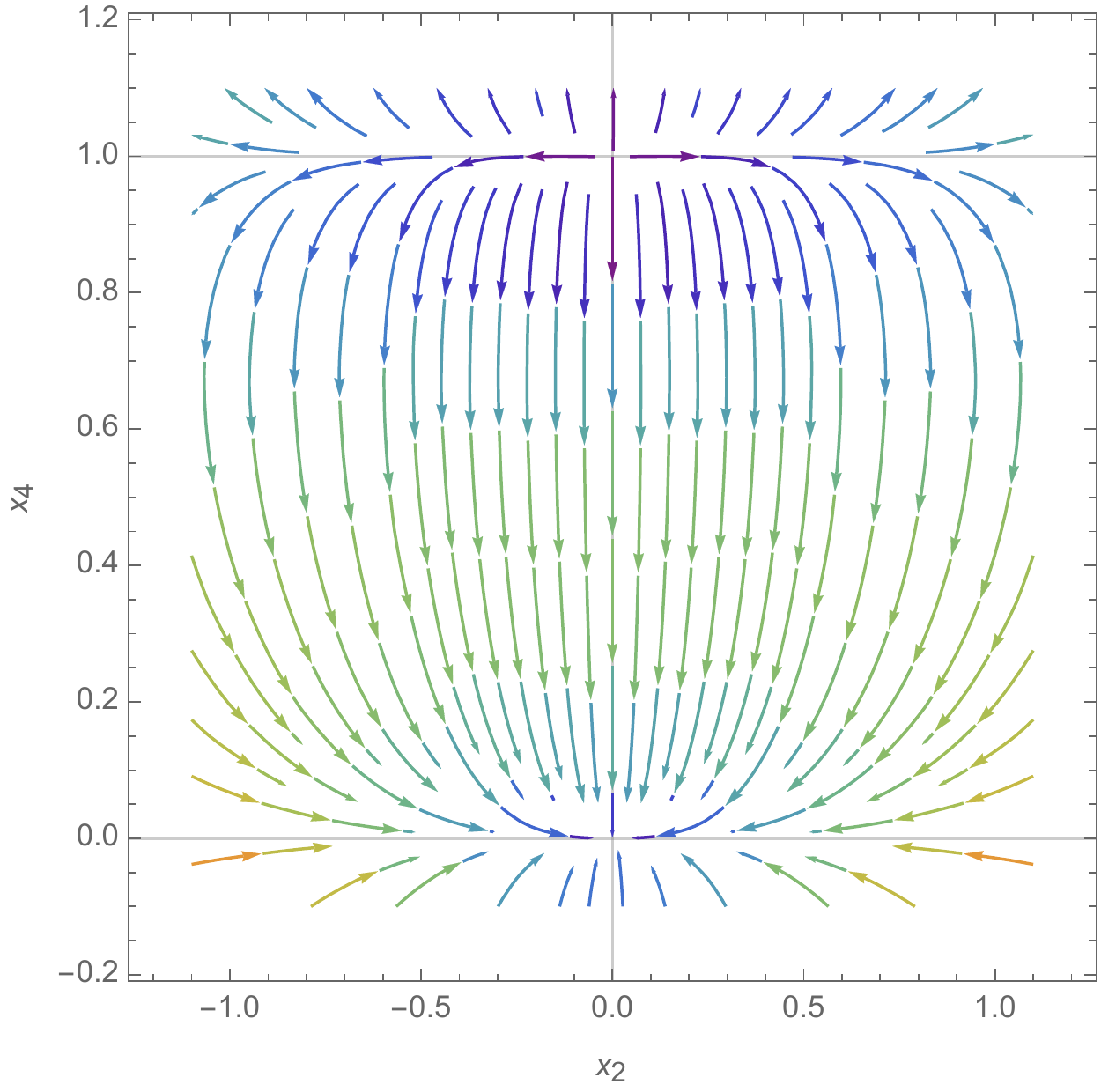}
\caption{Phase portraits for the velocity perturbations ($x_2 =\theta/H$) vs. matter density ($x_4 = \Omega_m$) in the $\Lambda$CDM model. We show the projection of the stream fields $x_2'-x_4'$ onto the planes $x_4-x_2$ (left) and $x_2-x_4$ (right), and their evolution for all $y_3$. The behavior of the perturbations from $x_4=0$ (DE dominated Universe) to $x_4=1$ (DM dominated Universe) is shown. The physical acceptable region for $\Omega_m=x_4$ starts to the right of the vertical line at point (0,0). The phase portrait shows how both a DE-only and DM-only dominated Universe are maximums for the energy. From the right panel we can se how a DM dominated Universe (point (0,1)) corresponds to a source and a DE dominated Universe (point (0,0)) to a sink for perturbations of the form $\theta/H$. For all values of $y_3$ the evolution is the same.}
\label{fig1-lcdm}
\end{figure*}

Figs.~\ref{fig1-lcdm}-\ref{fig4-lcdm} show the critical points of the stream fields $x_2'-x_4'$, $x_2'-x_5'$, $y_3'-x_4'$, $y_3'-x_5'$, and $x_4'-x_5'$, for the $\Lambda$CDM model, projected onto the corresponding dynamical variables. We have particularly chosen these stream fields and projections as they can clearly show the behavior of the corresponding perturbations with respect to the content of matter density ($x_4$) and the evolution of the scale factor ($x_5$).
\begin{figure*}[!]
\centering
\includegraphics[width=0.45\textwidth]{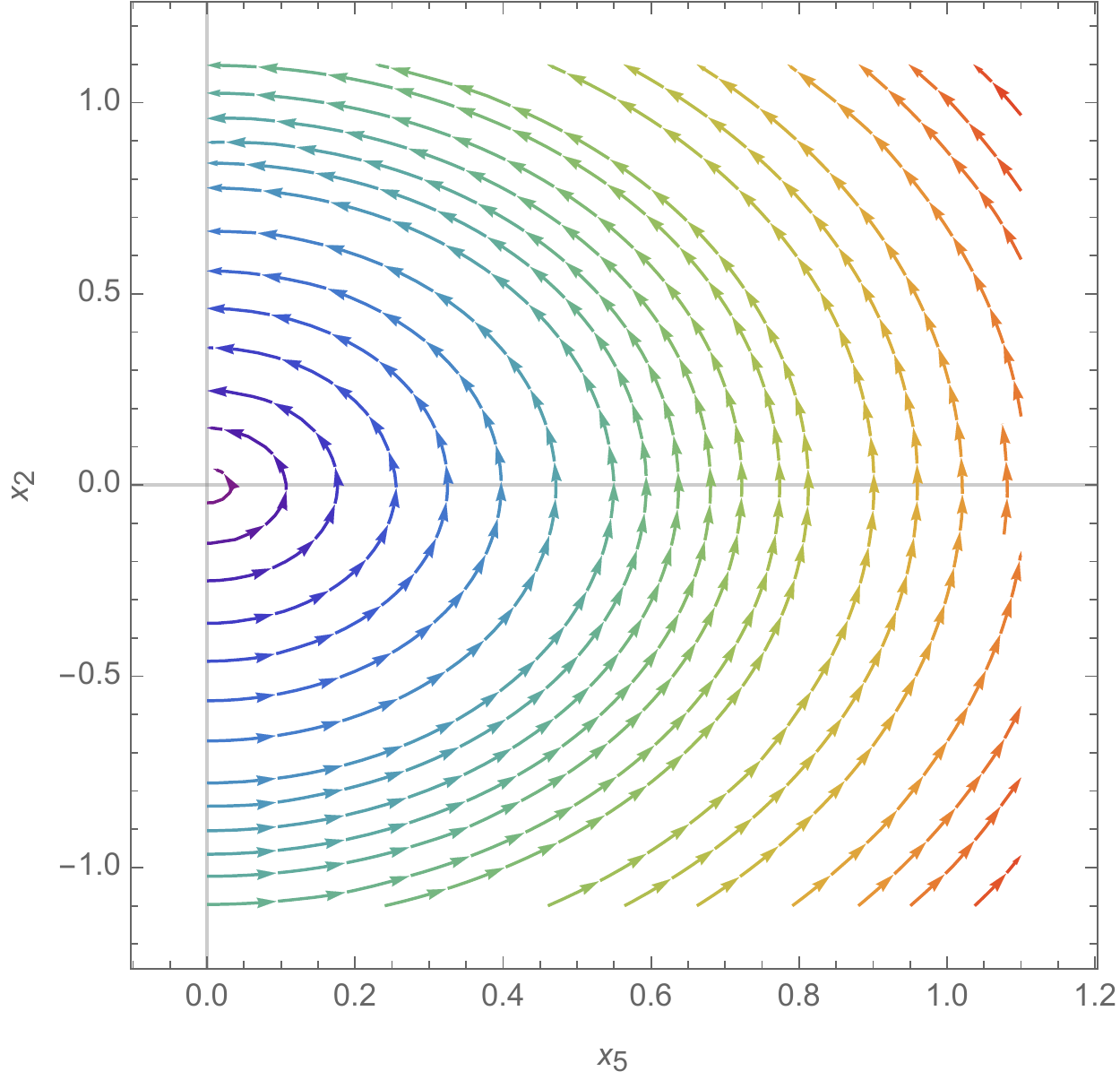}
\includegraphics[width=0.45\textwidth]{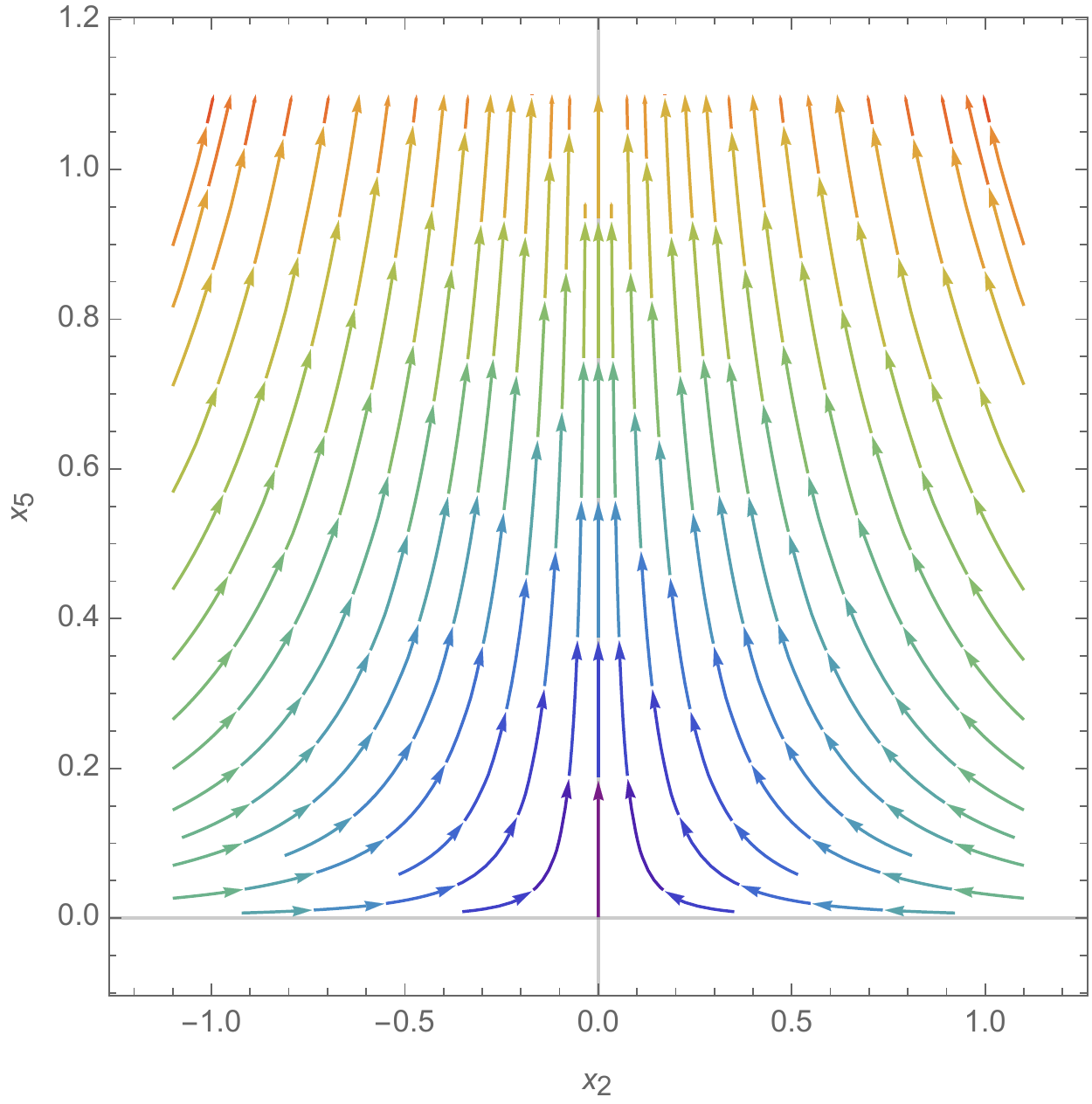}
\caption{Phase portraits for the velocity perturbations ($x_2 =\theta/H$) vs. scale factor ($x_5 =a^2$) in the $\Lambda$CDM model. We show the projection of the stream fields $x_2'-x_5'$ onto the planes $x_5-x_2$ (left) and $x_2-x_5$ (right) and their evolution for all $y_3$. In this case $x_4=0$ (DE dominated). The physical acceptable region for $a^2=x_5$ starts to the right of the vertical line at point (0,0). The phase portrait shows how in a DE-only dominated Universe the point (0,0) (Big Bang) is a minimum for the velocity perturbations (saddle point). For all values of $y_3$ the evolution is the same.}
\label{fig2-lcdm}
\end{figure*}
\begin{figure*}[!]
\includegraphics[width=0.32\textwidth]{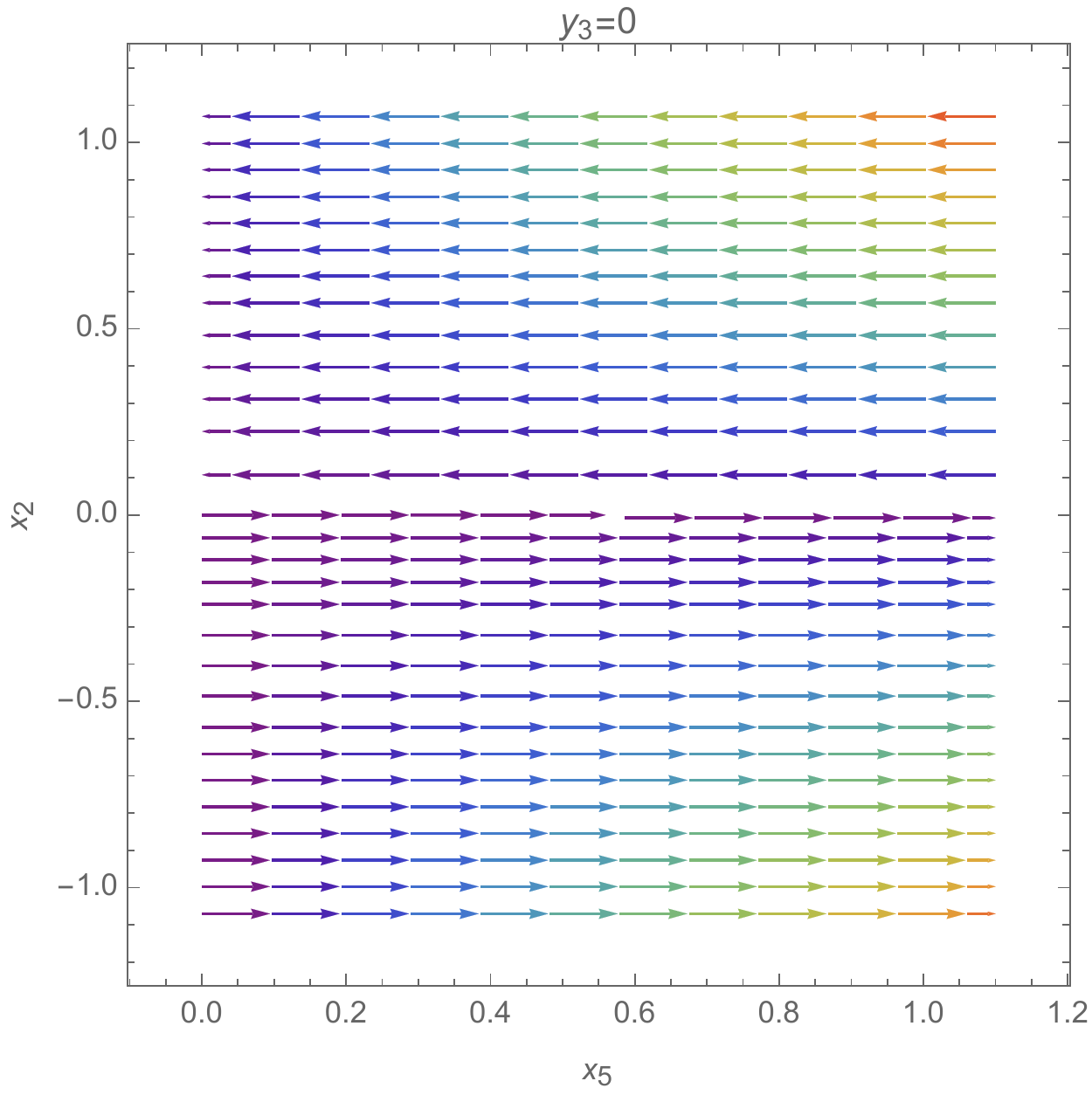}
\includegraphics[width=0.32\textwidth]{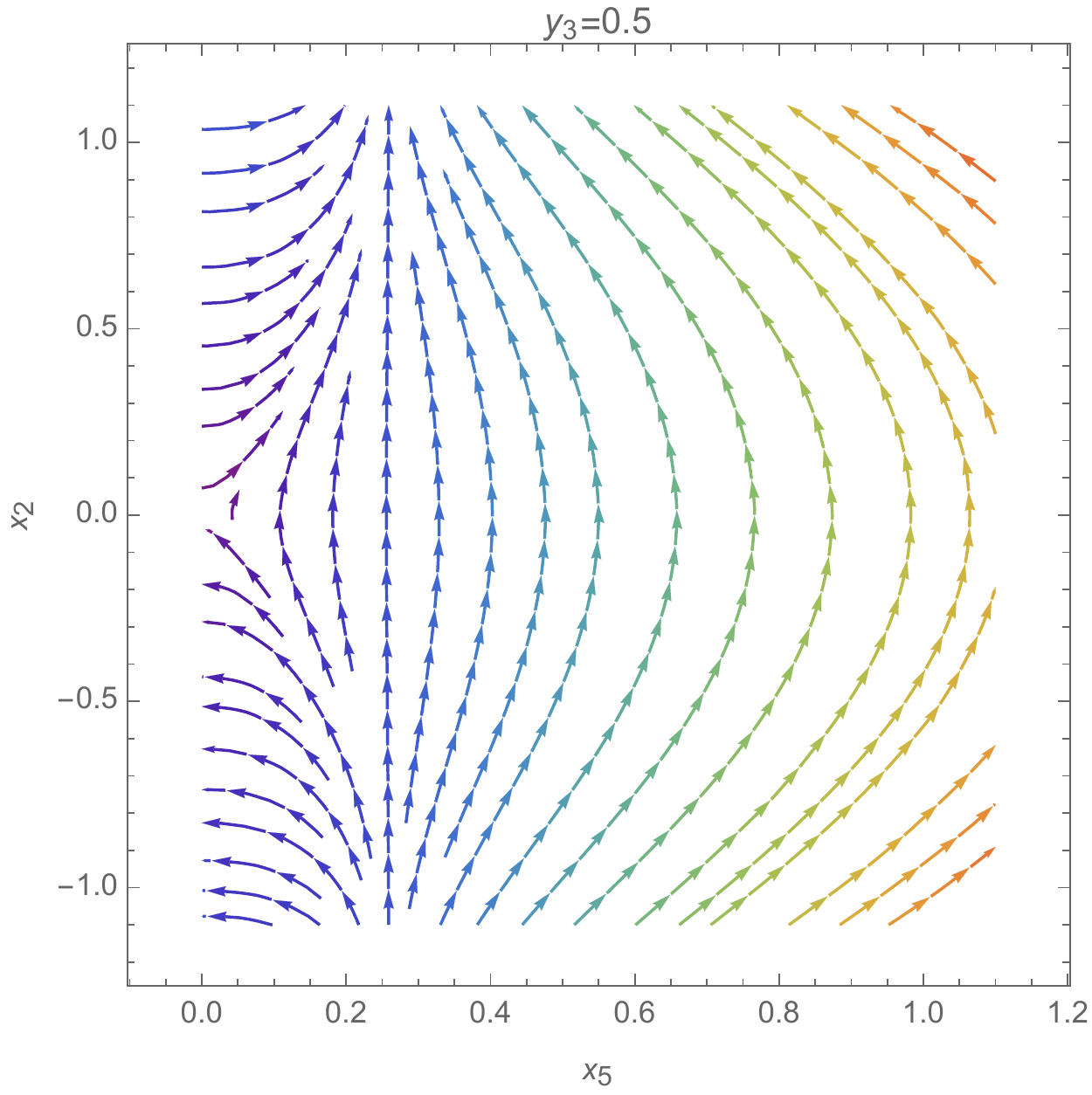}
\includegraphics[width=0.32\textwidth]{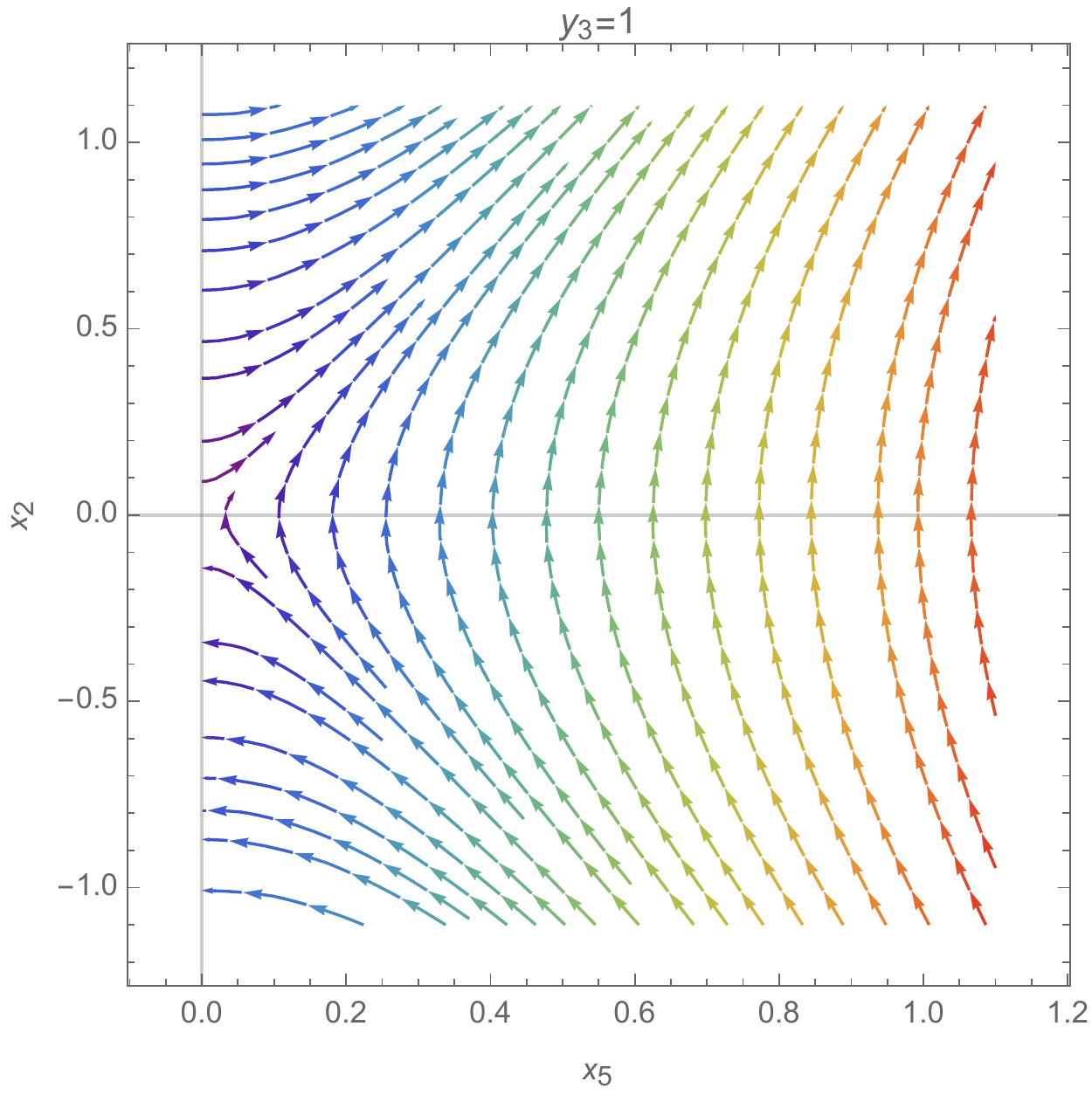}
\includegraphics[width=0.32\textwidth]{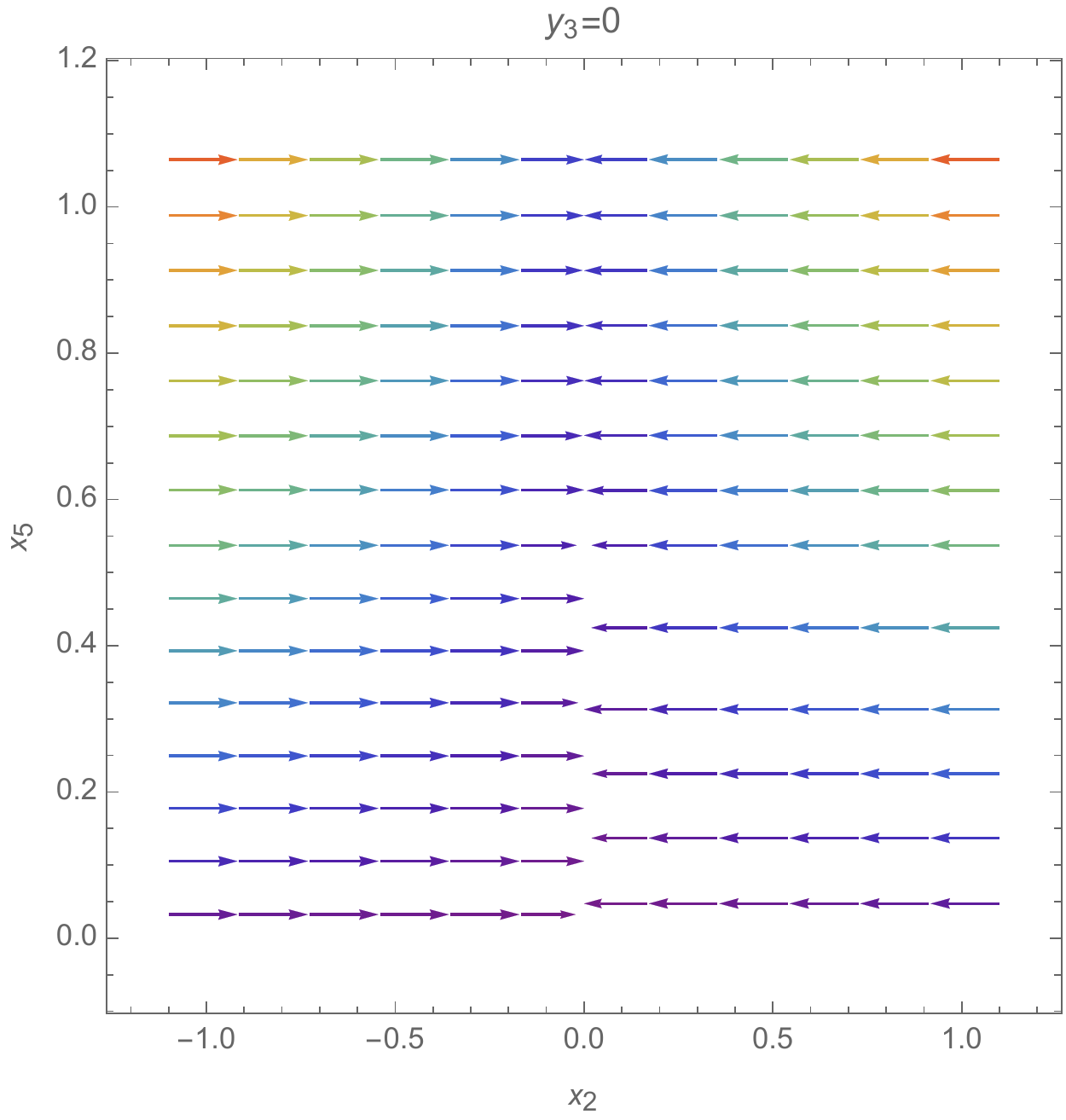}
\includegraphics[width=0.32\textwidth]{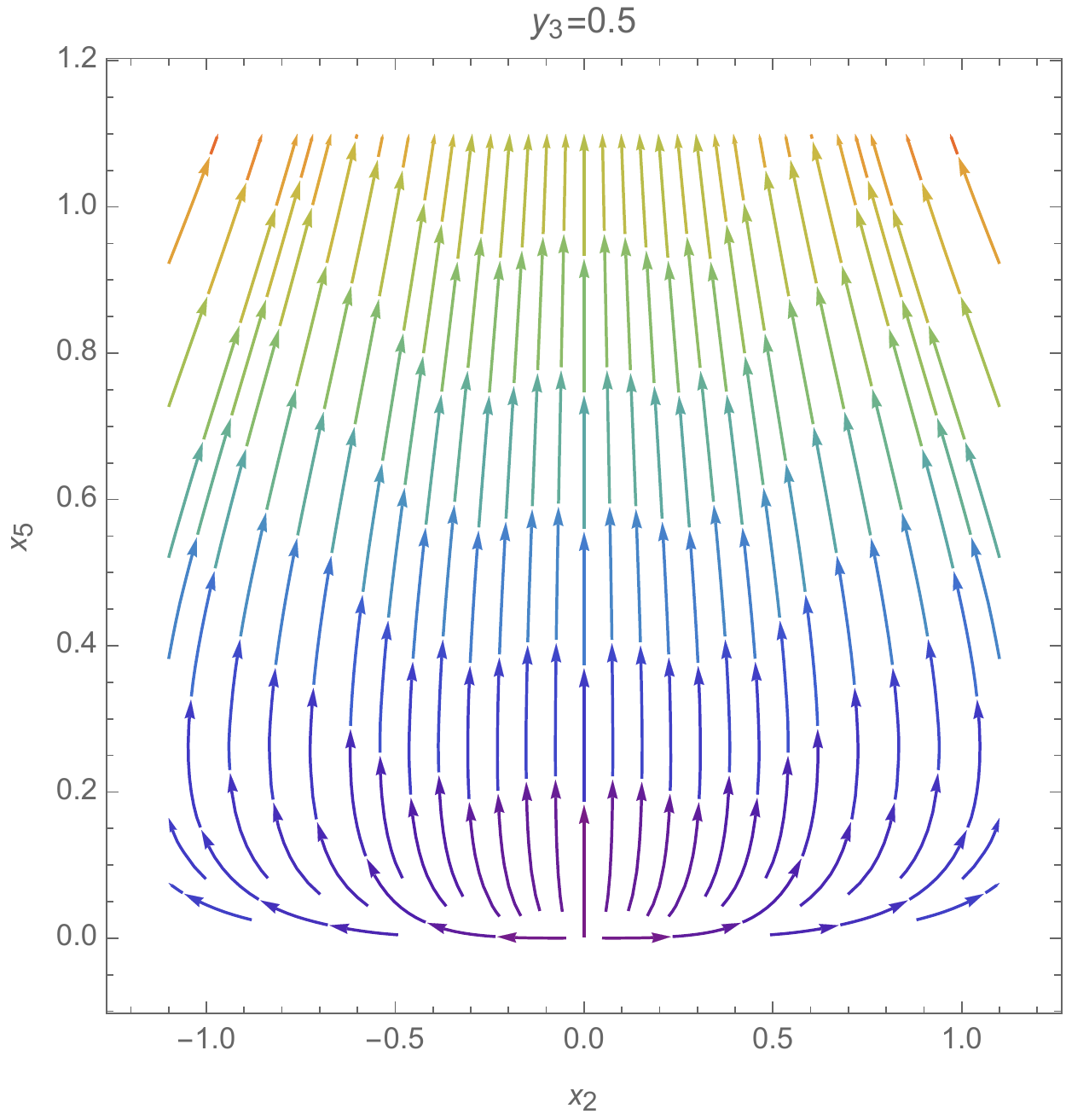}
\includegraphics[width=0.32\textwidth]{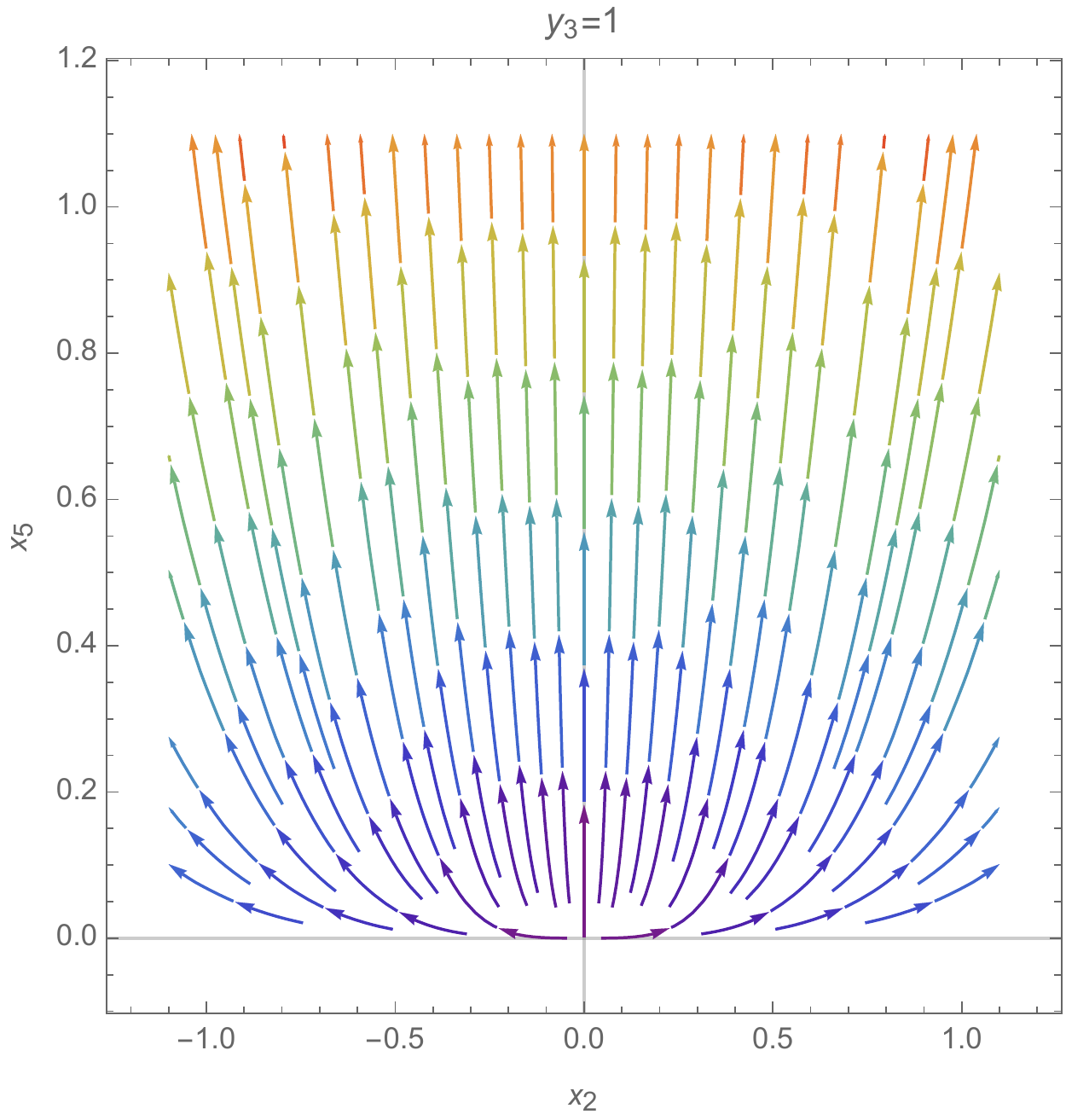}
\caption{Phase portraits for the velocity perturbations ($x_2 =\theta/H$) vs. scale factor ($x_5 =a^2$) in the $\Lambda$CDM model for a DM dominated Universe ($x_4=1$). We show the projection of the stream fields $x_2'-x_5'$ onto the planes $x_5-x_2$ (upper panel) and $x_2-x_5$ (bottom panel), and their evolution for different values of $y_3$. The physical acceptable region for $a^2=x_5$ starts to the right of the vertical line at point (0,0). The phase portrait shows how in a DM-only dominated Universe the point (0,0) (Big Bang) is a maximum for the velocity perturbations when the condition $y_3=1$ is reached ($H=0$). When $y_3=0$ the evolution is just stationary ($H\rightarrow\infty$ or $k=0$).}
\label{fig3-lcdm}
\end{figure*}

As mentioned above, a critical point of the system is a {\itshape saddle point} when it is neither a sink (all $\mathcal{R}e(\lambda)<0$) nor a source ($\mathcal{R}e(\lambda)>0$).

Physically, this illustrates the behavior for the perturbations from a DM to a DE dominated Universe. When the point $x_4=0$ is reached, the solution is completely DE dominated, acting as a cosmological constant. 
\begin{figure*}[!]
\centering
\includegraphics[width=0.45\textwidth]{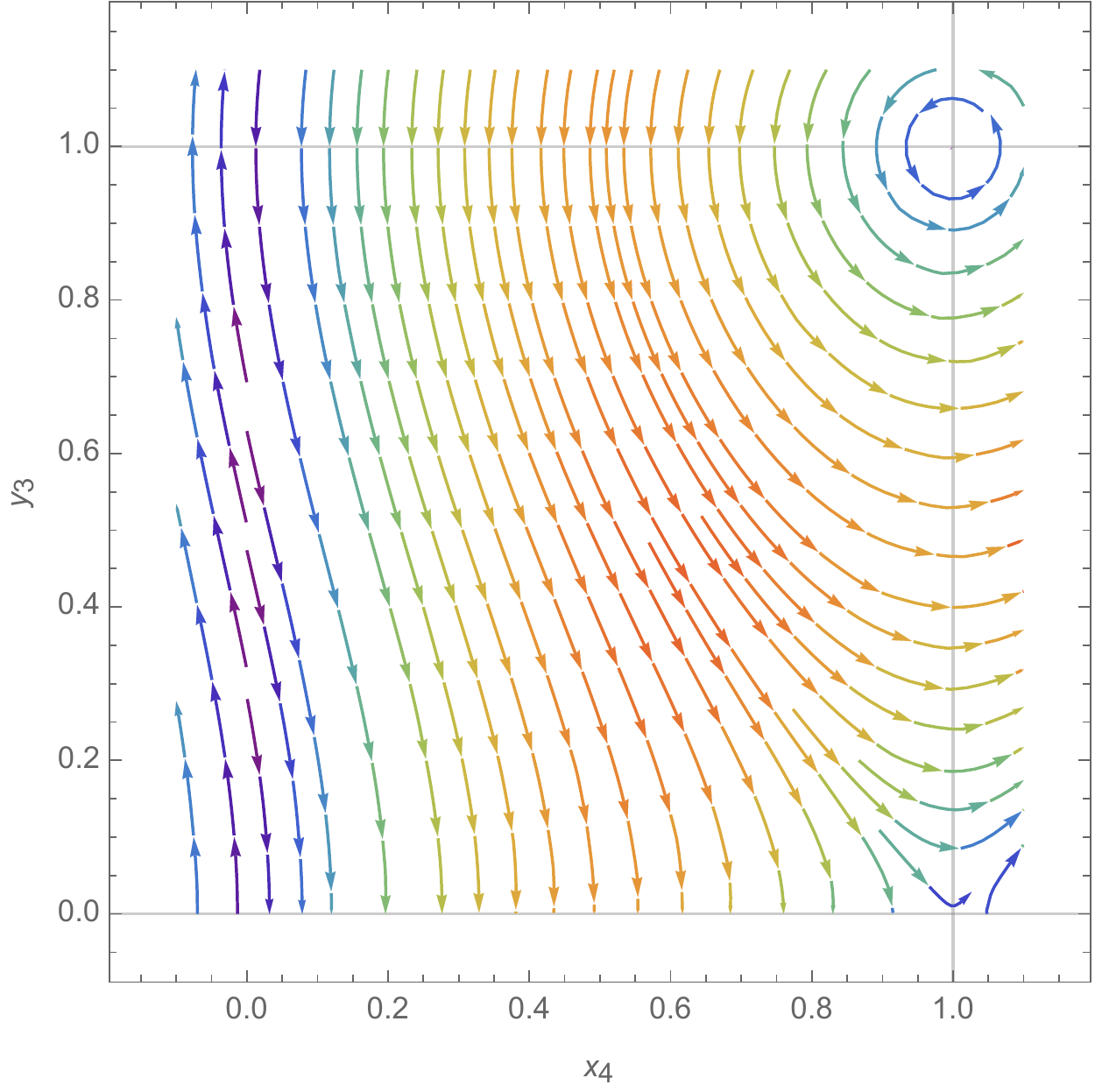}
\includegraphics[width=0.45\textwidth]{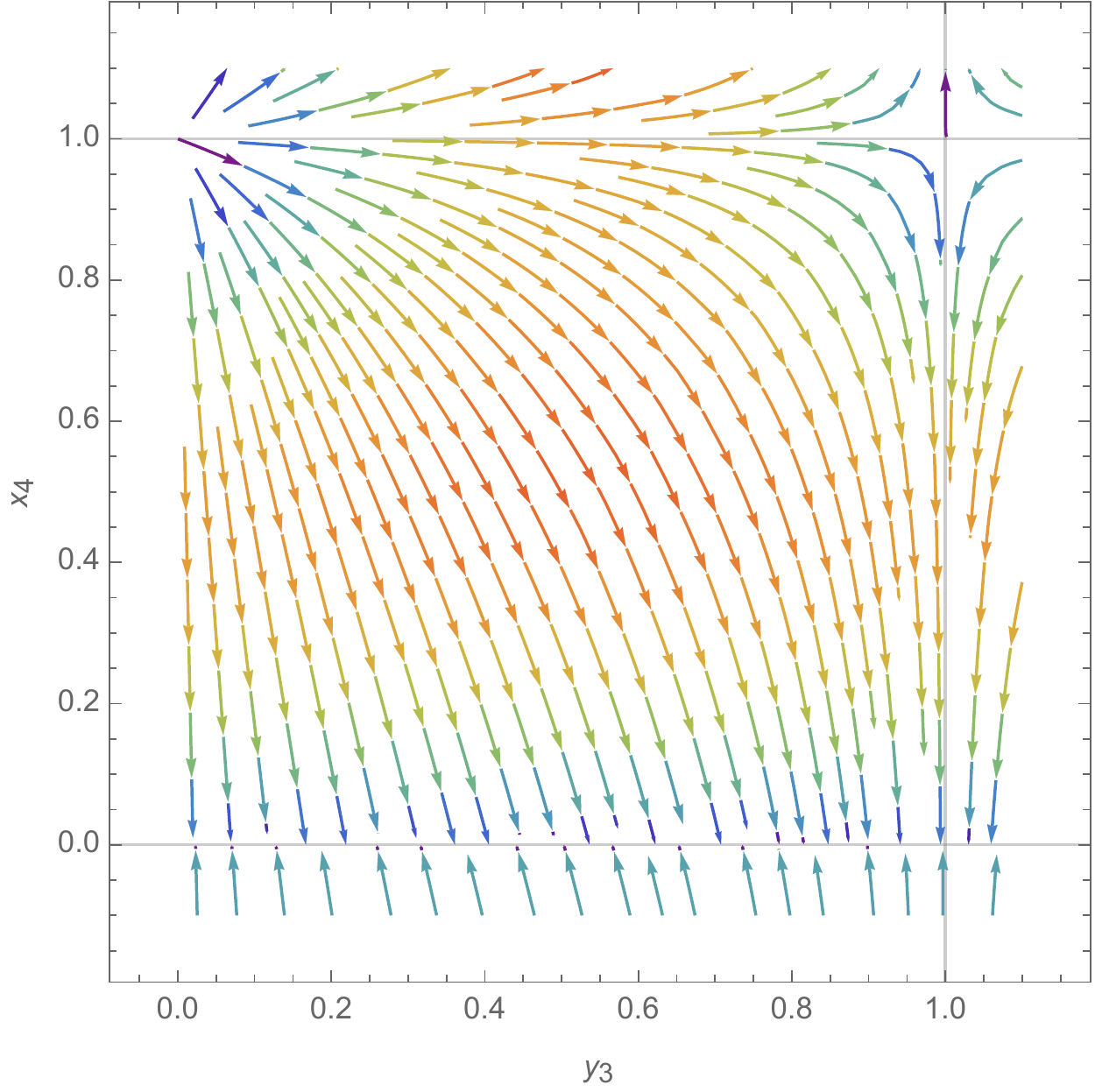}
\captionof{figure}{Phase portraits for $y_3=H^2/k^2$ vs. matter density ($x_4=\Omega_m$) in the $\Lambda$CDM model. The phase portrait shows the behavior of $y_3$ from $x_4=0$ (DE dominated Universe) to $x_4=1$ (DM dominated Universe). We show the projection of the stream fields $y_3'-x_4'$ onto the planes $x_4-y_3$ (left panel) and $y_3-x_4$ (right panel). Notice how the critical points appear only for a DM dominated universe. There is a maximum of energy at (1,0) corresponding to the Big Bang (source) and a minimum at (1,1) corresponding to a static universe (saddle point). 
The physical acceptable region for the matter density $\Omega_m$ is from $x_4=0$ (DE dominated) to $x_4=1$ (DM dominated), we show a wider range for clarity.}
\label{fig4-lcdm}
\end{figure*}
\begin{figure*}[!]
\centering
\includegraphics[width=0.45\textwidth]{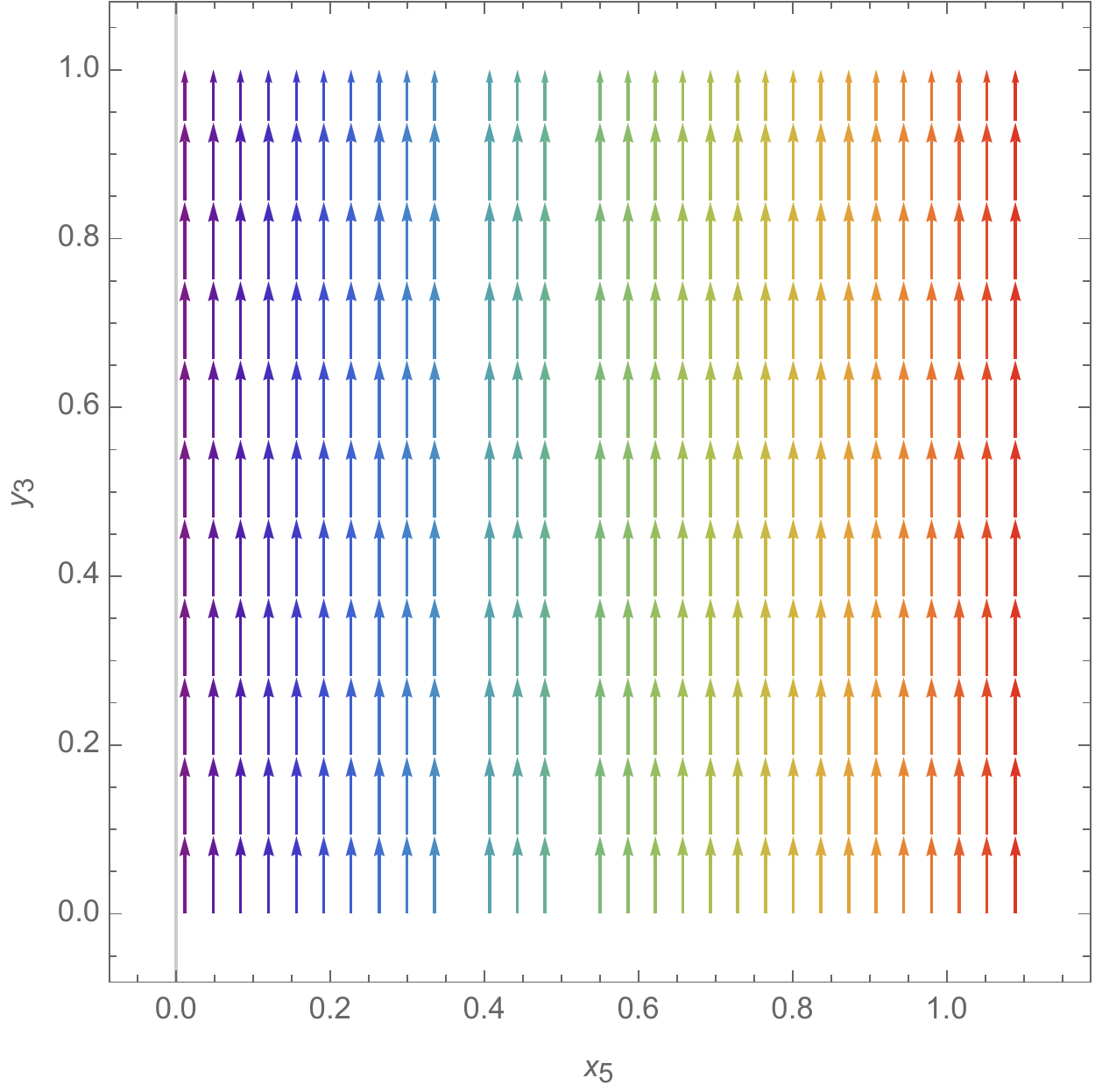}
\captionof{figure}{Phase portrait for $y_3=H^2/k^2$ vs. the scale factor ($x_5=a^2$) in the $\Lambda$CDM model. We show the projection of the stream fields $y_3'-x_5'$ onto the plane $x_5-y_3$ when $x_4=0$ (DE dominated). In this case the evolution is always stationary.} 
\label{fig5-lcdm}
\end{figure*}
\begin{figure*}[!]
\centering
\includegraphics[width=0.45\textwidth]{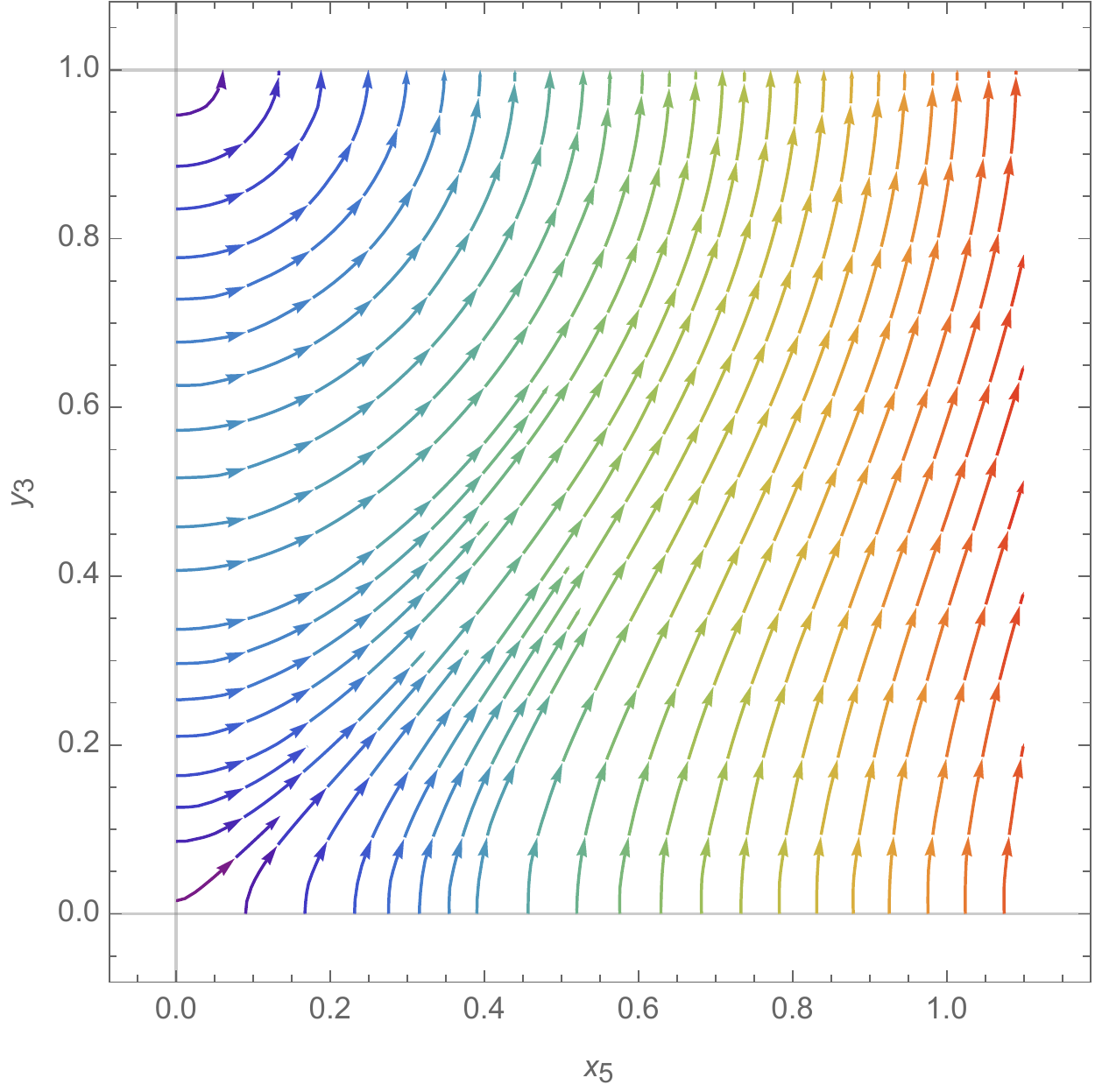}
\includegraphics[width=0.45\textwidth]{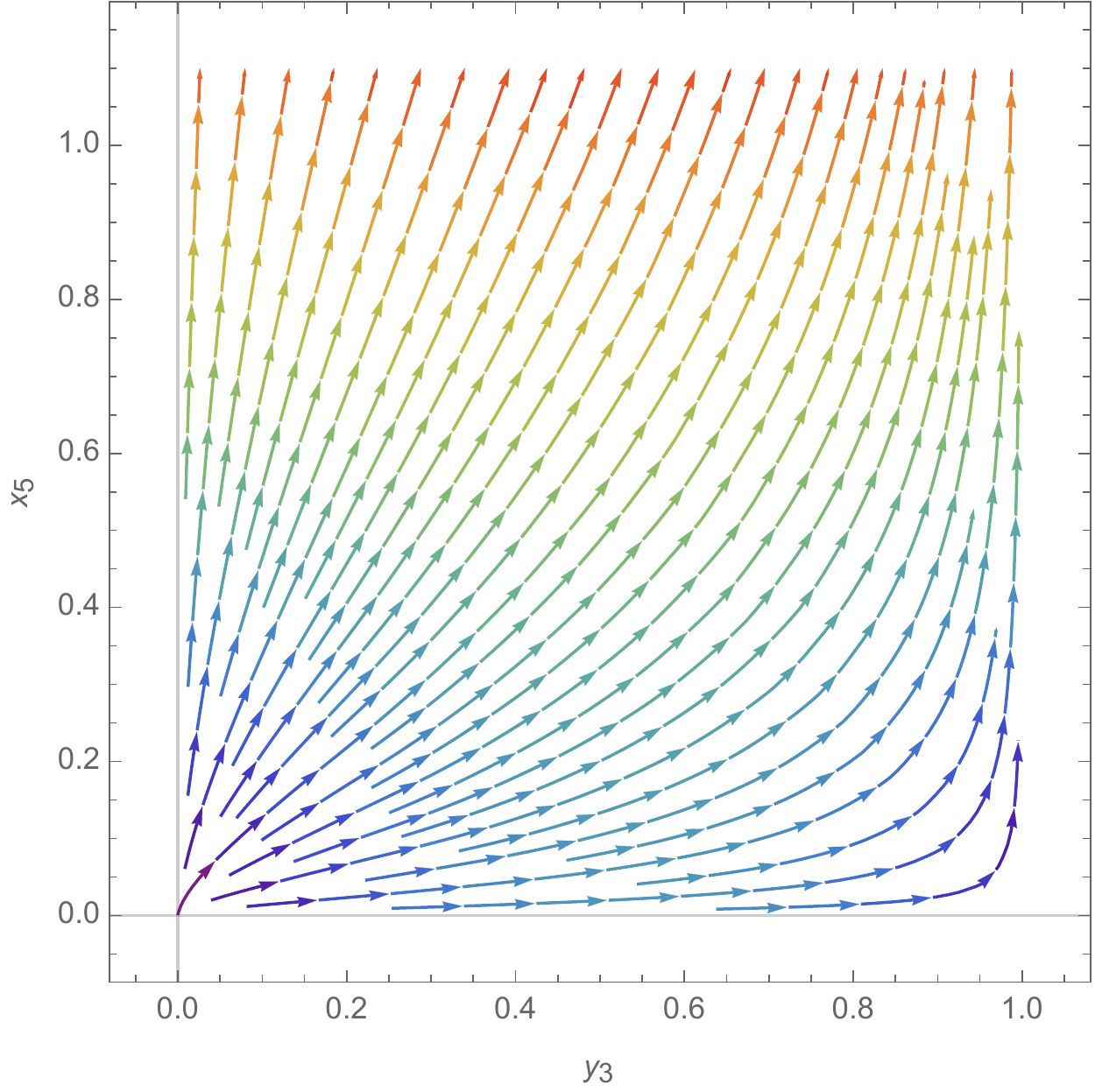}
\captionof{figure}{Phase portrait for $y_3=H^2/k^2$ vs. the scale factor ($x_5=a^2$) in the $\Lambda$CDM model. We show the projection of the stream fields $y_3'-x_5'$ onto the planes $x_5-y_3$ (left panel) and $y_3-x_5$ (right panel) for all $y_3$ for a DM dominated model ($x_4=1$). The phase portraits show the behavior of the parameter $y_3=H^2/k^2$ from $x_5=0$ ($a=0$) (initial singularity) to $x_5=1$ (today, $a=1$). It can be seen how the initial condition $a=0$ with $y_3=0$ corresponding to the point $(0,0)$ is a maximum of energy (source point corresponding to the Big Bang singularity). The point (0,1) corresponds to a minimum (saddle point corresponding to a static universe). The physical acceptable region starts to the right of the vertical line at point (0,0). We show a wider range for more clarity.}
\label{fig4-lcdm}
\end{figure*}
\begin{figure*}[!]
\centering
\includegraphics[width=0.45\textwidth]{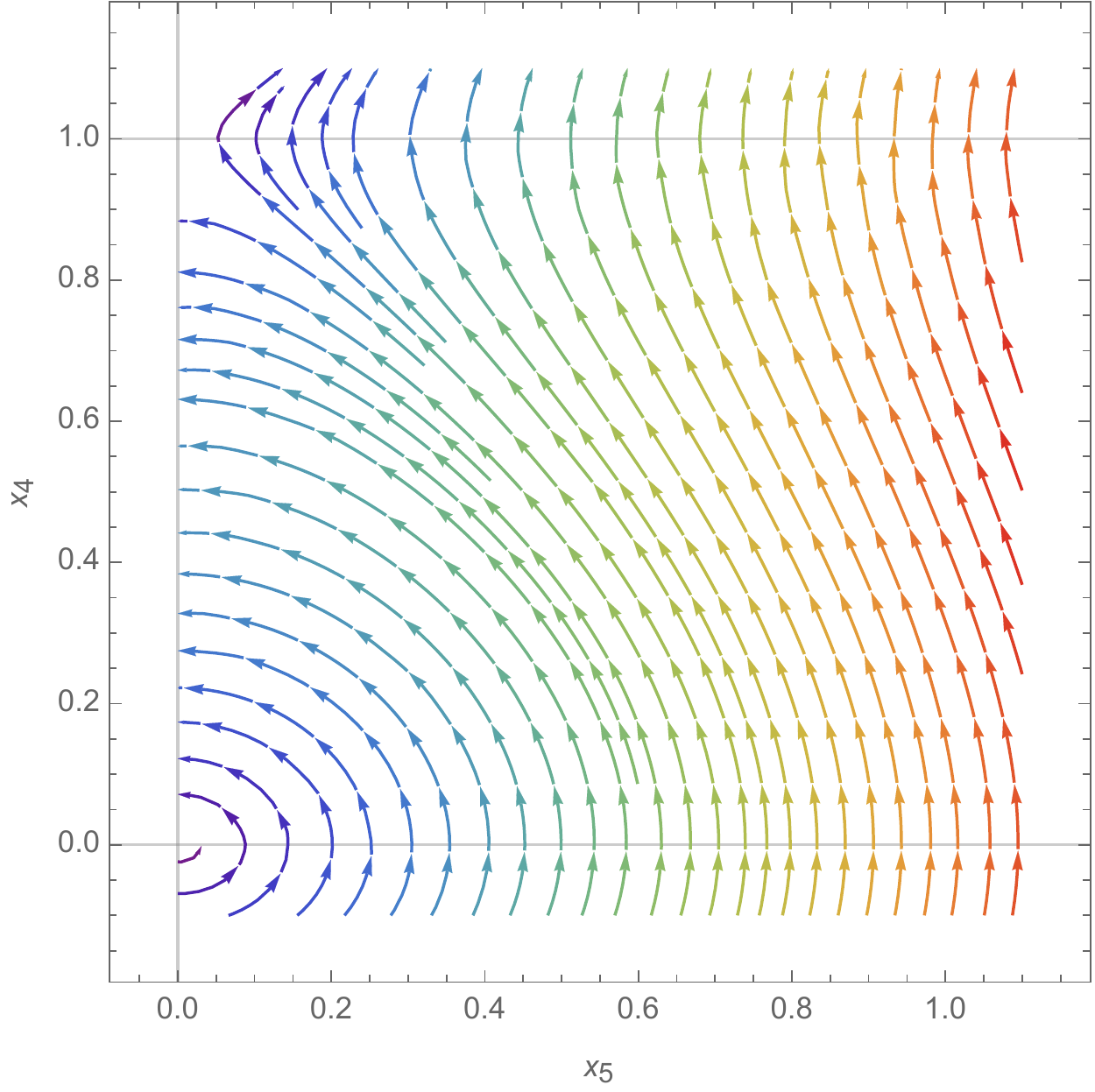}
\includegraphics[width=0.45\textwidth]{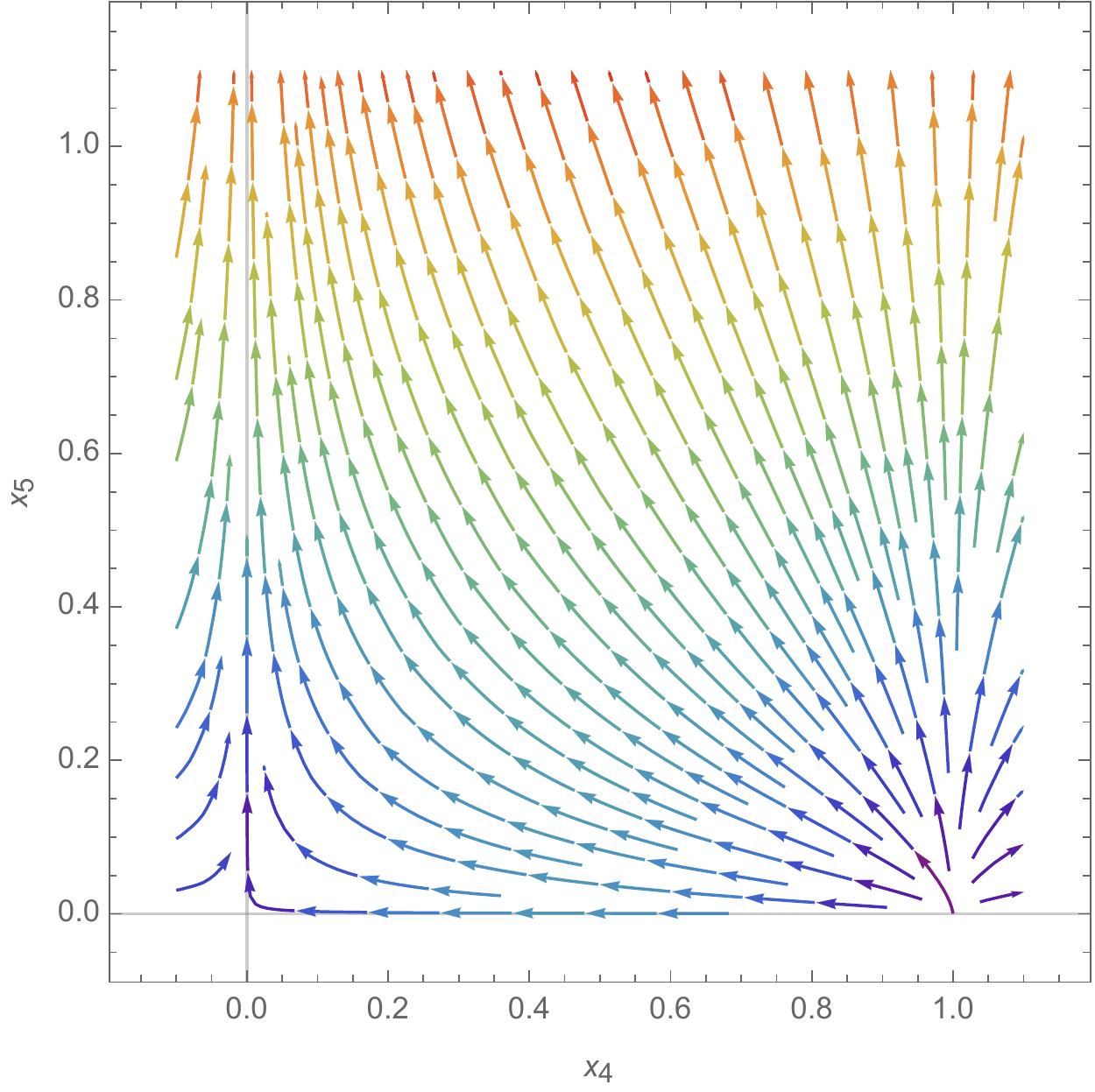}
\captionof{figure}{Phase portrait for the matter density ($x_4=\Omega_m$) vs the scale factor ($x_5=a^2$). We show the projection of the stream fields $x_4'-x_5'$ onto the planes $x_5-x_4$ (left panel) and $x_4-x_5$ (right panel) for all $y_3$. The phase portraits show the behavior of the parameter $\Omega_m$ from $x_5=0$ (initial singularity, $a=0$) to $x_5=1$ (today, $a=1$). It can be seen how the initial condition $a=0$ with $x_4=0$ (DE dominated universe) corresponds to a minimum for the DM density, corresponding to the point $(0,0)$ (saddle point). Finally, the point (0,1) corresponds to a maximum in energy (a source corresponding to the Big Bang in a DM dominated universe). The physical acceptable region starts to the right of the vertical line at point (0,0). We show a wider range for more clarity.}
\label{fig4-lcdm}
\end{figure*}
\subsection{Modified gravity models}
In this section we study some of the most significant linear MG models in cosmology, making use of the full DS given by Eqs.~\eqref{eq-x1}-\eqref{eq-x5} (see Table \ref{tab:ST-models}). In principle, cosmological observations of cosmic structure on linear scales can be used to measure $\mu$ and $\gamma$, and to constrain specific models.
It is overall required that GR holds at early times, meaning that $s>0$.
\subsubsection{Scale-Independent Parametrizations}
In the scale-independent case, the modifications of gravity are completely described by the dynamics of the perturbations at large scales, and the dynamical variable connected to $\gamma=\mu$ is given by
\begin{equation}
	\gamma=\mu=1+\beta(\sqrt {x_5})^s,
\label{gm}
\end{equation}
with $\beta$ and $s$ constants. This parametrization is purely phenomenological and it is motivated by the arguments given in Section~\ref{sip}.

From Eqs.~\eqref{eq-x1}-\eqref{eq-x5} it was found that $x_5=0$ corresponds to a critical point for the system, and we note that the scale-independent parametrization depends only on this variable. From Eq.~\eqref{gm} we then have, $\gamma=\mu=1$, at the critical points. With these values and comparing this system with the DS of the $\Lambda$CDM case (Eqs.~\eqref{lcdme1}-\eqref{lcdme5}), we can see that the physics of the scale-independent DS compared to that of $\Lambda$CDM is the same. The phase portraits for this case are exactly the same to those obtained for the $\Lambda$CDM model at the level of the perturbations, so we do not show them here again. In conclusion, from the DS point of view, there is no difference between $\Lambda$CDM and the scale-independent MG theories.
\subsubsection{General scale-dependent parametrizations}
From the point of view of model testing, scale-dependent MG introduces a richer phenomenology compared to the scale-independent case treated in the previous subsection. These models have two extra parameters plus the exponent giving the time dependence. More importantly, the natural coherence of $\Lambda$CDM perturbations is lost in these models since a characteristic scale is introduced, below which modifications to gravity show up. This key feature brings up interesting phenomenological features of LSS formation in the context of MG theories.

Now, as mentioned in Sec.~\ref{scdp}, an appropriate way to encompass all theories having a modification of gravity is through the Horndeski class, for which the dimensionless PPF parametrizations are given by the following equations, according to  Eqs.~\eqref{mu}-\eqref{gamma}:
\begin{eqnarray}
\mu(l_1,x_5)&=&\frac{1+\beta_1 l_1x_5^{s/2}}{1+l_1x_5^{s/2}},\\
\gamma(l_2,x_5)&=&\frac{1+\beta_2 l_2x_5^{s/2}}{1+l_2x_5^{s/2}},
\end{eqnarray}
where $\beta_1$ and $\beta_2$ are fixed dimensionless coupling constants and $s$ is a dimensionless constant that should be constrained by observations. In this parametrization, for convenience, we have defined the new auxiliary and dimensionless parameter
\begin{equation}
l_i\equiv\lambda_i^2 k^2\mbox{ for } i=1,2.
\end{equation}
For simple PPF models, the parameters $\lambda_i$ define characteristic length-scales for a given modified model: $\lambda_1$ corresponds to the scale below which the gravitational potential $\phi$ turns out to be modified. 
Similarly, $\lambda_2$ is the upper bound for scales of modified modes of the scalar curvature perturbation $\psi$. Notice that the $l_i$'s are the only free dimensionless parameters which depend on the wavenumber (scale of the perturbation), and correspond to the ratio of the wavelength of the perturbation to the characteristic scales of the model. 
\subsubsection{Parametrization in $f(R)$ and Chameleon-like models}
In this subsection we make use of the general DS given by Eqs.~\eqref{eq-x1}-\eqref{eq-x5} to find the special characteristics for the I and II-$f(R)$ and Chameleon-like models shown in Table \ref{tab:ST-models}. To this end, we have written $\lambda_2$ in terms of $\lambda_1$ as: $\lambda_2=\sqrt{\lambda_1^2\beta_1}$, so the number of parameters reduces to four ($\lambda_1$, $\beta_1$, $\beta_2$ and $s$), with only one free parameter given by the wavenumber $k$.
 
In what follows, the phase portraits shown will be only those of the variables that seem to be most affected by the modifications to GR. The rest behave in the same manner as in the $\Lambda$CDM model.
\begin{figure*}[!]
\includegraphics[width=0.32\textwidth]{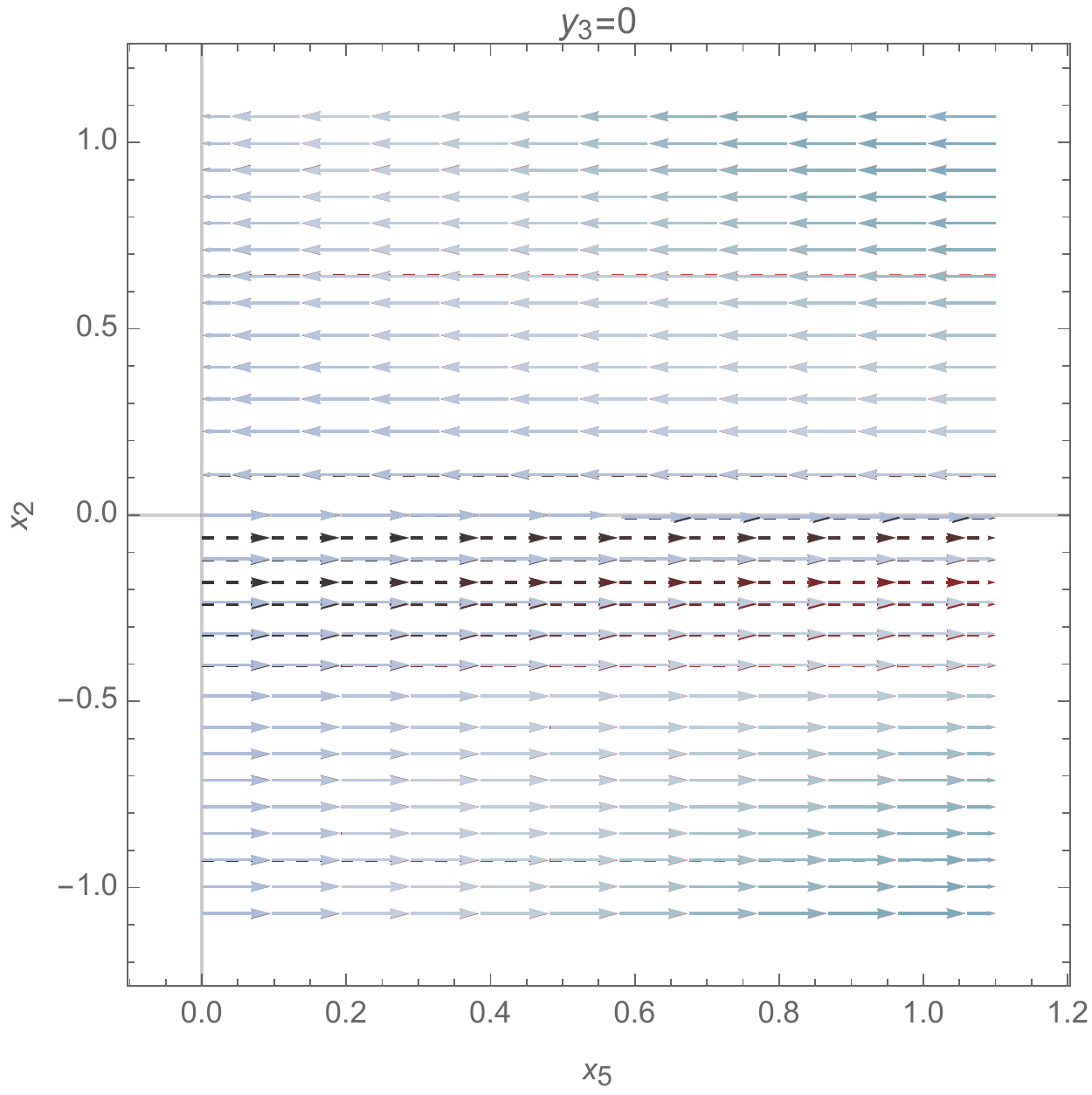}
\includegraphics[width=0.32\textwidth]{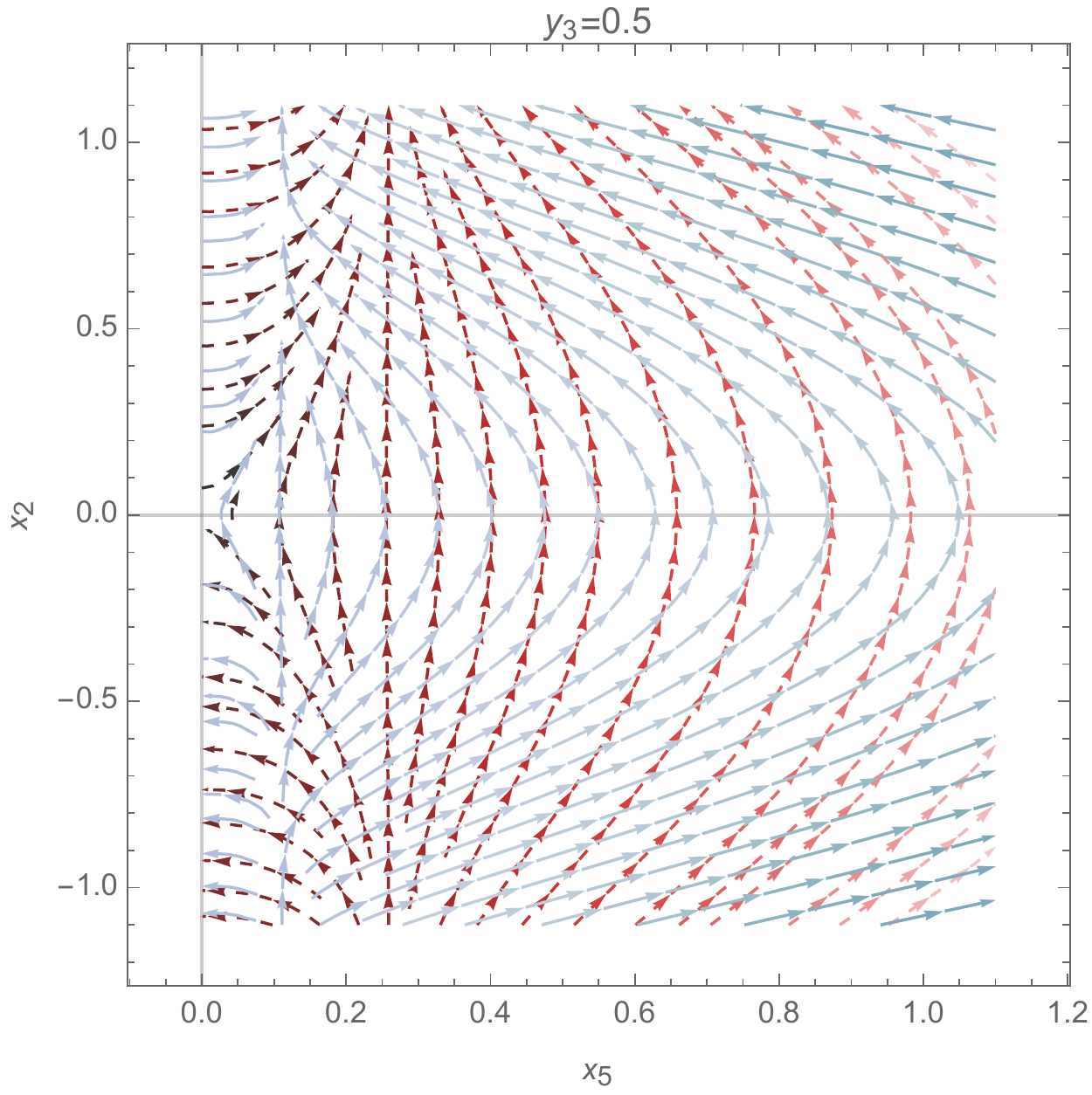}
\includegraphics[width=0.32\textwidth]{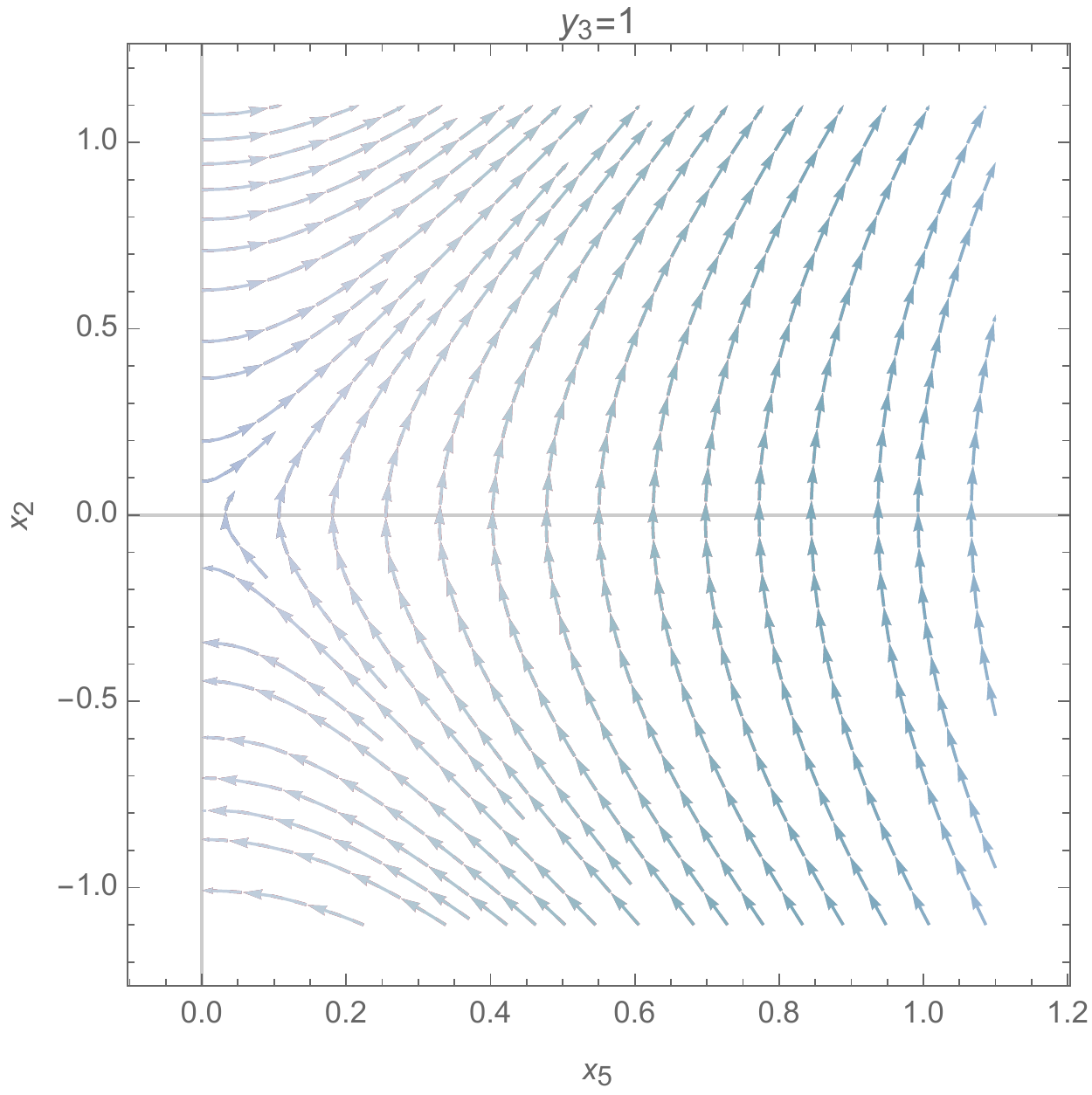}
\includegraphics[width=0.32\textwidth]{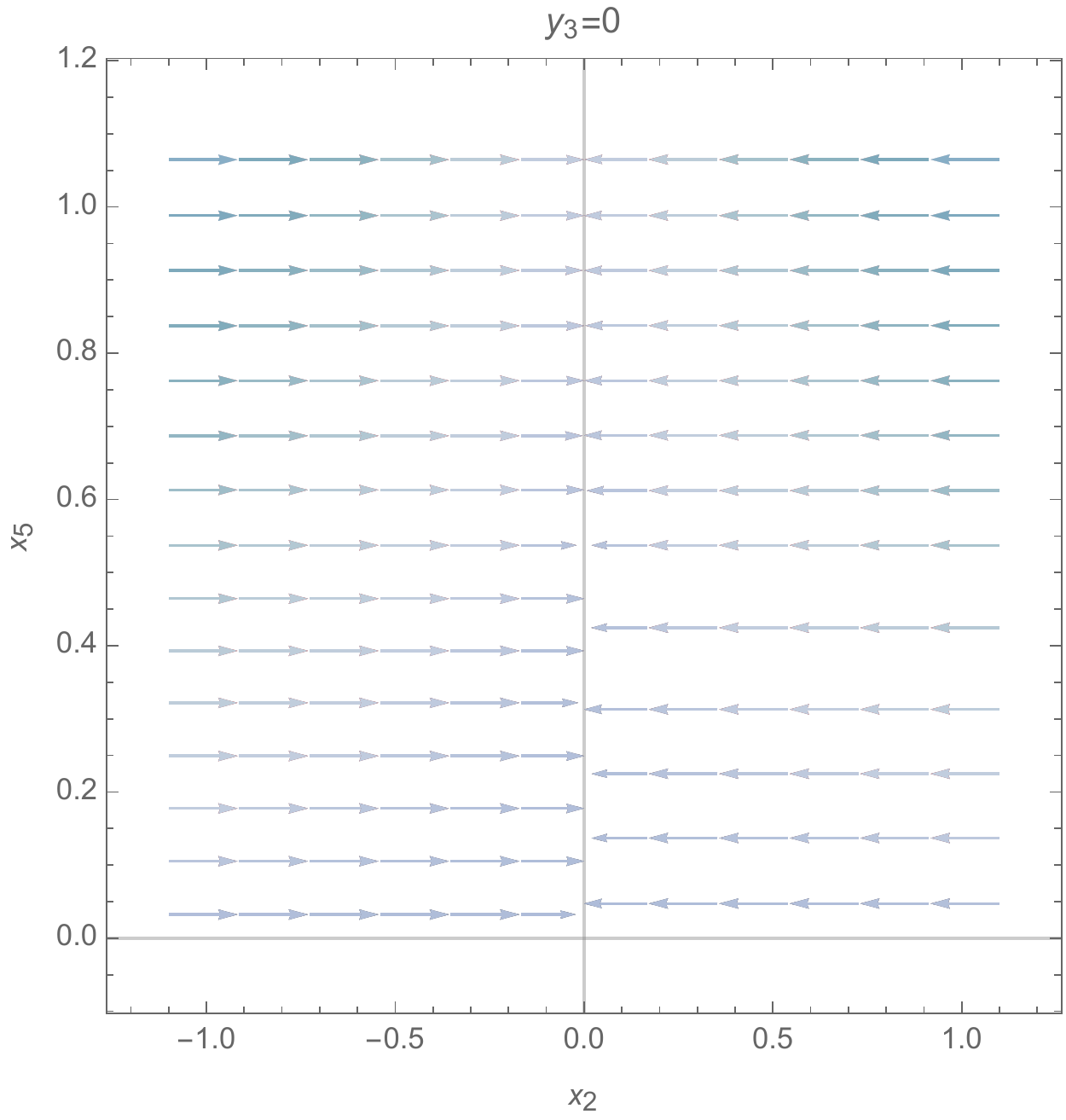}
\includegraphics[width=0.32\textwidth]{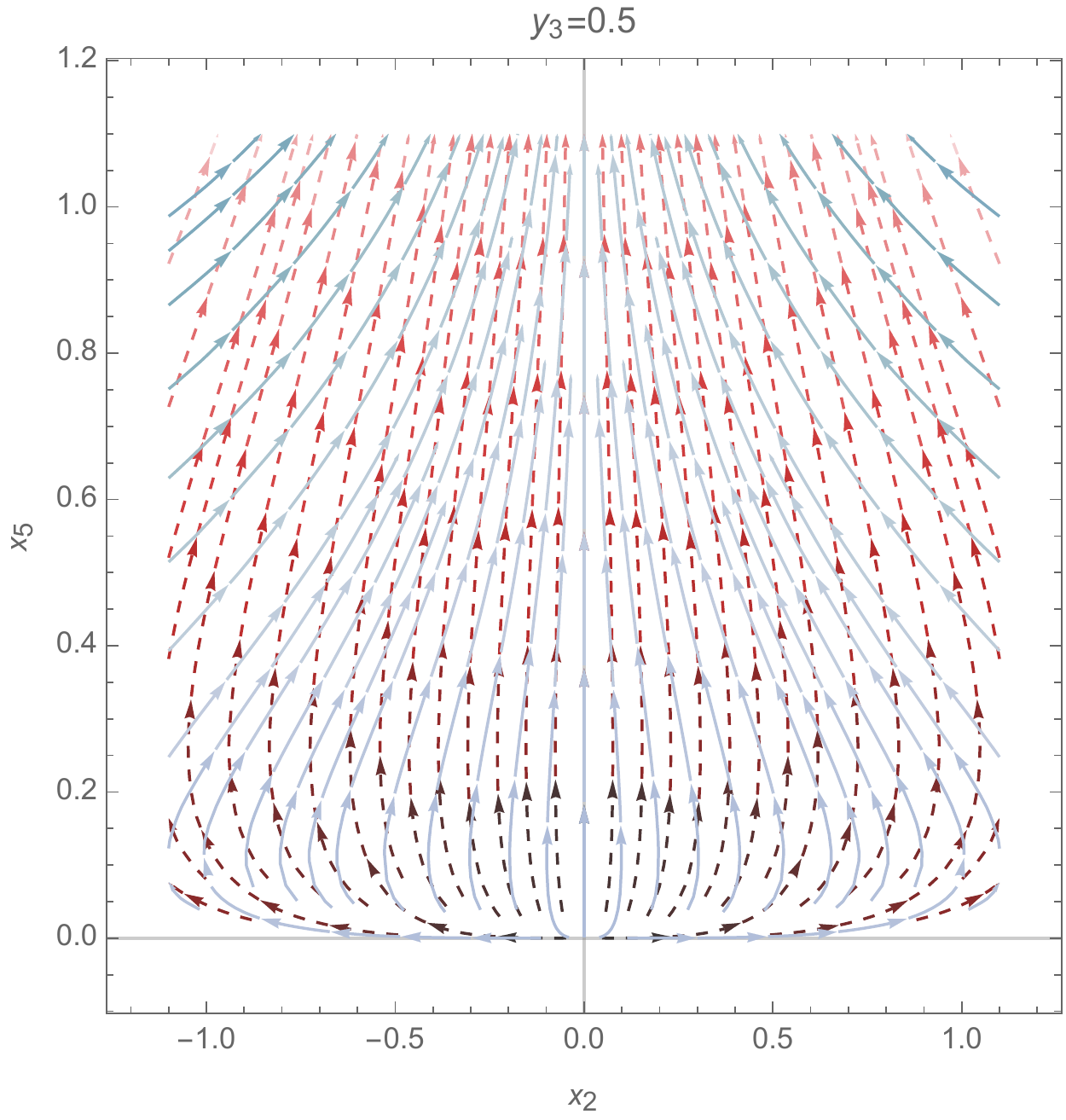}
\includegraphics[width=0.32\textwidth]{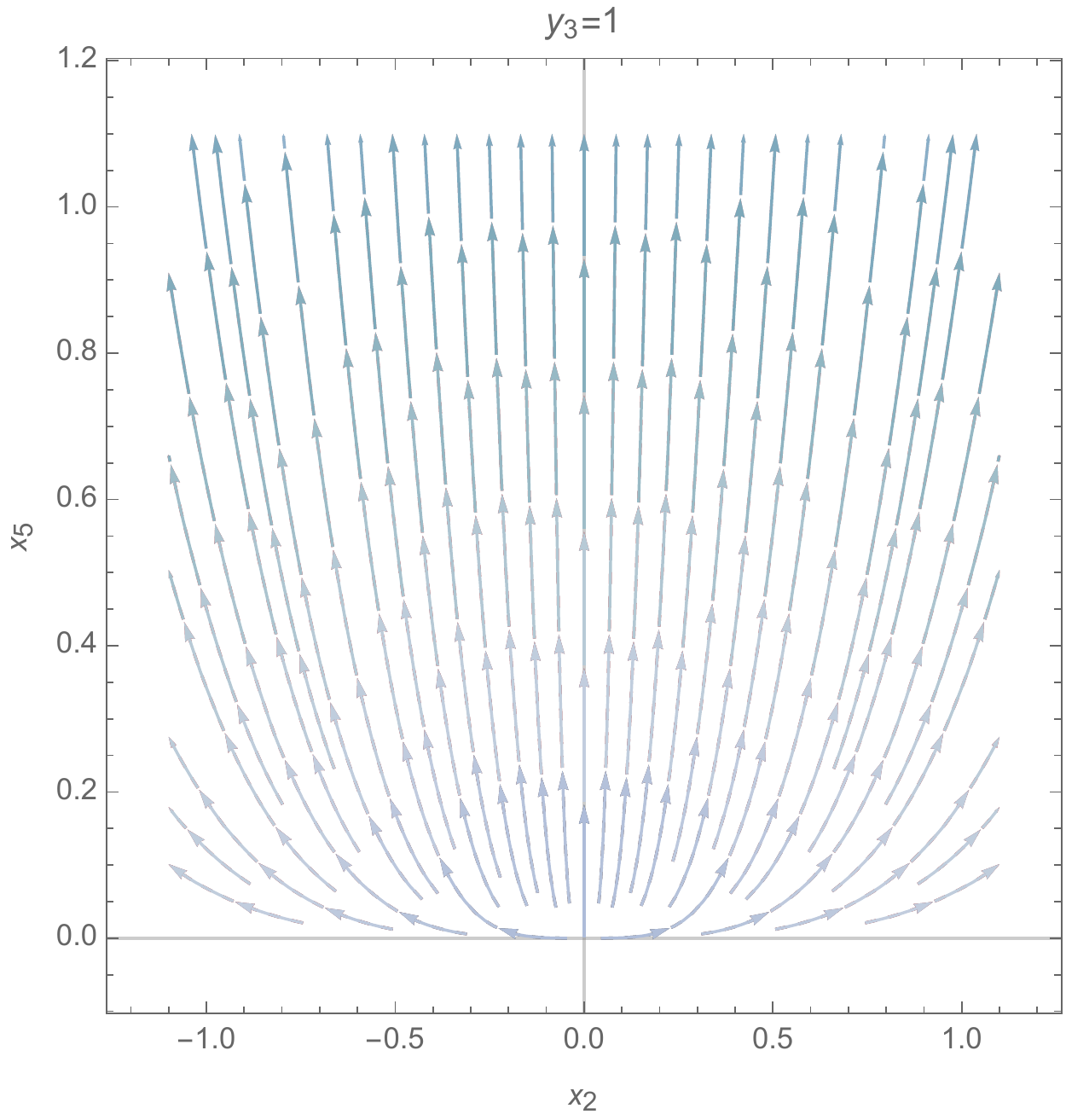}
\captionof{figure}{Phase portraits for the velocity perturbations ($x_2=\theta/H$) vs. scale factor ($x_5=a^2$) in MG models. Projection of the stream fields $x_2'-x_5'$ onto the plane $x_5-x_2$ (upper panel) and $x_2-x_5$ (bottom panel). The stream fields represent the MG models, $f(R)$ (blue) vs. the $\Lambda$CDM model (red), for $k=0.012h\text{Mpc}^{-1}$. The phase portraits show the behavior of the stream field of the velocity perturbation from $y_3=0$ to $y_3=1$. There is only one critical point at (0,0) corresponding to the Big Bang (source).}
\label{fig:phporx2}
\end{figure*}
\begin{figure*}
\centering
\includegraphics[width=0.32\textwidth]{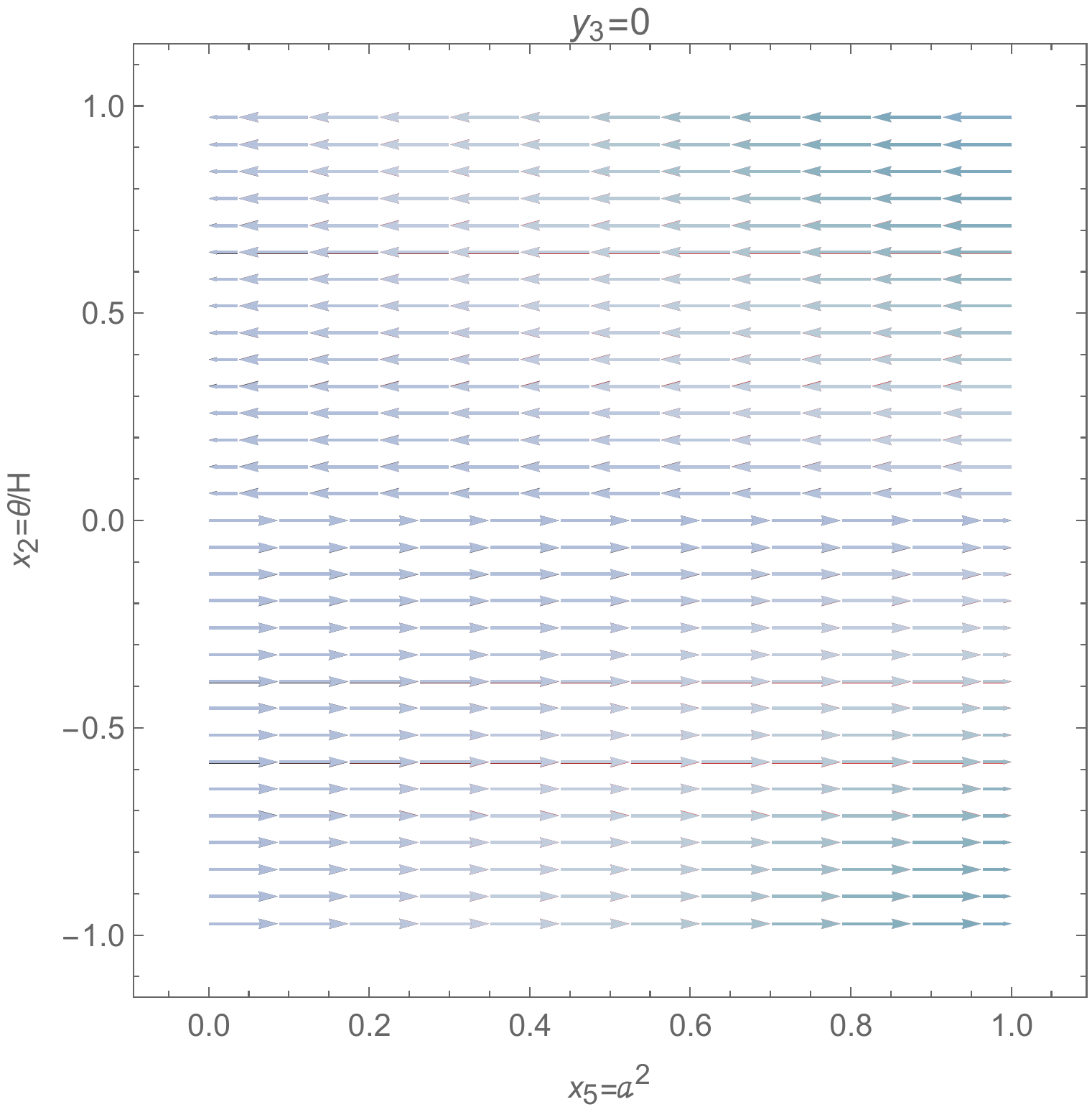}
\includegraphics[width=0.32\textwidth]{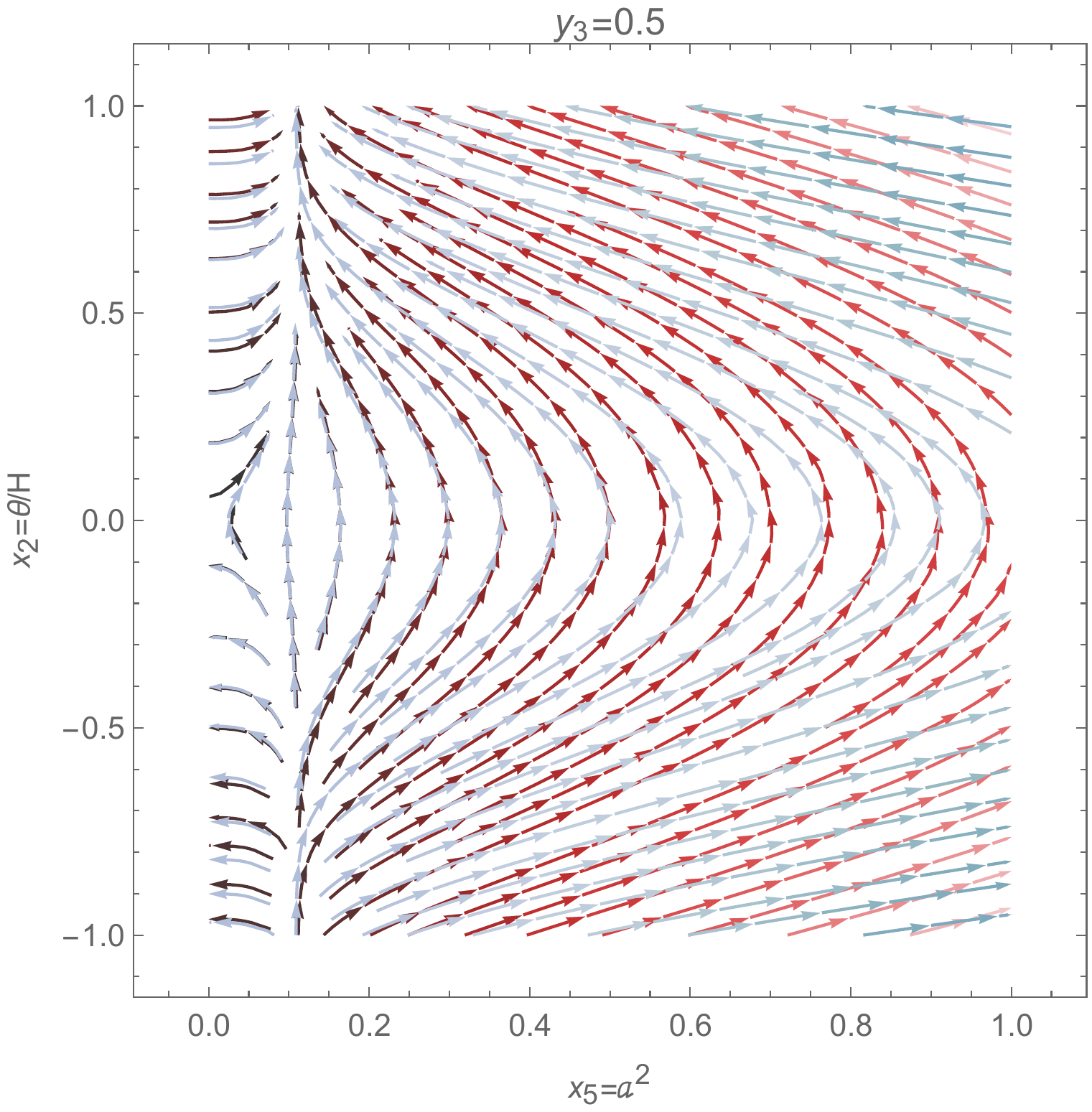}
\includegraphics[width=0.32\textwidth]{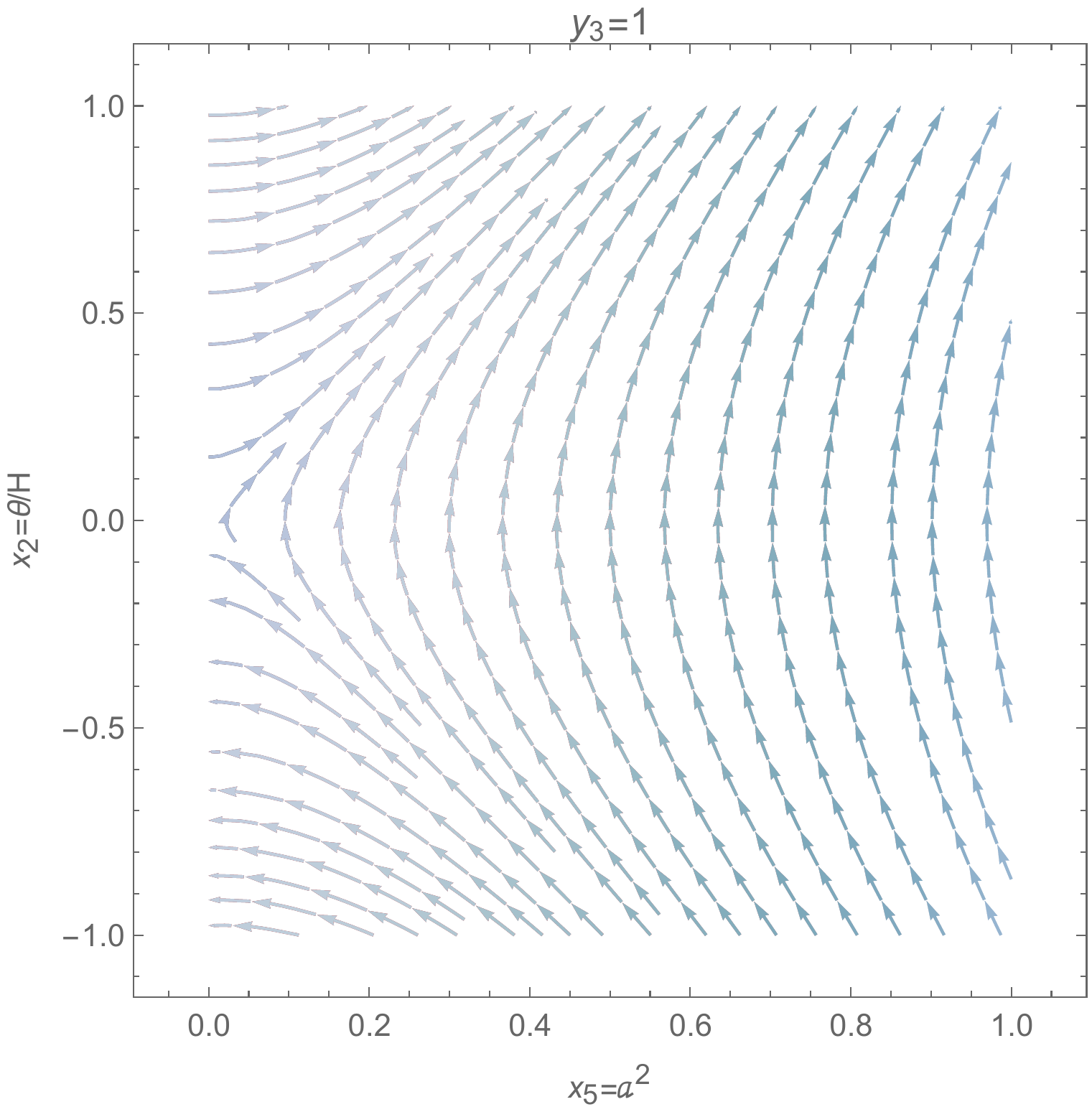}
\includegraphics[width=0.32\textwidth]{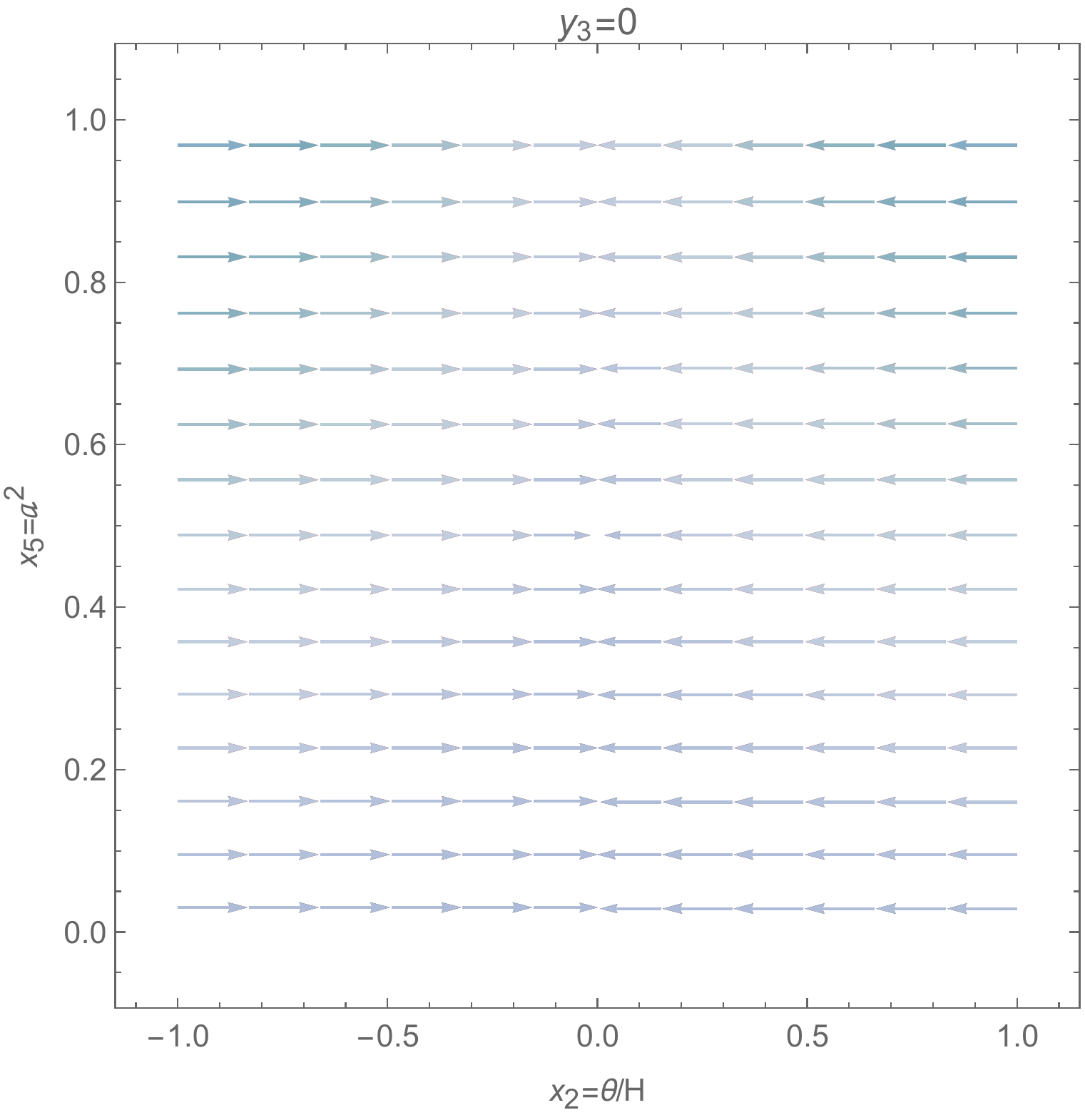}
\includegraphics[width=0.32\textwidth]{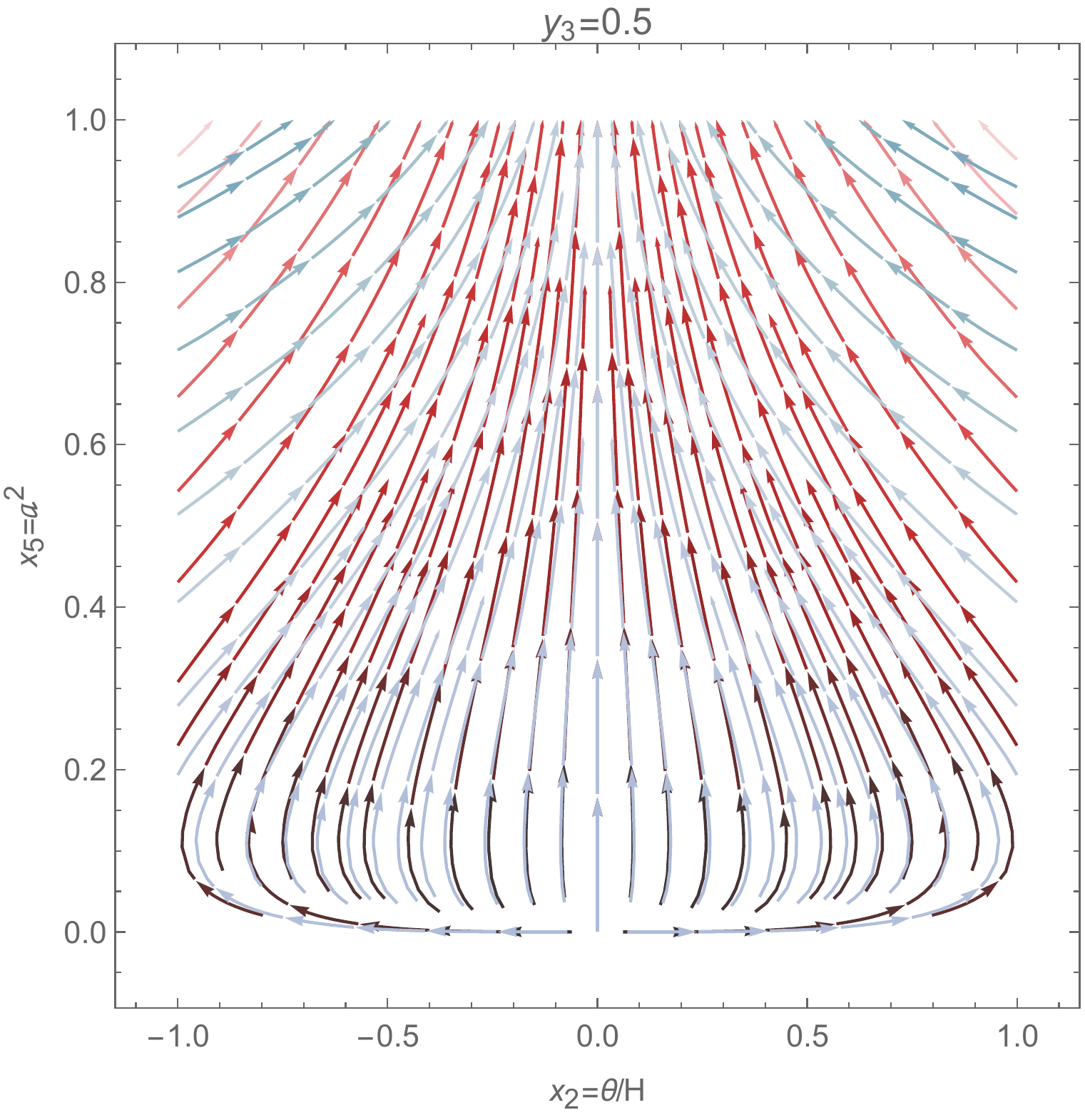}
\includegraphics[width=0.32\textwidth]{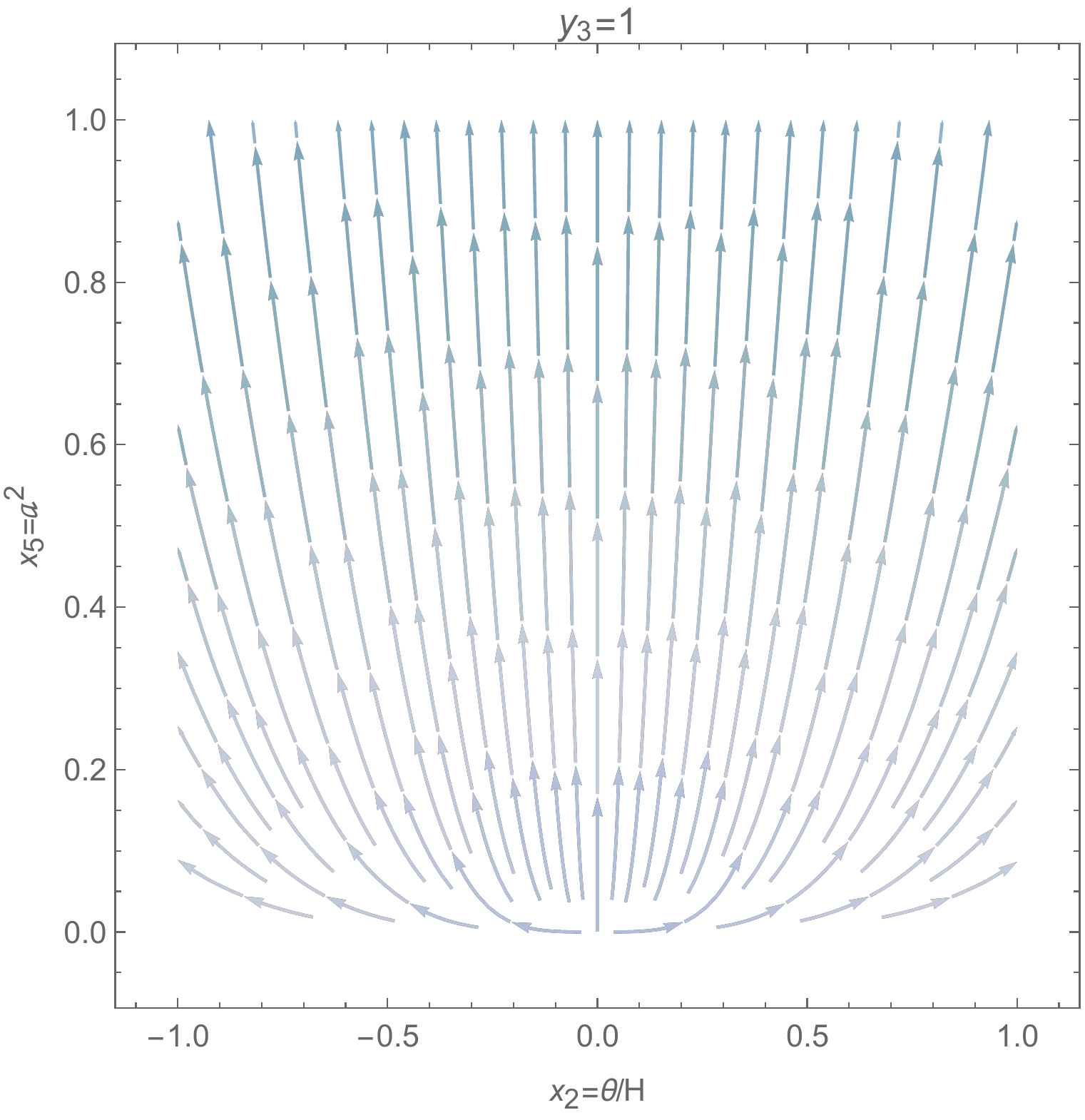}
\caption{Phase portraits for the velocity perturbations ($x_2=\theta/H$) vs. scale factor ($x_5=a^2$) in $f(R)$ models (blue) compared to $\Lambda$CDM (red), for $k=0.05h\text{Mpc}^{-1}$. Projection of the stream fields $x_2'-x_5'$ onto the planes $x_5-x_2$ (upper panel) and $x_2-x_5$ (bottom panel). The phase portraits show the behavior of the stream field of the velocity perturbations from $y_3=0$ to $y_3=1$. There is only one critical point at (0,0) corresponding to the Big Bang (source).}
\label{fig:lcdm-fR}
\end{figure*}
\begin{figure*}
\centering
\includegraphics[width=0.32\textwidth]{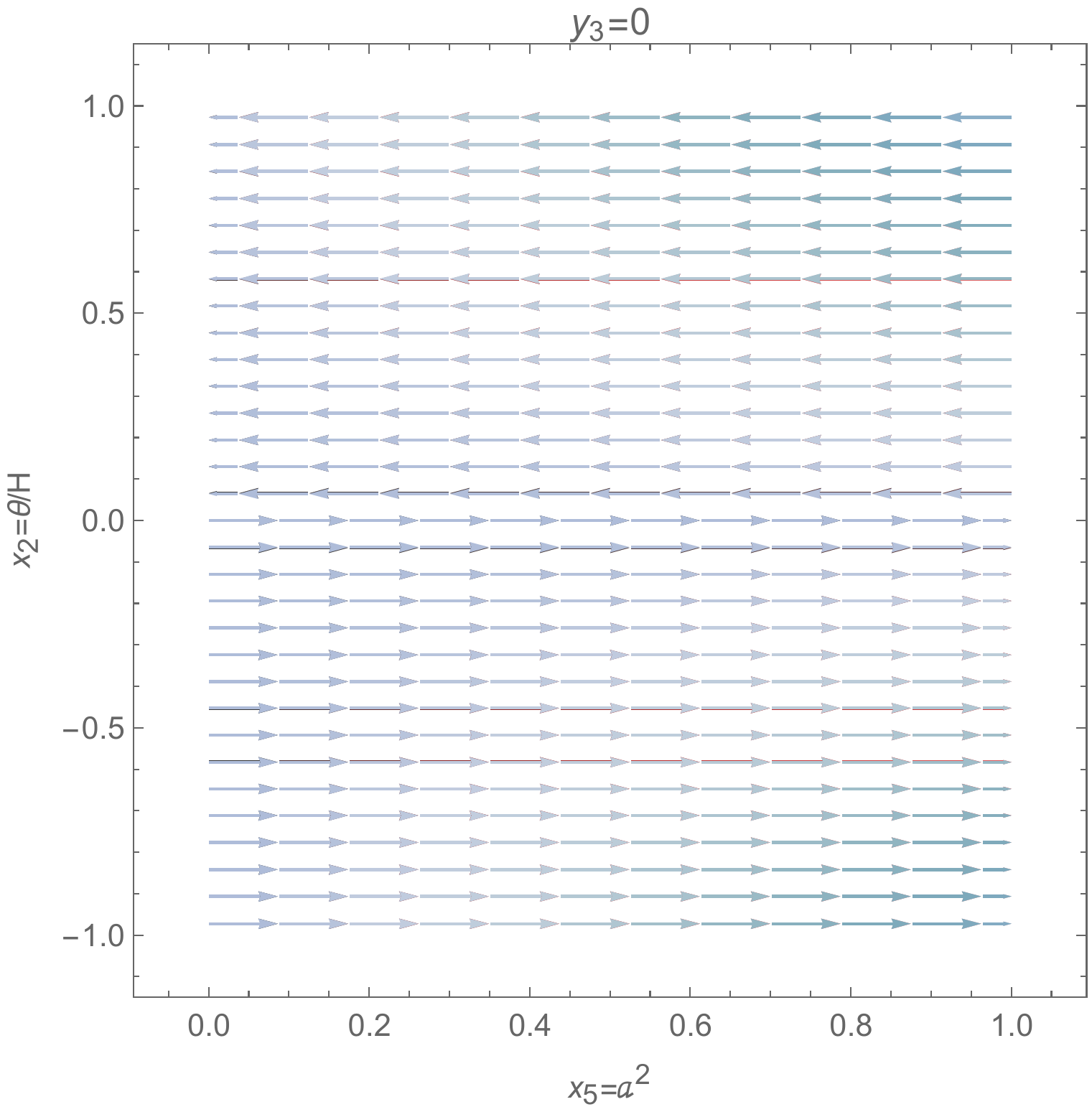}
\includegraphics[width=0.32\textwidth]{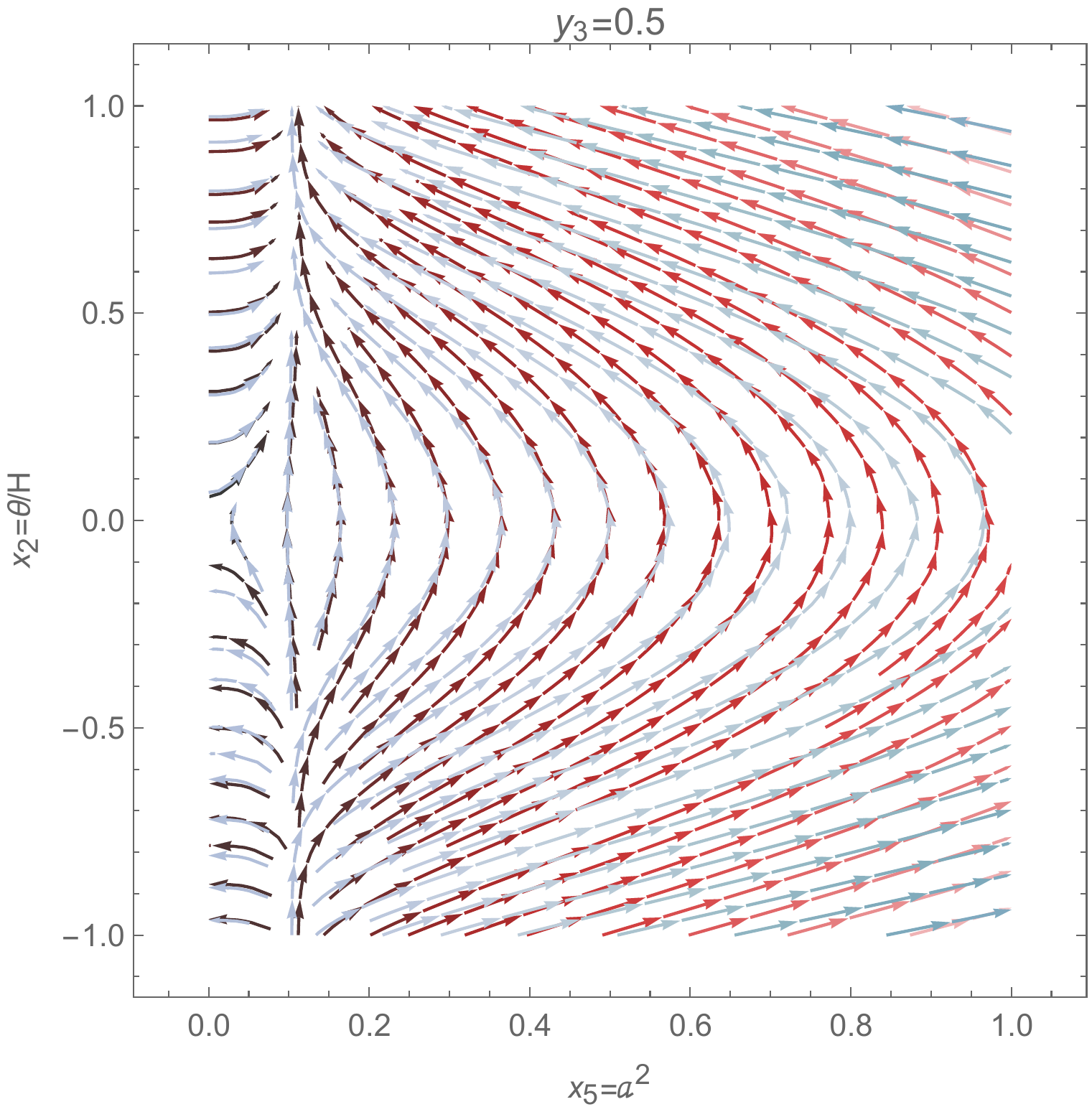}
\includegraphics[width=0.32\textwidth]{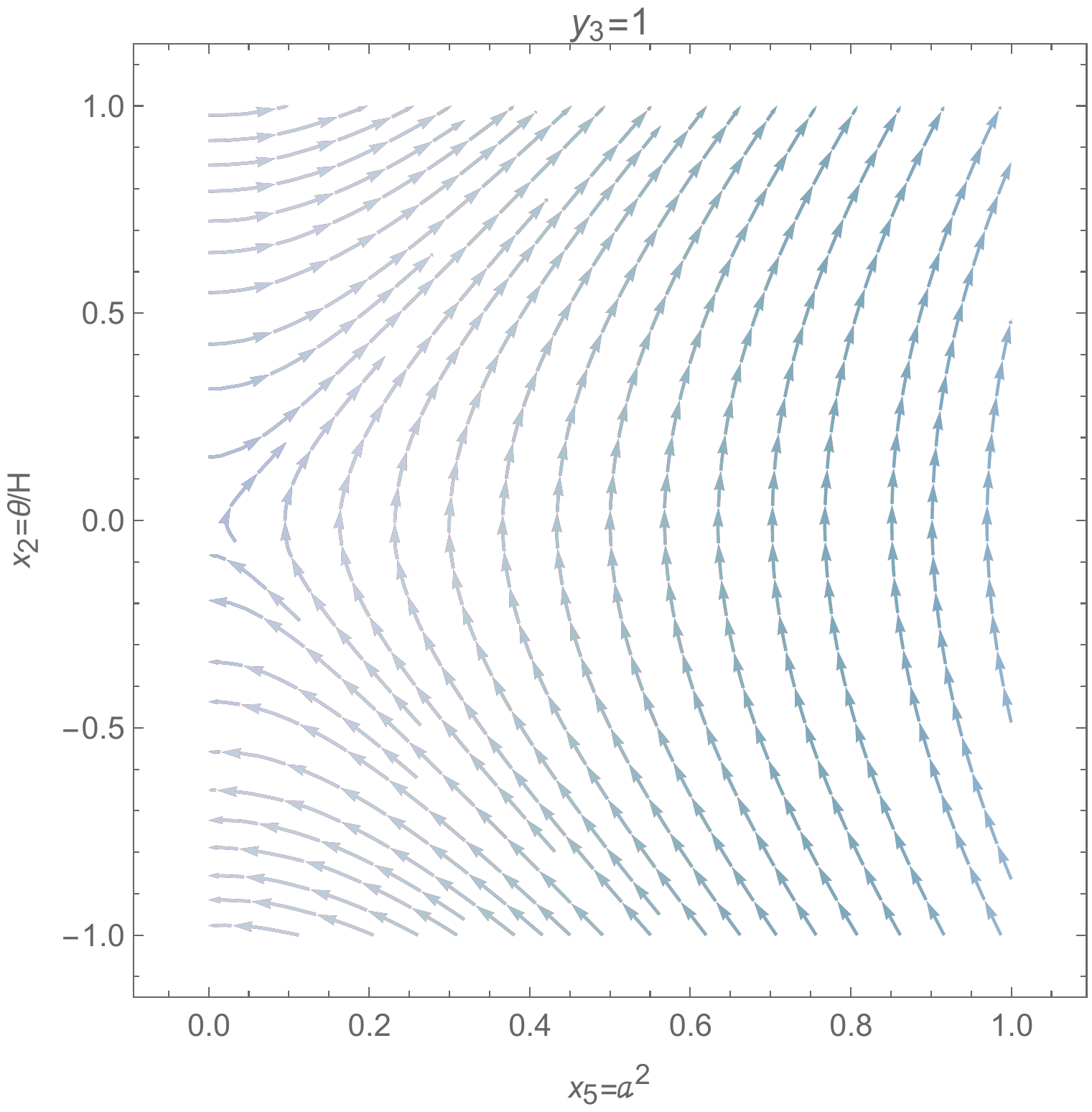}
\includegraphics[width=0.32\textwidth]{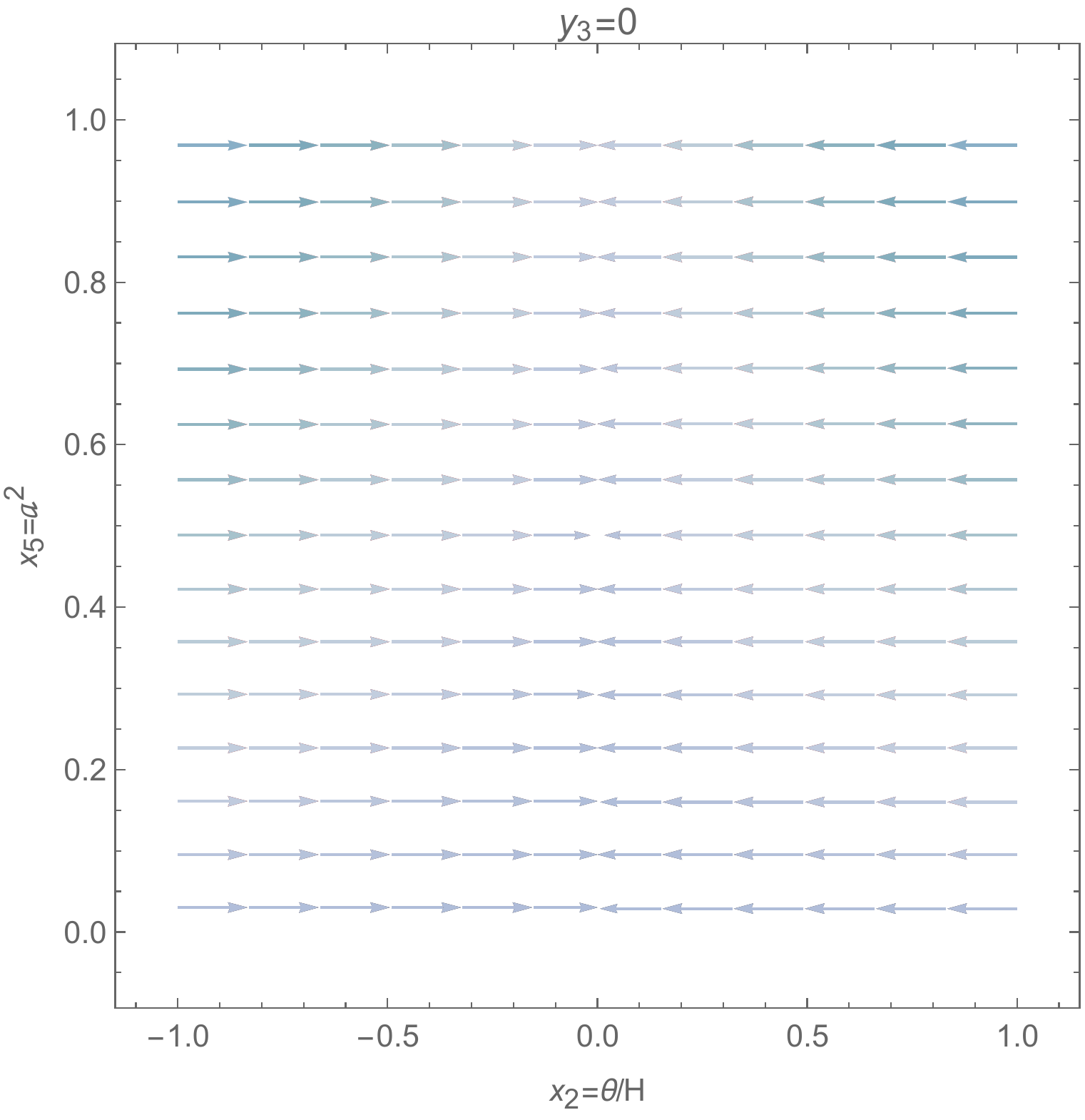}
\includegraphics[width=0.32\textwidth]{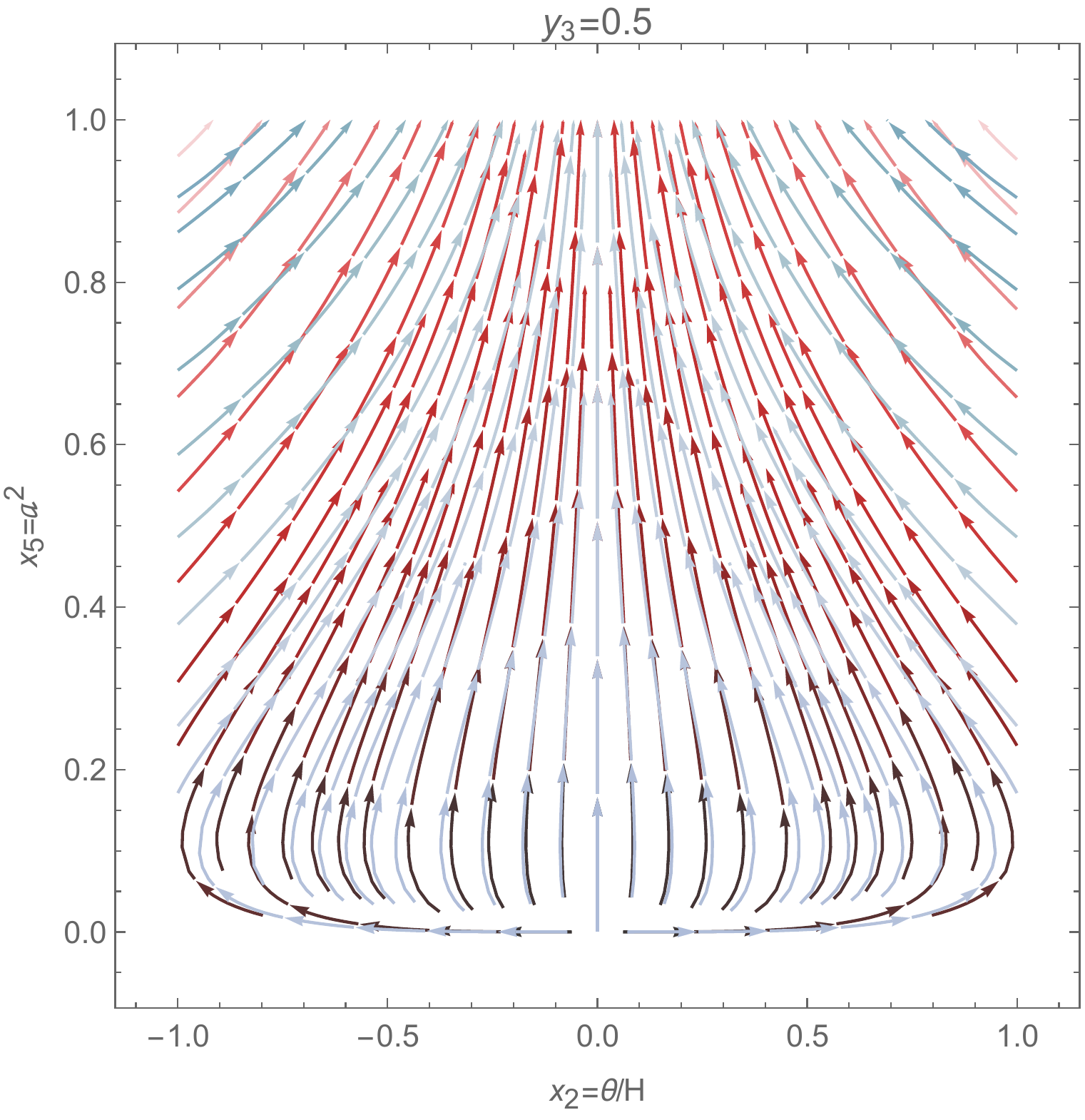}
\includegraphics[width=0.32\textwidth]{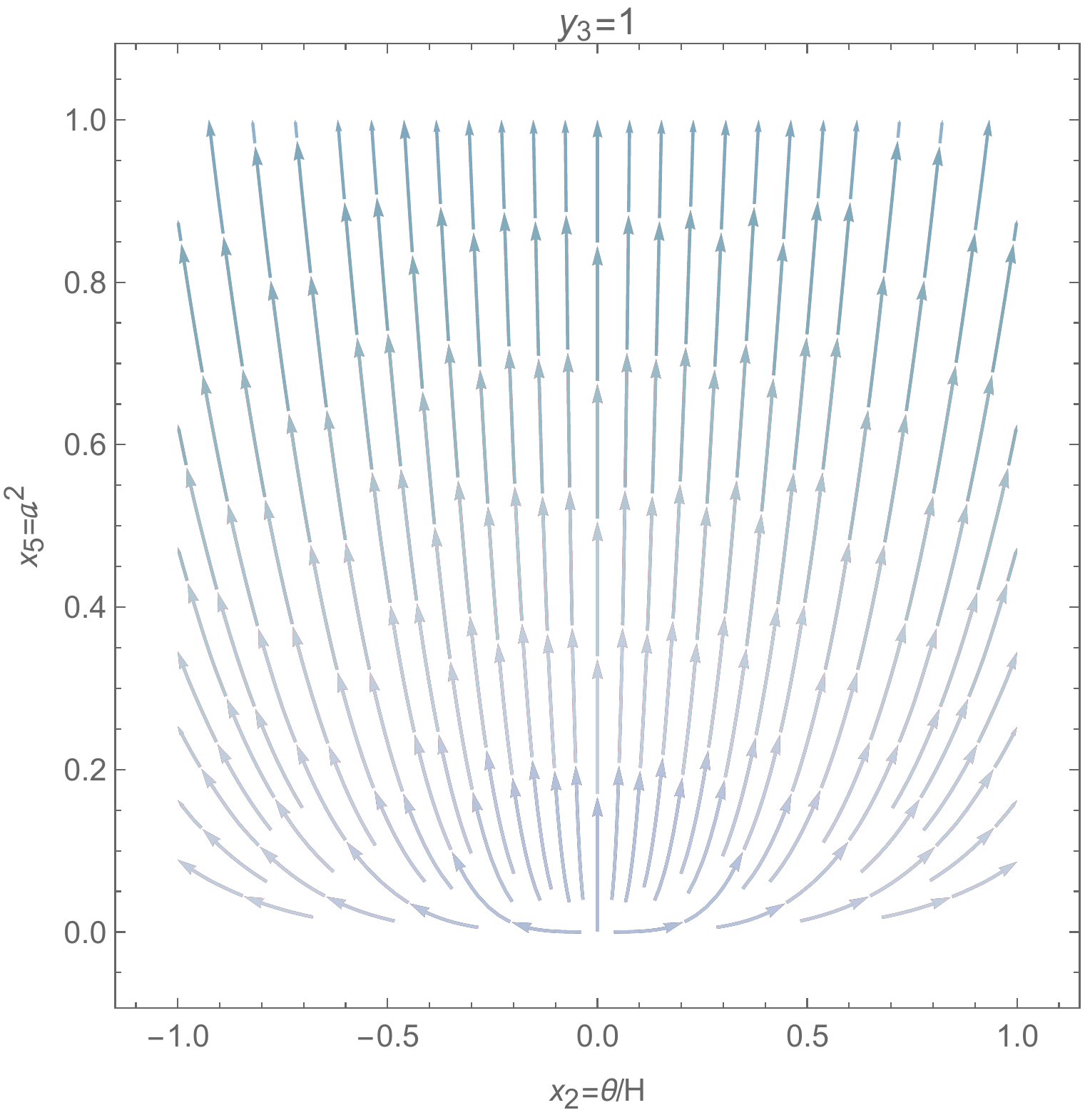}
\caption{Phase portraits for the velocity perturbations ($x_2=\theta/H$) vs. scale factor ($x_5=a^2$) in Chameleon-like models (blue) compared to $\Lambda$CDM (red), for $k=0.05h\text{Mpc}^{-1}$. Projection of the stream fields $x_2'-x_5'$ onto the planes $x_5-x_2$ (upper panel) and $x_2-x_5$ (bottom panel). The phase portraits show the behavior of the stream field of the velocity perturbations from $y_3=0$ to $y_3=1$. The critical point at (0,0) corresponds to the Big Bang (source).}
\label{fig:lcdm-cham}
\end{figure*}

In Figs. \ref{fig:phporx2} and \ref{fig:lcdm-fR} it is shown how the evolution of the stream-field of $x_2=\theta/H$ is completely different for $f(R)$ models to those from $\Lambda$CDM for different wavenumbers (in particular $k=0.012h\text{Mpc}^{-1}$ and $k=0.05h\text{Mpc}^{-1}$). In Fig. \ref{fig:lcdm-cham} the evolution of the stream-field of $x_2=\theta/H$ for Chameleon-lke models is shown for $k=0.05h\text{Mpc}^{-1}$. In this case the behavior of the stream field, although it does have differences compared to those from $\Lambda$CDM, they are very small compared to the differences obtained from $f(R)$ models. For the MG models considered in these phase portraits, the case $y_5=0$ seems to be stationary for the velocity field. There is only one critical point corresponding to the Big Bang. There are no critical points corresponding to the evolution of the perturbations in the future. The phase portraits correspond to $x_4=1$ (DM dominated). We can see how the point (0,0) is a source point for $x_2$, in all the cases there is a maximum in energy at this point when $y_5=1$ (which corresponds to $H=0$).
\subsection{Modified Gravity Solutions with Initial Conditions in a Vicinity of the Critical Points}
In this subsection we analyze the evolution of the perturbations and calculate the deviations of the MG models from $\Lambda$CDM, considering an arbitrary set of initial conditions that allows us to estimate the maximum extent of deviation.  
Specifically, we study solutions within both: Chameleon-like and $f(R)$ models at scales $k=0.012h\text{Mpc}^{-1}$ and $k=0.05h\text{Mpc}^{-1}$, with initial conditions close  to the different critical points previously derived. 

We solved the DS using as initial conditions random points laying in a squared vicinity centered at each critical point with size equal to $\epsilon=0.001$. Here we have fixed $\epsilon$ to the smallest possible value which at the same time is greater than the resolution of our solutions.  We name this set of points the \textit{squared vicinity} (SV) to a critical point. 
This setup  ensures an unbiased selection of initial conditions which shall allow us to explore how an initially random sample evolves within different models (see Table~\ref{tab:paramters}). 

\begin{figure*}[!]
\includegraphics[width=\textwidth]{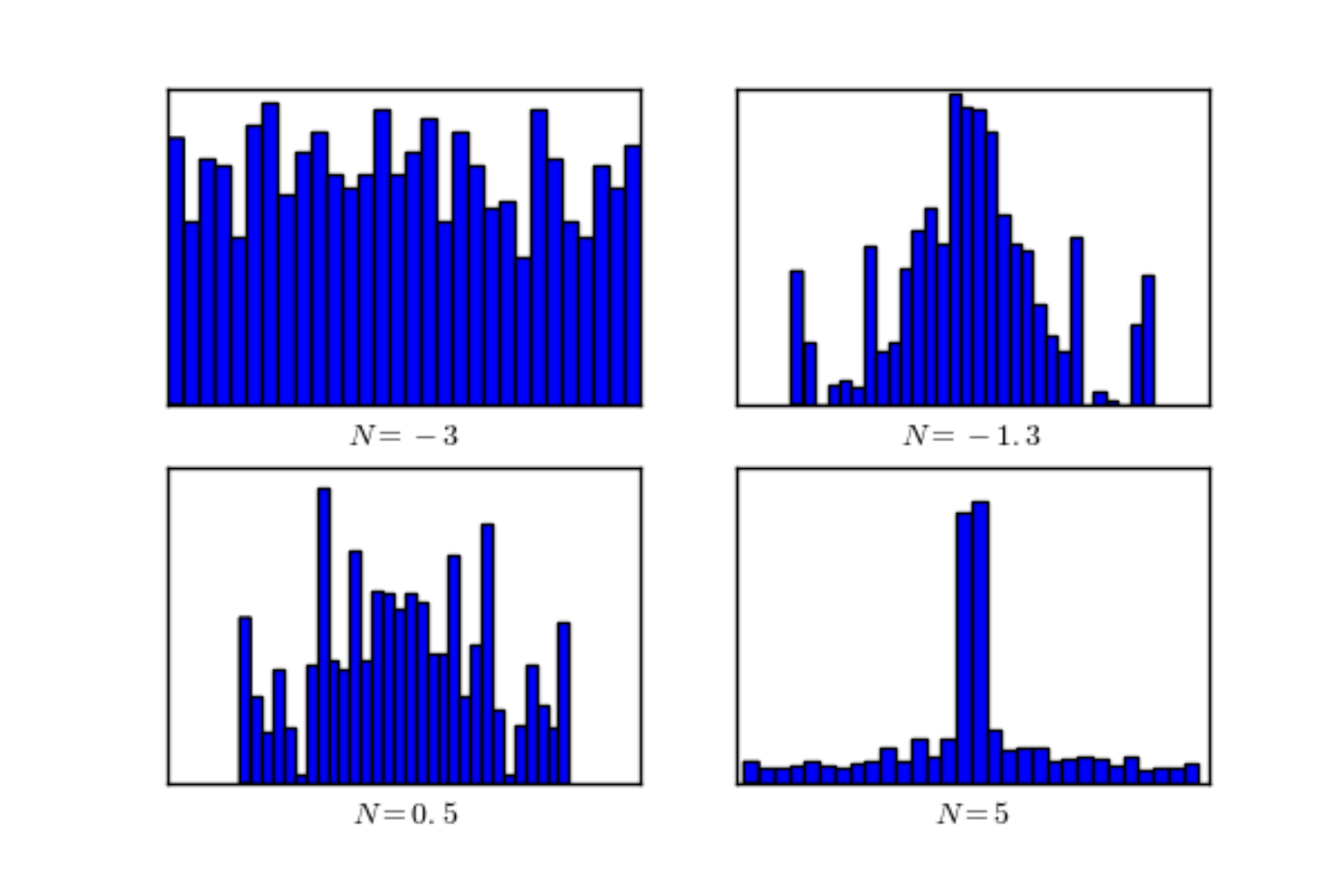}
\captionof{figure}{Distribution of points $x_2$ for different values of the e-folding parameter $N$.  The upper left panel shows the sample of initial conditions for $x_2$ randomly distributed in a vecinity of $P_2$ of size $\epsilon=0.001$. This particular direction corresponds to an ''AE'' vecinity. The upper right panel shows the sample of points corresponding to the solutions at a posterior cosmic time laying in the past. It can be clearly seen that the sample in no longer randomly distributed. As time passes by, the sample starts to move towards a single attractor point placed at zero.}
\label{fig:hist}
\end{figure*}

In order to get a better understanding about the behavior of the solutions around the critical points and to track possible bifurcations of dynamical flow-lines within specific regions of the phase space, we consider the evolution of a sample of initial points laying nearby $P_2$ and randomly distributed along a special direction associated with the only eigenvector of the Jacobian matrix with negative eigenvalue. We call them \textit{attractor eigenvectors} (AE). 
We have special interest in the subspace along the AE since their associated negative eigenvalue suggests a possible attractor behavior of the solutions projected onto them. This set of initial conditions is the \textit{AE-vicinity}, and as a case of study we consider solutions with initial points at the AE-vicinity for $P_2$. Fig.~\ref{fig:hist} illustrates how the initial randomly distributed sample of points for $x_2$ evolves as the e-folding parameter $N$ increases.  
For $N\sim -1$ clearly there is a bifurcation of the $x_2$ solutions which tend towards well bounded values. 
Later on, at $N\sim 0$ (corresponding to the present cosmic time) the bifurcation in three regions is still evident. 
At remote future times, all solutions are attracted towards an stationary point $x_2=0$ (at the center of the plot).     

In Table \ref{tab:paramters} we show the type of initial conditions considered for each critical point.
\begin{table*}[ht]
\begin{tabular}{|c|c|c|c|c|}
\hline
Critical point & AE & Corresponding &&\\&& Eigenvalues & Squared Vicinity & AE-Vicinity\\ 
\hline
$P_1=(0,0,0,1,0)$ & $ (0, 1, 0, 0, 0) $&$-4.345$&$(R,R,R,1-R,e^{2N_i})$&$(0,R,0,0,e^{2N_i})$\\
\hline
$P_2=(0,0,1,1,0)$ & $ (0, 1, 0, 0, 0) $&$-3$&$(R,R,1-R,1-R,e^{2N_i})$&$(0,R,0,0,e^{2N_i})$\\
&$ (0, 0, 0, 0, 1) $&$-1$&&\\
\hline
$P_3=(0,0,0,0,0)$ & $ (0, 1, 0, 0, 0) $&$-3$&$(R,R,R,R,e^{2N_i})$&$(0,R,0,R,e^{2N_i})$\\
&$ (0, 0, 0, 1, 0) $&$-1$&&\\
\hline
$P_4=(0,0,1,0,0)$ & $ (0, 1, 0, 0, 0) $&$-3$&$(R,R,1-R,R,e^{2N_i})$&$(0,R,0,0,e^{2N_i})$\\
\hline
\end{tabular}
\captionof{table}{Initial conditions around each critical point. The initial conditions were selected using the values $R = r\epsilon$, where $r$ is a random number between $(0,1)$ and $\epsilon=10^{-3}$. The integration starts at $\ln a=N_i=-4$, i.e., $x_5^\text{ini}=e^{-8}$. Only the initial condition having non-trivial AE are shown.}
\label{tab:paramters}
\end{table*}

An important remark to be made is that the solutions here shown are not necessarily those picked by observations, rather we aim to have a glance into the plethora of possible solutions laying nearby the critical points of the DS describing the dynamics of linear perturbations for parametrized models of gravity within the context of PPF, in order to be able to compare them to the $\Lambda$CDM model and understand the nature of the modifications to gravity.
\subsection{Signatures of Modified Gravity Through its Perturbations}
An important feature of $\Lambda$CDM perturbations in the linear regime -and therefore within the PPF context- is that they are coherent. 
Such quality clearly manifests in the DS, as the evolution of any mode is equivalent to any other. 
However, in typical MG theories this scale invariance is broken by construction when introducing the characteristic scale $\lambda_1$, which establishes a spatial turnover from which the modifications to the space-time geometry arise and become important. 
Therefore, linear perturbations in these models lose their coherence at some point regarding to the sort of phenomena they are intended to explain and this decoherence is manifested in their evolution by becoming scale-dependent.      
In this subsection we aim to study the scale-dependence of some modes for our considered parametrized $f(R)$ and Chameleon-like models close to the critical points and compare the outcomes from both models.
Secondly,  we quantify the percentage level at which the solutions deviate from the $\Lambda$CDM model close to each critical point.
\subsubsection{Quantitative comparison of MG gravity models against $\Lambda$CDM}\label{percentage comparison} 
In this section we show the deviations of the dimensionless perturbations $x_1=\delta$ and  $x_2=\theta/H$ from the $\Lambda$CDM case nearby the critical points. For this purpose and sake of clarity, we define the percentage deviation as
\begin{equation}
	\Delta_i = \left(\frac{x_i}{x_i^{(0)}} -1\right) 100\%,
\end{equation}
where $x_i^{(0)}$ is the dimensionless perturbation of the $\Lambda$CDM model. 

As we are treating with linear perturbations, the values of the embedded parameters of $\mu(a, \beta_1, \lambda_1, s, k)$ and $\gamma(a, \beta_2, \lambda_2, s, k)$ corresponding to this regime must be fixed.
We choose the values of $\beta_1$, $\beta_2$, $\lambda_1$, $\lambda_2$ and $s$ for the models $f(R)$ II and Chameleon-like II shown in Table \ref{tab:ST-models}. Then, for the mode $k$, we choose two common pivot modes: $k=0.012h\text{Mpc}^{-1}$ and $k= 0.05h\text{Mpc}^{-1}$; both are the limit cases where the linear regime is valid. Finally, to recover $\Lambda$CDM, we know that the particular cases $\lambda_1=\lambda_2=0$ boil down into $\mu = \gamma =1$.

\begin{figure*}[!]
\centering
\includegraphics[width=0.99\textwidth]{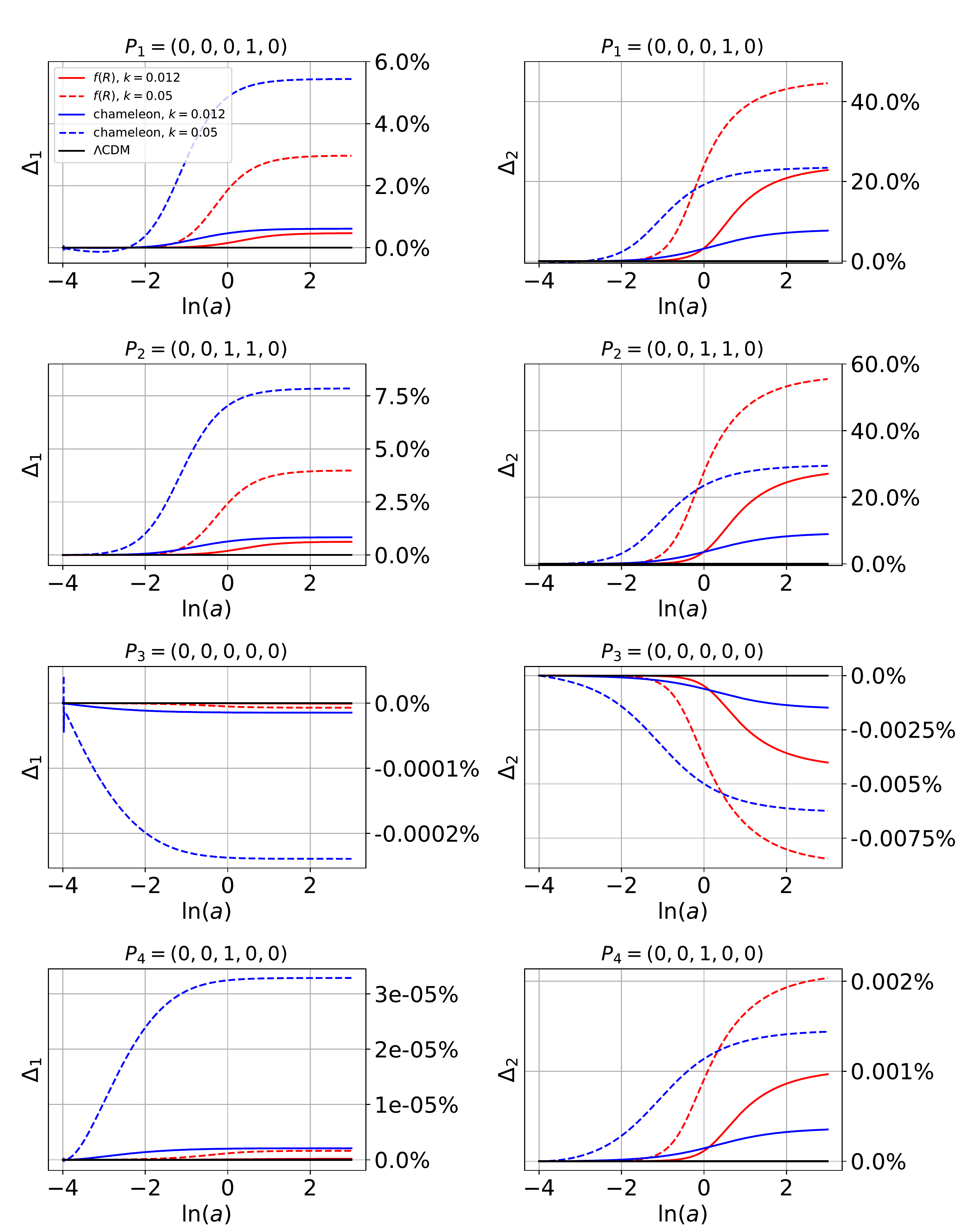} \captionof{figure}{Percentage deviations of the dimensionless perturbations close to the critical points in function of the scale factor. The $f(R)$-II model is shown in red, Chameleon-like II in blue and $\Lambda$CDM in black. Solid lines are for $k=0.012h\text{Mpc}^{-1}$ and dashed lines for $k=0.05h\text{Mpc}^{-1}$. For the initial conditions see Table~\ref{tab:paramters}.}
\label{fig:percentage}
\end{figure*}

Taking into account all this, we solve the differential equations of the DS numerically and plot the percentage deviations from the $\Lambda$CDM model using initial conditions close to the four critical points previously calculated. In Fig.~\ref{fig:percentage} the percentage deviations of the dimensionless perturbations from $\Lambda$CDM are plotted. 

The main features of $\Delta_1$ and $\Delta_2$ for each critical point are:
\begin{itemize}

\item For $\Delta_1$ in all critical points and values of $k$, the Chameleon-like models deviate more from $\Lambda$CDM than those coming from $f(R)$. 
Notice how the deviations are negative for $P_3$ and in all the other cases positive.  

\item For $\Delta_1$ in $P_1$ and $P_2$, both plots are qualitative the same, the deviations in $P_2$ are slightly bigger than in $P_1$. 
The deviations of $\Delta_1$ for early times start being negligible, then increase, first the Chameleon-like then $f(R)$, to then finally becoming almost constant. 
For the case of $k=0.05h\text{Mpc}^{-1}$, the gap between the final value of the perturbation in the future between the models is clear, however, for $k=0.012h\text{Mpc}^{-1}$ the final deviation for both models is almost the same. In this case the deviations are significant, the largest case being $\sim 7.5\%$ and the smallest $\sim 1\%$.

\item For $\Delta_1$ in $P_3$ and $P_4$, the deviations begin almost at the start of the integration interval.  
As mentioned before, for $P_3$ they are negative, i.e. the growth of perturbations in $\delta$ is slower than in $\Lambda$CDM.
In both cases, Chameleon-like models deviate more than those of $f(R)$ and the maximum deviation in the future (Chameleon-like with $k=0.05h\text{Mpc}^{-1}$) is around of $10^{-4}\% - 10^{-5}\%$. 
In contrast, for $f(R)$ they are almost equal to zero.
In these particular cases, $\Delta_1$ in $P_3$ and $P_4$, the deviations from the standard cosmological model are very small, being a challenge their possible observation. 

\item For $\Delta_2$ in all critical points, the deviations are small at early times and then start to grow (diminish for $P_3$). Chameleon-like models depart earlier than $f(R)$ models but for late times the deviations in $f(R)$ are larger for the same $k$ than those from Chameleon-like models. 
In both MG models larger values of $k$ (smaller wavelengths) arise larger deviations.
As in $\Delta_1$ the deviations are negative for $P_3$ and in all the other cases positive.  

\item For $\Delta_2$ in $P_1$ and $P_2$, both plots are qualitative the same. 
The deviations are clearly significant, the largest cases is $\sim 60\%$ and the smallest $\sim 10\%$ in the future.
This is the most notorious case that differs from $\Lambda$CDM. Independently of the model or $k$ the growth of perturbation is clearly larger than in the standard case.

\item For $\Delta_2$ in $P_3$ and $P_4$, the largest deviations are around $\sim10^{-3}\%$, larger than its $\Delta_1$ pair but still very small.

\item In all cases the Chameleon-like model deviations occur earlier than those coming from $f(R)$, owing to the shift in values of $s$ between models that consequently shifts the starting stage of the modification. 

\item In all cases, the deviations are bigger for $k=0.05h\text{Mpc}^{-1}$ than to the counterpart for $k=0.012h\text{Mpc}^{-1}$.
\end{itemize}

This analysis is mainly qualitative looking for the main physical differences between the models, quantitatively the main purpose is to calculate the orders of magnitude of the deviations. The approach clearly shows that the DS machinery provides a powerful tool to test the sensitivity of perturbations to variations of a given parametric model. At the same time it is helpful to determine which kind of phenomena may possibly show up detectable signatures of a given model.  

Finally, let us make the analysis regarding the scale-dependence of the gravitational potentials. By construction, either in parametrized Chameleon-like or $f(R)$ models, the main modifications to $\Lambda$CDM are introduced through the perturbations in the metric and in a scale-dependent way in the perturbed variables (as we have seen in the results of the previous sections).
 
\begin{figure*}
\includegraphics[width=0.99\textwidth]{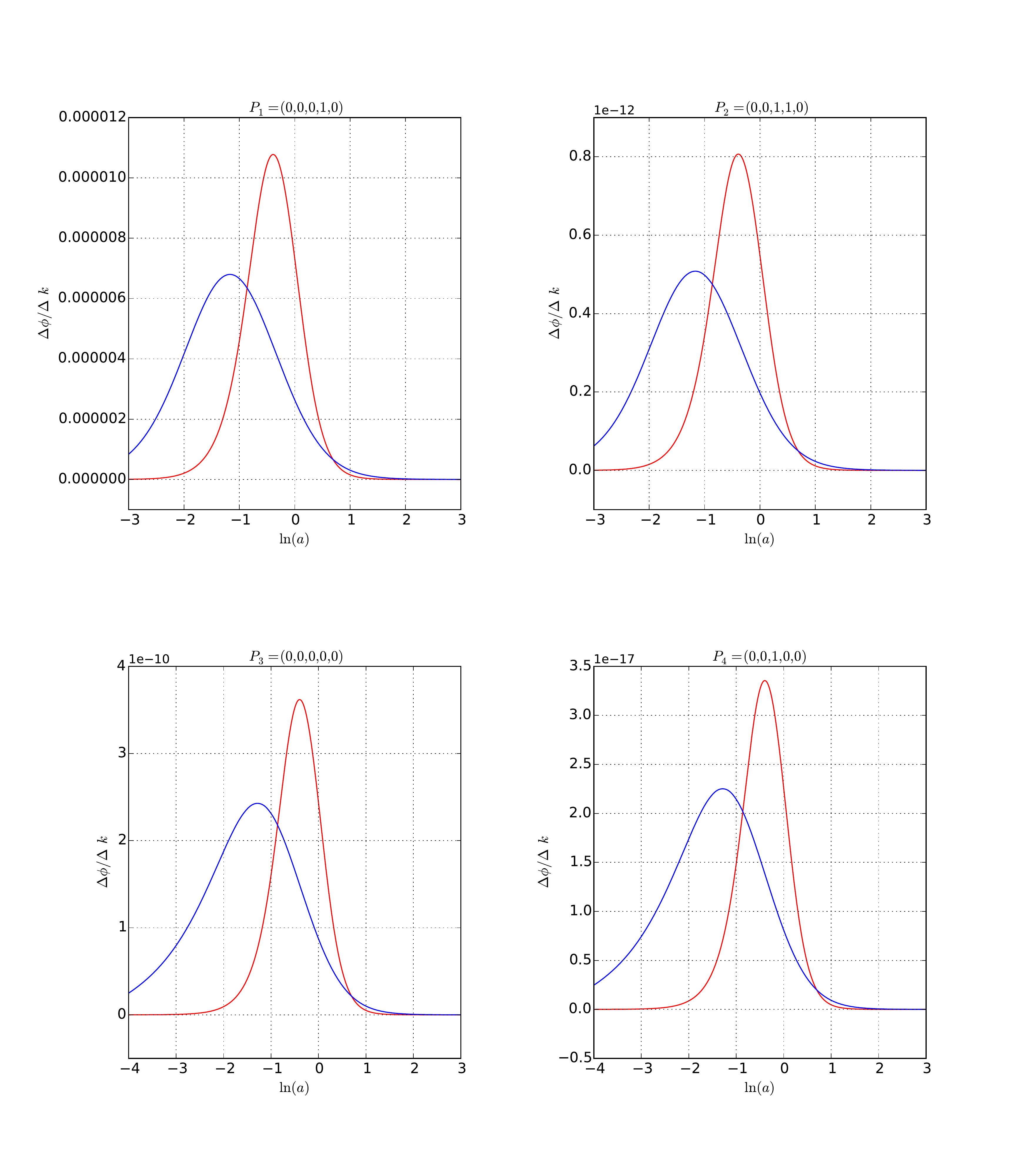} 
\captionof{figure}{Variation of the gravitational potential $\phi(\delta_m,\theta_m)$ with respect to the variation of the wavenumber for $f(R)$ (red) and Chameleon-like (blue) models at each critical point for an arbitrary initial condition.}
\label{fig:x4k}
\end{figure*}

As mentioned before, both dynamical variables describing the density contrast and peculiar velocity perturbations depend on the scale within different models with different initial conditions nearby each critical point. Then, the gravitational potentials, which are derived physical quantities from the dynamical variables via the constraint Einstein equations, naturally inherit such scale dependence. Fig.~\ref{fig:x4k} shows the variation of the gravitational potential $\phi$ with respect to the variation of the wavenumber at different e-folds at the different critical points for the different models. As it can be seen from Fig.\ref{fig:x4k}, it happens that the maximum variation occurs in the past, before for Chameleon-like models than for $f(R)$. In the latter the scale-dependence is clearly larger than in the former which is consistent with our previous results. We find that solutions with initial conditions in the SV of $P_1$ have the maximum variations with respect to the scale. 
\section{Conclusions}
\label{C} 
In this work a DS analysis is carried out for the first time for linear perturbed cosmologies. We analyzed different MG models in the QS limit and employing the PPF formalism that assumes a background behavior identical to that of the $\Lambda$CDM model and parametrizes linear perturbations. We analyzed both $k$-independent and $k$-dependent MG models.

A set of variables that allows us to transform our system of differential equations into a finite-dimensionless system has been introduced. The most well known way to do this is by parametrizing the system through the scale factor or the Hubble parameter. Some particular cases were then studied, finally arriving to the more general models of MG within the PPF scheme. 

The machinery set up in this work is general and easily applicable to other cases. Critical points, their existence and their corresponding eigenvalues are shown in Tables~\ref{tab-cdm} and~\ref{tablcdm} for different models which include $f(R)$ and Chameleon-like models of MG and $\Lambda$CDM with respect to which the numerical analysis and comparison is made. After finding the critical points, an stability analysis was carried out for each one of them and their stability and cosmological characteristics where displayed.

In Sec.~\ref{percentage comparison}, the percentage deviations estimated for different initial conditions and scales of the dimensionless perturbations from the $\Lambda$CDM case nearby the critical points of the DS were shown. The differences in the evolution of the perturbation between different critical points are subtle, therefore deviations between $\Lambda$CDM and the MG models considered mainly appear when varying the scale $k$ of the perturbation and the MG model considered.

The deviations in the density contrast, $x_1=\delta$, evolve from zero in the past to an stationary value in the future. Small-scale perturbations (i.e. $k=0.05h\text{Mpc}^{-1}$) present larger deviations with respect to $\Lambda$CDM compared to those with smaller wavenumbers (larger scales, $k=0.012h\text{Mpc}^{-1}$). 

Chameleon-like perturbations start deviating earlier than those of $f(R)$. An extreme case appears around the critical points $P_3$ and $P_4$ (DE dominated universes) in $x_1$, for which there is almost no difference with $\Lambda$CDM at large scales ($k=0.012h\text{Mpc}^{-1}$). The largest deviations appear around the critical points $P_1$ and $P_2$ (DM dominated universes) being around a $10\%$ difference compared to $\Lambda$CDM.

$f(R)$ perturbations deviate at later times than those of Chameleon-like models. One of its extreme cases also appears in $P_3$ and $P_4$ for $x_1$, for which there is also almost no difference with $\Lambda$CDM at small scales ($k=0.05h\text{Mpc}^{-1}$). The largest deviations in this case also appear around the points $P_1$ and $P_2$ having around a $5\%$ difference compared to the $\Lambda$CDM model.

Deviations in the velocity divergence over the Hubble parameter, $x_2=\theta/H$, evolve from zero and seem to increase with the same rate for the different scales considered. In comparison to perturbations in $x_1$, it is not clear if they reach an stationary value in the future. Small-scale deviations ($k=0.05h\text{Mpc}^{-1}$) departure first and reach higher values. Similar to the case of $x_1$, for $P_3$ and $P_4$ the deviations from the $\Lambda$CDM model are small (but bigger than $x_1$) being around $10^{-3}\%$ and the largest deviations also appear around $P_1$ and $P_2$ having a difference of $60\%$ for both MG models.

In addition, we studied the behavoir of a random distribution of solutions of $x_2$ with  initial conditions in AE vecinity of $P_2$. The initially homogeneous sample of points seems to bifurcate at early times towards three mains regions and they all tend to a stationary point at zero in the future. We illustrated this peculiar behavoir in Figure \ref{fig:hist}.

In summary, for the couple of MG models studied in this work inside the PPF formalism ($f(R)$ and Chameleon-like) we obtained that the most sensitive perturbation at linear order from which a mayor difference appears compared to that same perturbation coming from the $\Lambda$CDM model is found in the variable $x_2$, corresponding to the perturbations in the peculiar velocity. Such deviations manifests themselves most importantly for Chameleon-like models at earlier times, becoming more important at late time for $f(R)$ models, see Fig.\ref{fig:percentage}. On the contrary, perturbations like the density contrast ($x_1$) do not show significant deviations compared to those coming from the $\Lambda$CDM model. The largest difference being around $10\%$ in Chameleon-like models for scales $k=0.05h\text{Mpc}^{-1}$.

Finally, when analysing the variations of the gravitational potential $\phi$ with respect to the scale in the couple of MG models considered, we found that the ratio $\Delta\phi/\Delta k$ results very small within the QS approximation, the largest of them found around the critical point $P_3$ with an amplitude of $\sim10^{-5}$. When analysing such perturbations compared to those of the $\Lambda$CDM model through the vector fields of the phase portraits, no significant difference was found.

The parametrizations here used are valid from horizon scales down to the scales at which nonlinearities become meaningful. This spans a wide range of observables aimed by the next generation of experiments \citep{desi,des,lsst}, including weak lensing (of galaxies and CMB), redshift space distortions, peculiar velocities surveys, the ISW effect and associated cross-correlations and galaxy clusters \citep{thomas}. The $(\beta,s)$ parameters of scale-independent MG given in Section~\ref{sip}, and the parameters ($\lambda_i,\beta_i$) of the scale-dependent MG models presented in Section~\ref{scdp}, are convenient for testing observational data. If measurements of galaxy clustering, peculiar velocities, and weak lensing are all consistent with $\lambda_i=\beta_i=0$, for example, then MG and exotic DE models can both be ruled out. If measurements need nonzero parameters, however, DE and MG continue to be possible solutions until extra hypothesis are made to differentiate between them (for other constraints see e. g.  \cite{Linder:2016wqw}).

The presented analysis of the percentage deviations is a useful tool because it shows the scale and MG model dependence of the scalar perturbations giving valuable clues for subsequent investigations which pursue some detailed information about scalar perturbations in MG models. In our results we conclude that measurements of peculiar velocities are the most important trackers to discriminate between models of MG to $\Lambda$CDM. Also, from the analysis of the phase portraits it is found that scale-independent MG models like Brans-Dicke evolve essentially in the same way as $\Lambda$CDM does as far as their linear perturbation is concerned. And although scale-dependent models do show differences with respect to the $\Lambda$CDM model, $f(R)$ is found to be the model with mayor differences in the evolution of its perturbations. The aim of this paper was to show a first general qualitative analysis in this theme; details such as parameter constraints according to observations is outside of the objectives and scope of this formalism.
\acknowledgments
This work was partially supported by Consejo Nacional de Ciencia y Tecnolog\'ia (CONACyT), Mexico, under grants: CB-2014-01 
No.~240512, Projects No.~269652 and 283151, and Fronteras Project 281. Also by Xiuhcoatl and Abacus clusters at Cinvestav, IPN. 
D.T. acknowledges financial support from CONACyT Mexico postdoctoral fellowship. A.A acknowledges financial support from CONACyT Mexico within ``Retenci\'on y Repatriaci\'on'' Program and from Vicerrector\'ia de Investigaci\'on y Estudios de Posgrado BUAP.


\begin{thebibliography}{99}
\bibitem{riess2} A. G. Riess, A. V. Filippenko, P. Challis {\itshape et al.}, Astron. J. {\bf 116}, 1009 (1998).
\bibitem{perlmutter2} S. Perlmutter, G. Aldering, G. Goldhaber {\itshape et al.}, Astrophys. J. {\bf 517}, 565 (1999).
\bibitem{betoule} M. Betoule, R. Kessler, J. Guy {\itshape et al.}, Astron. Astrophys. {\bf 568}, A22 (2014).
\bibitem{komatsu} E. Komatsu, K. M. Smith, J. Dunkley {\itshape et al.}, Astrophys. J. Suppl. {\bf 192}, 2 (2011).
\bibitem{sullivan} M. Sullivan, J. Guy, A. Conley {\itshape et al.}, ApJ {\bf 737}, 102 (2011).
\bibitem{reid} B. A. Reid, W. J. Percival, D. J. Eisenstein {\itshape et al.}, Mon. Not. R. Astron. Soc. {\bf 404}, 60 (2010).
\bibitem{sahni} V. Sahni and A. A. Starobinsky, Int. J. Mod. Phys. D {\bf 9}, 373 (2000).
\bibitem{peebles} P. J. E. Peebles and B. Ratra, Rev. Mod. Phys. {\bf 75}, 559 (2003).
\bibitem{padmanabhan} T. Padmanabhan, Phys. Rep. {\bf 380}, 235 (2003).
\bibitem{planck} P. A. R. Ade {\itshape et al.}, Astron. Astrophys. {\bf 594}, A16 (2016). https://pla.esac.esa.int/pla.
\bibitem{copeland1} E. Copeland, M. Sami and S. Tsujikawa, Int. J. Mod. Phys. D {\bf 15}, 1753 (2006).
\bibitem{amendola2} L. Amendola and S. Tsujikawa, {\itshape Dark Energy: Theory and Observations}, Cambridge University Press (2010).
\bibitem{Will} C. M. Will, Living Rev. Rel. {\bf 17}, 4 (2014).
\bibitem{Berti} E. Berti, E. Barausse, V. Cardoso {\itshape et al.}, Class. Quantum Grav. {\bf 32}, 243001 (2015).
\bibitem{moresco} M. Moresco, A. Cimatti, R. Jimenez {\it et al.}, J. Cosm. Astroparticle Phys. {\bf 1208}, 006 (2012).
\bibitem{gil} H. Gil-Mar\'in {\it et al.}, Mon. Not. Roy. Astron. Soc. {\bf 460}, 4210 (2016).
\bibitem{percival} W. J. Percival, L. Samushia, A. J. Ross {\it et al.}, Phil. Trans. R. Soc. A {\bf 369}, 5058 (2018).
\bibitem{brans} C. Brans and R. Dicke, Phys. Rev. {\bf 124}, 925 (1961).
\bibitem{amendola1} L. Amendola, Phys. Rev. D {\bf 60}, 043501 (1999).
\bibitem{tsujikawa} S. Tsujikawa, K. Uddin, S. Mizuno {\itshape et al.}, Phys. Rev. D {\bf 77}, 103009 (2008).
\bibitem{defelice} A. De Felice and S. Tsujikawa, Living Rev. Rel. {\bf 13}, 3 (2010).
\bibitem{nojiri} S. Nojiri and S. D. Odintsov,
Phys. Rep. {\bf 505}, 59 (2011).
\bibitem{nojiri2} S. Nojiri and S. D. Odintsov, ECONF C0602061:06, Int. J. Geom. Meth. Mod. Phys. {\bf 4}, 115 (2007).
\bibitem{Horndeski:1974wa} G. W. Horndeski, Int. J. Theor. Phys. {\bf 10}, 363 (1974).
\bibitem{Deffayet:2009mn} C. Deffayet, S. Deser and G. Esposito-Farese, Phys. Rev. D {\bf 80}, 064015 (2009).
\bibitem{Charmousis:2011bf} C. Charmousis, E. J. Copeland, A. Padilla, {\itshape et al.}, Phys. Rev. Lett. {\bf 108}, 051101 (2012).
\bibitem{Koyama:2015vza} K. Koyama, Rept. Prog. Phys. {\bf 79}, 046902 (2016).
\bibitem{capozziello} S. Capozziello, V. F. Cardone and A. Troisi, Phys. Rev. D {\bf 71}, 043503 (2005).
\bibitem{nojiri1} S. Nojiri and S. D. Odintsov, Phys. Rev. D {\bf 74}, 086005 (2006).
\bibitem{song} Y. S. Song, W. Hu and I. Sawicki, Phys. Rev. D {\bf 75}, 044004 (2007).
\bibitem{des} The Dark Energy Survey Data Release 1 DES Collaboration, T.M.C. Abbott (Cerro-Tololo InterAmerican Obs.) {\itshape et al.}, FERMILAB-PUB-17-603-AE-E 28 (2018). http://www.darkenergysurvey.org/
\bibitem{desi} The DESI Experiment Part I: Science, Targeting and Survey Design, arXiv:1611:00036. http://desi.lbl.gov.
\bibitem{euclid} http://sci.esa.int/euclid. Cosmology and fundamental physics with the Euclid satellite,
Euclid Theory Working Group (Luca Amendola et al.). 2012. 224 pp. 
Published in Living Rev.Rel. 16 (2013) 6.
\bibitem{lsst} LSST Dark Energy Science Collaboration, arXiv:1211.0310 (2012). http://www.lsst.org.
\bibitem{hu} W. Hu and I. Sawicki, Phys. Rev. D {\bf 76}, 104043 (2007).
\bibitem{Baker:2012zs} T. Baker, P. G. Ferreira and C. Skordis, Phys. Rev. D {\bf 87}, 024015 (2013).
\bibitem{bertschinger1} E. Bertschinger and P. Zukin, Phys. Rev. D {\bf 78}, 024015 (2008).
\bibitem{linder1} E. V. Linder and R. N. Cahn. Astroparticle Phys. {\bf 28(4)}, 481 (2007).
\bibitem{carloni} S. Carloni, P. K. S. Dunsby, S. Capozziello {\itshape et al.}, Class. Quantum Grav. {\bf 22}, 4839 (2005).
\bibitem{carloni1} S. Carloni, J. Leach, S. Capozziello {\itshape et al.}, Class. Quantum Grav. {\bf 25}, 035008 (2008).
\bibitem{bahamonde} S. Bahamonde, C. G. Boehmer, S. Carloni {\itshape et al.}, Preprint arXiv: gr-qc 1712.03107.
\bibitem{wainwright} J. Wainwright and G. F. R. Ellis, {\itshape Dynamical Systems in Cosmology}, Cambridge University Press (1997).
\bibitem{coley} A. A. Coley, {\itshape Dynamical Systems and Cosmology}, Astrophysics and Space Science Library, Springer (2003).
\bibitem{wainwright1} J. Wainwright and W. C. Lim, J. Hyperbol. Diff. Eq. {\bf 2}, 437 (2005).
\bibitem{clifton} T. Clifton, P. G. Ferreira, A. Padilla {\itshape et al.}, Phys. Rept. {\bf 513}, 1 (2012).
\bibitem{bean} R. Bean and M. Tangmatitham, Phys. Rev. D {\bf 81}, 083534 (2010).
\bibitem{hojjati} A. Hojjati, L. Pogosian and G.-B. Zhao, J. Cosm. Astroparticle Phys. {\bf 1108}, 005 (2011).
\bibitem{amin} M. A. Amin, R. V. Wagoner and R. D. Blandford,Mon. Not. R. Astron. Soc. {\bf 390}, 131 (2008).
\bibitem{silvestri} A. Silvestri, L. Pogosian and R. V. Buniy, Phys. Rev. D {\bf 87}, 104015 (2013).
\bibitem{defelice1} A. De Felice, T. Kobayashi and S. Tsujikawa, Phys. Lett. B {\bf 706}, 123 (2011).
\bibitem{zhao} G.-B. Zhao, L. Pogosian, A. Silvestri {\itshape et al.}, Phys. Rev. D {\bf 79}, 083513 (2009).
\bibitem{szydlowski} M. Szydlowski and O. Hrycyna, Gen. Relativ. Gravit. {\bf 38}, 121 (2006).
\bibitem{strogatz} S. H. Strogatz, {\itshape Nonlinear Dynamics And Chaos: With Applications To Physics, Biology, Chemistry And Engineering}, Perseus Books (2001).
\bibitem{copeland} E. J. Copeland, M. Sami and S. Tsujikawa, Int. J. Mod. Phys. D {\bf 15}, 1753 (2006).
\bibitem{yoo} J. Yoo, A. L. Fitzpatrick and M. Zaldarriaga, Phys. Rev. D {\bf 80}, 083514 (2009).
\bibitem{yoo1} J. Yoo, Phys. Rev. D {\bf 82}, 083508 (2010).
\bibitem{challinor} A. Challinor and A. Lewis, Phys. Rev. D {\bf 84}, 043516 (2011).
\bibitem{thomas} D. B. Thomas and C. R. Contaldi, J. Cosm. Astroparticle Phys. {\bf 12}, 013 (2011).
\bibitem{Linder:2016wqw} E.~V.~Linder, Phys.\ Rev.\ D {\bf 95}, no. 2, 023518 (2017).
\end{thebibliography}
\end{document}